\documentclass[twocolumn]{revtex4-2}
\usepackage[utf8]{inputenc}
\usepackage{graphicx}
\usepackage{subfigure}
\usepackage{float}
\usepackage{braket}
\usepackage{amsmath}
\usepackage{hyperref}
\usepackage{color}

\begin{document}
\title{High-throughput discovery of high-temperature conventional superconductors}

\author{Alice M. Shipley}
\email{ams277@cam.ac.uk}
\thanks{These authors contributed equally.}
\affiliation
{
    Theory of Condensed Matter Group,
    Cavendish Laboratory,
    J.~J.~Thomson Avenue,
    Cambridge CB3 0HE,
    United Kingdom
}

\author{Michael J. Hutcheon}
\email{mjh261@cam.ac.uk}
\thanks{These authors contributed equally.}
\affiliation
{
    Theory of Condensed Matter Group,
    Cavendish Laboratory,
    J.~J.~Thomson Avenue,
    Cambridge CB3 0HE,
    United Kingdom
}

\author{Richard J. Needs}
\affiliation
{
    Theory of Condensed Matter Group,
    Cavendish Laboratory,
    J.~J.~Thomson Avenue,
    Cambridge CB3 0HE,
    United Kingdom
}

\author{Chris J. Pickard}
\email{cjp20@cam.ac.uk}
\affiliation
{
    Department of Materials Science and Metallurgy,
    27 Charles Babbage Rd,
    Cambridge CB3 0FS,
    United Kingdom
}
\affiliation
{
    Advanced Institute for Materials Research, 
    Tohoku University, 2-1-1 Katahira, 
    Aoba, Sendai, 980-8577, 
    Japan
}

\date{\today}

\begin{abstract}
We survey the landscape of binary hydrides across the entire periodic table from 10 to 500\ GPa using a crystal structure prediction method. Building a critical temperature ($T_c$) model, with inputs arising from density of states calculations and Gaspari-Gyorffy theory, allows us to predict which energetically competitive candidates are most promising for high-$T_c$ superconductivity. Implementing optimisations, which lead to an order of magnitude speed-up for electron-phonon calculations, then allows us to perform an unprecedented number of ``high-throughput" calculations of $T_c$ based on these predictions and to refine the model in an iterative manner. Converged electron-phonon calculations are performed for 121 of the best candidates from the final model. From these, we identify 36 above-100\ K dynamically stable superconductors. To the best of our knowledge, superconductivity has not been previously studied in 27 of these. Of the 36, 18 exhibit superconductivity above 200 K, including structures of NaH$_6$ (248-279\ K) and CaH$_6$ (216-253\ K) at the relatively low pressure of 100\ GPa.
\end{abstract}

\maketitle

\section{Introduction}
A number of dense metal hydrides have been shown to be conventional superconductors with high critical temperatures ($T_c$) \cite{duan2017structure, zurek2019high, flores2019perspective, boeri2019, oganov2019, pickard2019}. However, electron-phonon calculations, used to determine the $T_c$ of such materials from first principles, remain computationally expensive. It is not necessarily clear \textit{before} performing these calculations which particular systems and structures might exhibit high-temperature superconductivity. Coupled with this, a huge variety of hydride stoichiometries and structures are either stable or metastable under pressure, even when the discussion is limited solely to binaries. This means that exhaustive theoretical investigation of this class of materials is a huge challenge. A combined searching and screening protocol could therefore provide a desirable and efficient way to discover new high-$T_c$ superconductors.

In this work, we use a crystal structure prediction method to explore binary hydrides of elements from across the entire periodic table over a 10 to 500\ GPa pressure range. Stable and metastable structures are screened using a model built to predict $T_c$ from inputs including electronic density of states (DOS) ratios and electron-phonon coupling estimates from Gaspari-Gyorffy theory \cite{gaspari1972}. High-throughput electron-phonon calculations are then performed for a large number of structures, selected based on the predictions of this model. This provides more $T_c$ results which can be fed back into the training set, allowing us to improve the model iteratively. The results of this high-throughput model-training stage allow us to identify the most promising candidate systems at each pressure; more thorough structure searching is then carried out for each of these systems and fully converged electron-phonon calculations are performed for the best predicted candidates. A large number of high-$T_c$ superconductors are efficiently identified, including many not reported in previous work.

\section{Methodology}
\subsection{Initial structure searching}
The structure searching calculations in this work were performed using \textit{ab initio} random structure searching (AIRSS) \cite{pickard2006, pickard2011} and the plane-wave pseudopotential code \textsc{castep} \cite{castep2005}. The Perdew-Burke-Ernzerhof (PBE) generalised gradient approximation \cite{pbe1996}, \textsc{castep} QC5 pseudopotentials, a 340\ eV plane-wave cut-off and a $\mathbf{k}$-point spacing of $2\pi\times$0.07\ \r{A}$^{-1}$ were used. For the initial searches, sp-AIRSS \cite{monserrat2018v} was utilised and structures with 8-48 symmetry operations were generated. This served the dual purpose of (1) reducing the computational cost of the searches and subsequent calculations during the training phase and (2) putting additional focus on high-symmetry structures (which may be metastable or stabilised at non-zero temperatures). Throughout this work, we aim to consider the broadest set of compositions possible. To this end, we construct a convex hull for binary hydrides of the form $X_nH_m$ for every element $X$ in the periodic table at 10, 100, 200, 300 and 500\ GPa in order to assess their stability. Structures on or near these static-lattice convex hulls were then selected for further investigation at the relevant pressure. In particular, in the training phase, stoichiometries within 80\ meV/formula unit of the hull were selected (for each of these stoichiometries just the lowest energy structure was chosen). In order to discuss stability more clearly in this work, we introduce two quantities: $E_{stoic}$, the distance of the given stoichiometry from the static-lattice convex hull, and $E_{struc}$, the distance of the given structure from the lowest energy structure of the same stoichiometry. Here, for example, we are selecting structures with $E_{stoic}\leq$ 80\ meV/formula unit and $E_{struc} = 0$.

\subsection{Electron-phonon coupling estimates}
McMillan \cite{mcmillan1968} showed that for strong-coupled superconductors the electron-phonon coupling constant, $\lambda$, can be expressed as
\begin{equation}
\label{eq:couplconst}
\lambda=\frac{N(E_F)\braket{I^2}}{M\braket{\omega^2}}
\end{equation}
where $N(E_F)$ is the DOS at the Fermi level, $\braket{I^2}$ is the Fermi surface averaged electron-phonon matrix element, $M$ is the atomic mass, and $\braket{\omega^2}$ is the average squared phonon frequency. The numerator in this expression can be relabelled as $\eta$, the so-called Hopfield parameter. In situations where the vibrational modes can be well-separated into those of different atomic character (as we typically see in binary hydrides) we can write
\begin{equation}
\label{eq:totlambda}
\lambda=\sum_j\lambda_j=\sum_j\frac{\eta_j}{M_j\braket{\omega_j^2}}
\end{equation}
where $j$ is the atom type.

Hopfield established that, in a local phonon representation, electron-phonon interactions mainly consist of scatterings that change the electronic angular momentum $l$ \cite{hopfield1969}; building on this observation, Gaspari-Gyorffy theory \cite{gaspari1972} then provides a way to approximate $\braket{I^2}$ in Eq.\ \ref{eq:couplconst}. Beginning from a multiple-scattering Green's function formalism, adopting the rigid muffin-tin approximation \cite{slater1937wave} allows a number of approximations concerning the potential and wavefunction to be applied \cite{papaconstantopoulos2015}. These approximations reduce the expression for the Hopfield parameter to a combination of electronic scattering phase shifts and (partial) electronic densities of states \cite{papaconstantopoulos2015}. Recent work has emerged using Gaspari-Gyorffy theory for metal hydrides under high pressure \cite{papaconstantopoulos2015, chang2019}.

The final form of the rigid-ion Gaspari-Gyorffy formula is
\begin{multline}
\label{eq:finalGG}
\braket{I^2}=\frac{E_F}{\pi^2N^2(E_F)} \\ \sum_l\frac{2(l+1)\sin^2(\delta_{l+1}-\delta_l)N_l(E_F)N_{l+1}(E_F)}{N_l^{(1)}N_{l+1}^{(1)}}
\end{multline}
where $N_l^{(1)}$ is the free-scatterer DOS - the DOS at the Fermi level for a single muffin-tin potential in a zero-potential background - and $\delta_l$ are the scattering phase shifts.

The scattering phase shifts can be obtained by matching the logarithmic derivative for solutions inside and outside the muffin-tin at the muffin-tin radius, $r=R_{MT}$, as detailed in Ref.\ \cite{sakurai2017}. The free scatterer DOS can also be easily computed from properties of the radial wavefunction. From a self-consistent DOS calculation, partitioned by both atom and angular momentum $l$, we therefore have all the components needed to approximate $\braket{I^2}$ for each atom type. These quantities can then be multiplied by $N(E_F)$ to give each $\eta_j$.



Gaspari-Gyorffy theory was implemented in the \textsc{elk} code \cite{elk_code} in Ref.\ \cite{hutcheon2020}; this modified version of the code is used in this work to calculate $\eta_H$ and $\eta_X$. The traditional way to use these approximate Hopfield parameters (or ``Gaspari-Gyorffy electron-phonon coupling estimates") is to calculate the average phonon frequency and use this to estimate $\lambda$ via Eq.\ \ref{eq:couplconst} (or, in our case, Eq.\ \ref{eq:totlambda}). This $\lambda$ value can in turn be used to obtain $T_c$ from the McMillan \cite{mcmillan1968} or Allen-Dynes \cite{AD_formula} equations. However, the phonon calculation required to obtain $\braket{\omega^2}$ means that this approach would be too expensive to be useful in a high-throughput screening scenario. Therefore, in this work, we do not obtain direct estimates of $\lambda$ or $T_c$ from Gaspari-Gyorffy theory and instead use the $\eta_j$ values themselves as descriptors in our $T_c$ model.

\subsection{Electron-phonon coupling calculations}
At several stages during this work electron-phonon coupling calculations are performed using density functional perturbation theory (DFPT) in the \textsc{quantum espresso} code \cite{QE-2009, QE-2017} and the results treated using Migdal-Eliashberg theory \cite{eliashberg1960} to obtain $T_c$ values from first principles. Within the theory, an electron in an occupied state $\ket{n,\mathbf{k}}$ is coupled to an unoccupied state $\ket{m,\mathbf{k}+\mathbf{q}}$ by a phonon with momentum $\mathbf{q}$ and frequency $\omega_{\mathbf{q},\nu}$. At the temperatures of interest for superconductivity, we will have $\omega_{\mathbf{q}\nu} \sim k_bT \ll 1$\ eV, so the initial and final electronic states will lie very close to the Fermi energy. However, on the finite $\mathbf{k}$-point grids used in DFT calculations, Kohn-Sham states will rarely lie this close to the Fermi level. To overcome this, we follow the method used in Ref.\ \cite{wierzbowska2005} and smear out the electronic energies using Gaussians with a characteristic smearing width $\sigma$, chosen according to the procedure outlined in Ref.\ \cite{shipley2020}.

Both the initial ($\mathbf{k}$) and final ($\mathbf{k}+\mathbf{q}$) states must be present in the $\mathbf{k}$-point grid. To ensure that this is the case, we take our $\mathbf{k}$-point grids to be fixed multiples of the $\mathbf{q}$-point grids, increasing the multiplicative factor until convergence is achieved (typically this results in $\mathbf{k}$-point grids larger than $30\times30\times30$). Coupling strengths are then interpolated onto a finer $\mathbf{q}$-point grid (8 times larger in each direction) to construct the Eliashberg function, $\alpha^2F(\omega)$. From $\alpha^2F(\omega)$, we obtain the superconducting critical temperature either by direct solution of the Eliashberg equations using the \textsc{elk} code \cite{elk_code} or using the Allen-Dynes equation \cite{AD_formula}. We compare these two methods in Sec.\ \ref{sec:allen_dynes_vs_eliashberg}.

Throughout this work, we profiled the electron-phonon coupling calculations in the \textsc{quantum espresso} code and implemented optimizations \footnote{The optimizations to \textsc{quantum espresso} resulting from this work have been submitted to the developers (see \url{https://gitlab.com/miicck/q-e})}. In particular, through optimized symmetrization of the electron-phonon matrix elements and evaluation of Gaussian smearing, we reduced the time to calculate $T_c$ for our $Fm\bar{3}m$-LaH$_{10}$ test system by a factor of 8. Combined with the selective ability of our screening method, this near-order-of-magnitude speedup enables us to perform the present wide-ranging study.

\section{\texorpdfstring{$T_c$}{Tc} model and training phase}\label{sec:Tcmodel}
Our $T_c$ model is a Gaussian Process Regression model, initially trained using a set of 160 structures and corresponding $T_c$ values from the literature (collected from Refs. \cite{semenok2018actinium, hou2015high, fu2016high, abe2011crystalline, yu2014exploration, hu2013pressure, ma2015high, wang2012superconductive, salke2019synthesis, yu2015pressure, heil2018absence, esfahani2017superconductivity, gao2008superconducting, zhong2012structural, duan2015enhancement, shamp2015superconducting, duan2014pressure, errea2015high, zhang2015phase, liu2015prediction, zhong2016tellurium, zhou2012ab, kruglov2020, gao2013theoretical, zhou2019high, liu2015structures, zhou2019superconducting, liu2016stability, ma2015unexpected, ye2018high, jin2010superconducting, esfahani2016superconductivity, zhuang2017pressure, li2016crystal, kvashnin2018high, semenok2019superconductivity, kruglov2018uranium, li2017superconductivity, zheng2018structural, liu2017high, li2017phase, abe2018high, semenok2018distribution, gu2017high, shipley2020, hutcheon2020}). The model inputs are the Gaspari-Gyorffy electron-phonon coupling estimates for hydrogen and element $X$ ($\eta_H$ and $\eta_X$), the mass of atom $X$ in atomic units ($M_X$), the total DOS at the Fermi energy ($N(E_F)$, normalised by cell volume), and the hydrogen DOS divided by the total DOS at the Fermi energy ($N_H(E_F)/N(E_F)$). Throughout this work, whether the structure originated from the literature or from our own searches, these inputs were computed using the modified version of \textsc{elk} \cite{elk_code} from Ref.\ \cite{hutcheon2020}. In cases where a literature structure was not given at the same pressure as $T_c$ was reported, the structure was relaxed at the correct pressure using \textsc{castep} \cite{castep2005}. Refs. \cite{bi2018search, duan2017structure, flores2019perspective, zurek2019high} and the data tables within were found to be helpful for identifying additional points to include in the original literature set.

The model was trained via optimisation of model parameters and hyperparameters in \textsc{matlab} \cite{matlab_GPR} and was tested using nested cross-validation, with an inner loop used to optimise the parameters and an outer loop used to monitor the fit of the resulting model to unseen data. The correlation was evaluated in this way, repeated over several different random splittings of the data each time.

The overall process used in the training phase was iterative; structures at a given pressure were selected for further study based on the predictions of the model, with $T_c$ values for the best predicted structures calculated explicitly using DFPT and fed back into the model's training set for use in the next iteration (provided the structure was found to be dynamically stable). The model was then retrained and predictions were made for a set of search structures at the next pressure until all pressures had been considered. To make performing a large number of electron-phonon calculations feasible, a relatively sparse $\mathbf{q}$-point sampling was used in the training stage, chosen to reproduce the known result for $Fm\bar{3}m$-LaH$_{10}$ \cite{drozdov2019superconductivity, shipley2020}. This corresponds to a $\mathbf{q}$-point spacing of $2\pi\times$0.15\r{A}$^{-1}$ (a $2 \times 2 \times 2$ grid for $Fm\bar{3}m$-LaH$_{10}$ at 200\ GPa). In keeping with the initial literature training set, the Allen-Dynes equation was used to estimate $T_c$ throughout the training phase.

\begin{figure}
    \centering
    \includegraphics[width=\columnwidth]{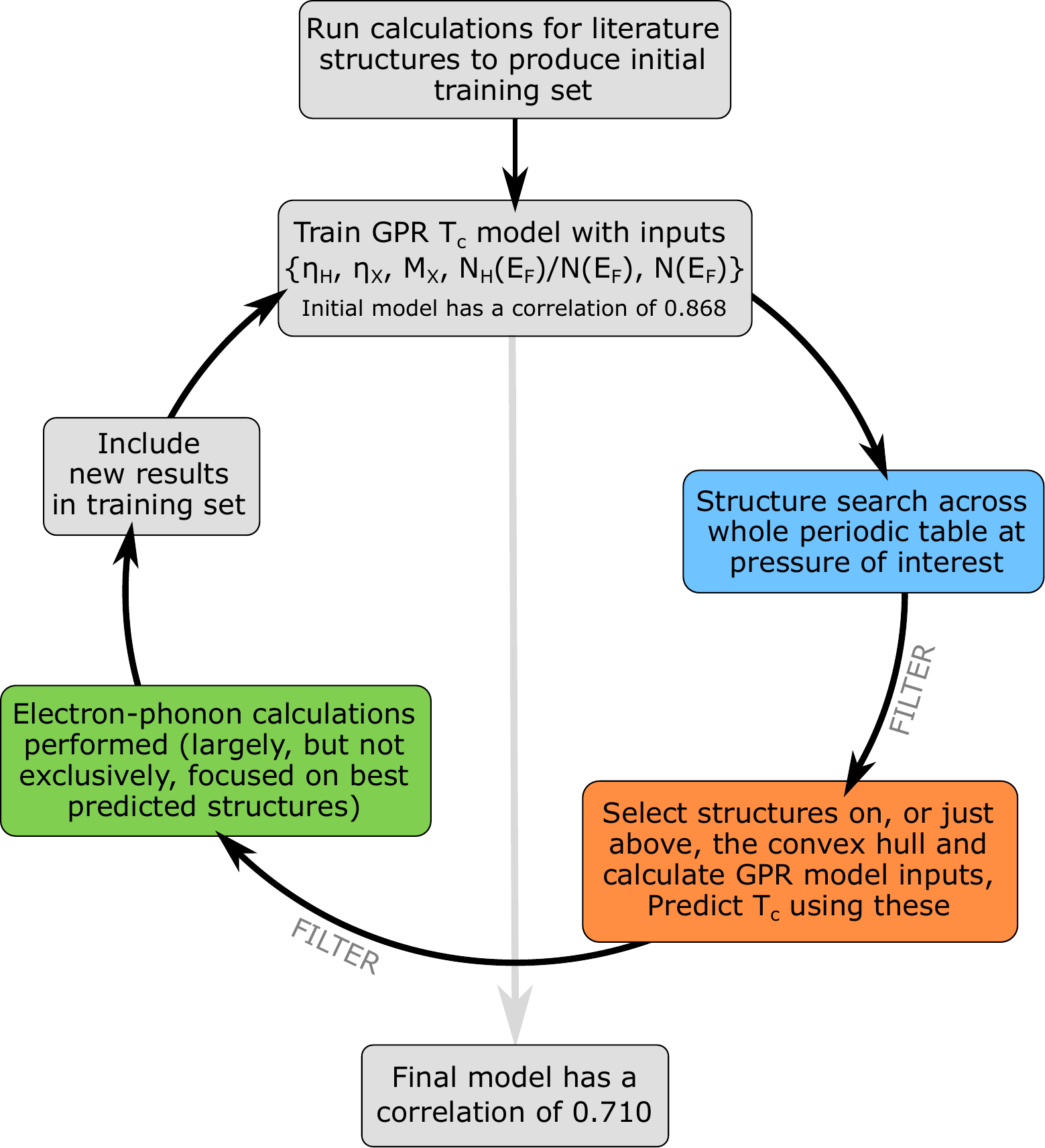}
    \caption{A flowchart summarising our methodology.}
    \label{fig:flowchart}

    \vspace{0.5cm}

    \centering
    \includegraphics[width=\columnwidth]{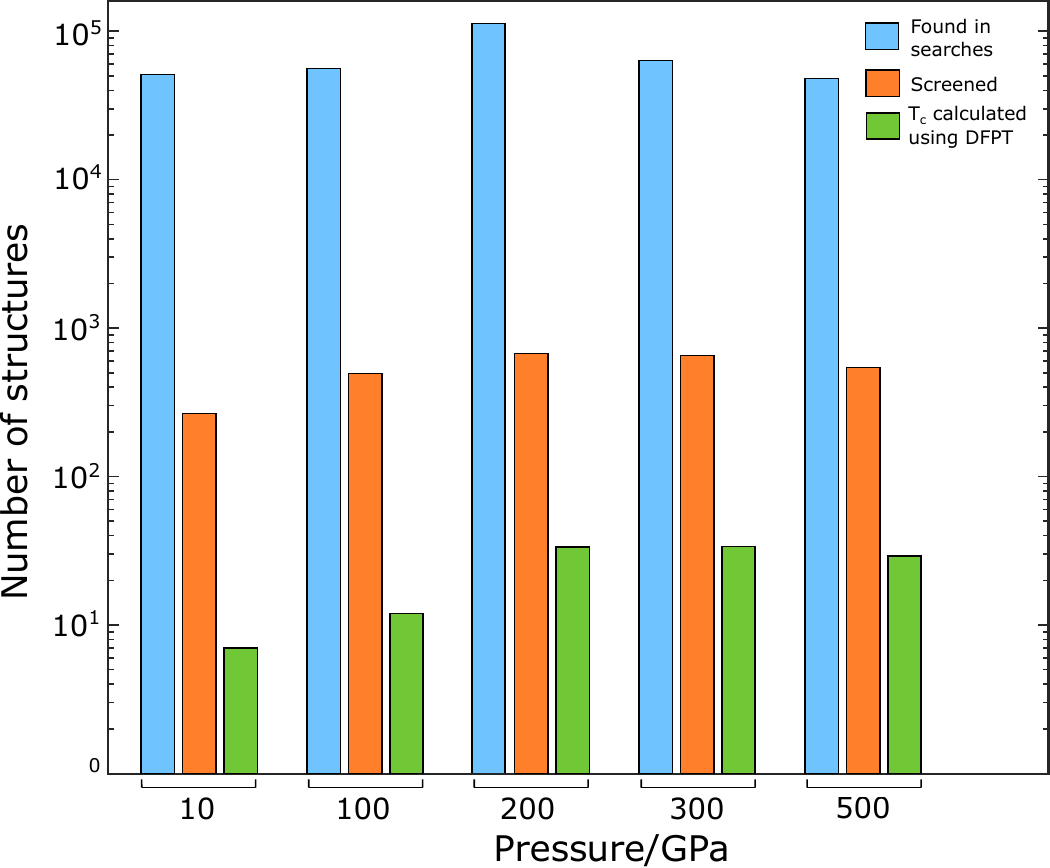}
    \caption{A summary of the number of structures studied at each stage of the training process - \textit{note the logarithmic scale}.}
    \label{fig:barchart}
\end{figure}

Our method is summarised in Fig.\ \ref{fig:flowchart}. Fig.\ \ref{fig:barchart} shows the number of structures considered in total at each stage of the training process - given the relatively large cost of electron-phonon calculations (even in high-throughput operation), this figure highlights the importance of the stability filtering and model-based screening steps in our workflow. In total, 119 new DFPT data points were added to the training set in this work. Predictions for all previously-considered search structures were recalculated using the final model to ensure nothing of interest had been missed in earlier iterations. Most of the electron-phonon calculations performed in the training phase were for structures with a high predicted $T_c$ according to our model, however, occasionally these calculations were performed for structures with mid-range or low $T_c$ predictions in order to improve the behaviour of the model. The correlation of the predicted $T_c$ values with the calculated $T_c$ values across the training set decreased slightly on addition of more data (see Fig.\ \ref{fig:flowchart}). This is not surprising for two reasons: (1) as mentioned previously, the results we add to the training set are computed in a high-throughput manner and therefore will be somewhat under-converged compared to typical values found in the literature, and (2) the final training set contains a wider range of elements and pressures than the original literature-based training set.

During the training of our model, we were able to efficiently rediscover a number of binary hydrides (with relatively high $T_c$ values) which had been reported previously, including the structures $Im\bar{3}m$-H$_3$S \cite{duan2014pressure,drozdov2015conventional,errea2015high}, $Im\bar{3}m$-LaH$_6$ \cite{peng2017hydrogen}, $I4/mmm$-AcH$_{12}$ \cite{semenok2018actinium}, $Im\bar{3}m$-SeH$_3$ \cite{zhang2015phase}, $R\bar{3}m$-SrH$_6$ \cite{bi2018search}, $R\bar{3}m$-LiH$_6$ \cite{xie2014lithium,howie2012high}, $Fm\bar{3}m$-LaH$_{10}$ \cite{liu2017potential, peng2017hydrogen,shipley2020,drozdov2019superconductivity,somayazulu2019evidence,kruglov2020}, $Fm\bar{3}m$-YH$_{10}$ \cite{peng2017hydrogen,shipley2020}, $Im\bar{3}m$-ScH$_6$ \cite{ye2018high,abe2017hydrogen,peng2017hydrogen}, $P6_3/mmc$-ThH$_9$ \cite{semenok2019superconductivity}, $R\bar{3}m$-SrH$_{10}$ \cite{tanaka2017electron}, $Pm\bar{3}m$-SiH$_3$ \cite{jin2010superconducting}, $C2/m$-LaH$_7$ \cite{kruglov2020}, $Im\bar{3}m$-CaH$_6$ \cite{wang2012superconductive}, $Im\bar{3}m$-MgH$_6$ \cite{feng2015compressed}, $Fm\bar{3}m$-ThH$_{10}$ \cite{kvashnin2018high,semenok2019superconductivity}, and the stoichiometries KH$_6$ \cite{zhou2012ab}, LaH$_8$ \cite{liu2017potential,kruglov2020}, BaH$_{12}$ \cite{semenok2018distribution}, LaH$_5$ \cite{kruglov2020}, AcH$_{10}$ \cite{semenok2018actinium}, LiH$_8$ \cite{xie2014lithium}, LaH$_{11}$ \cite{kruglov2020}, MgH$_{12}$ \cite{lonie2013metallization}, YH$_9$ \cite{peng2017hydrogen, kong2019superconductivity} and ScH$_{12}$ \cite{ye2018high}. Although found previously, only half of these 26 stoichiometries appeared in our original training set; the other half were re-predicted, highlighting the capabilities of our method. We also identified a number of other systems with the potential to exhibit high-$T_c$ superconductivity. The most promising systems overall are studied further in Sec \ref{focused_search}.

\section{Focused searches and final results}\label{focused_search}
From the $T_c$ results obtained during training, the most promising candidate systems could be identified and studied in more depth. More focused structure searches were performed for hydrides of Na, Ca, La, Ac, and K at 100\ GPa, hydrides of La, Ac, S, Mg, and Na at 200\ GPa, hydrides of Li, Sr, K, Mg, Na, and Sc at 300\ GPa, and hydrides of Li, Sr, Mg, Na, Yb, Y, and Ca at 500\ GPa. For these calculations, the earlier symmetry constraints were relaxed, but all other parameters remained the same. These composition choices reflect a bias towards elements from the left-hand side of the periodic table amongst the most promising candidates from the training stage. However, this is not to say that only the compositions that we study further in this work are capable of supporting high-$T_c$ superconductivity. Future studies, with a different focus, could potentially identify more candidates. No additional searches were performed at 10\ GPa as the highest $T_c$ calculated at this pressure in the training phase was only 12-15\ K (belonging to $Fm\bar{3}m$-CaH$_2$) and a large proportion of the 245 structures screened were only weakly metallic (in fact, 48 of them were insulating).

\begin{figure}
    \centering
    \includegraphics[width=\columnwidth]{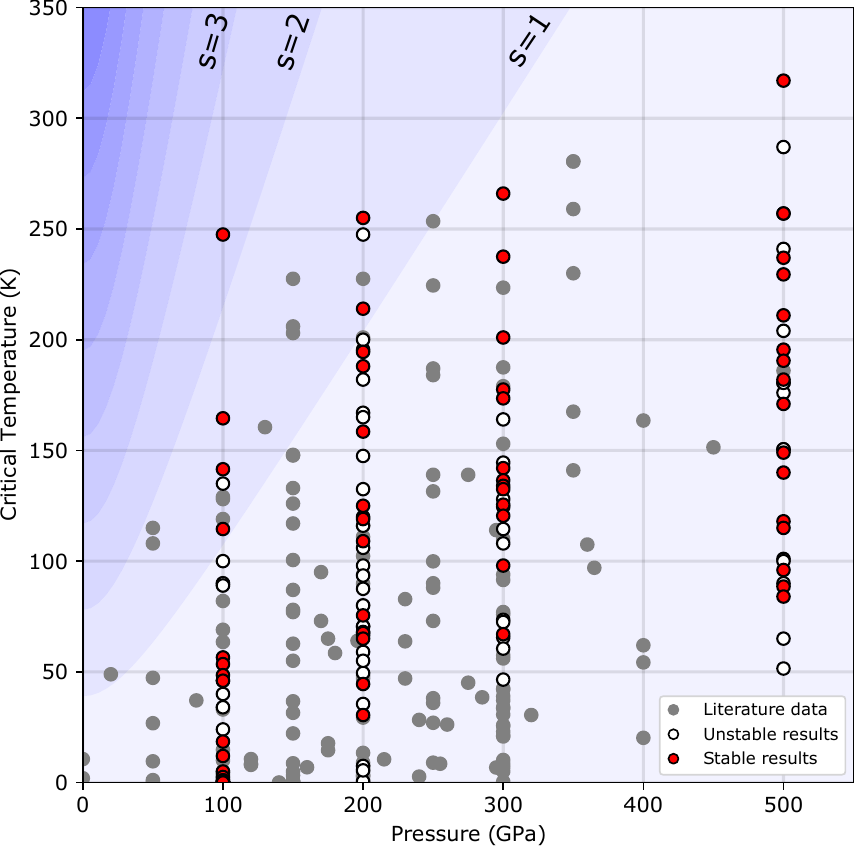}
    \caption{Critical temperatures obtained from converged DFPT calculations for the most promising candidates according to our screening process (calculated using the Allen-Dynes equation to facilitate comparison with the literature). Both dynamically stable and dynamically unstable results are shown. The background is shaded according to the figure of merit $S$, introduced in Ref.\ \cite{pickard2019}.}
    \label{fig:convergedresults}
    
    \vspace{0.25cm}
    
    \centering
    \includegraphics[width=\columnwidth]{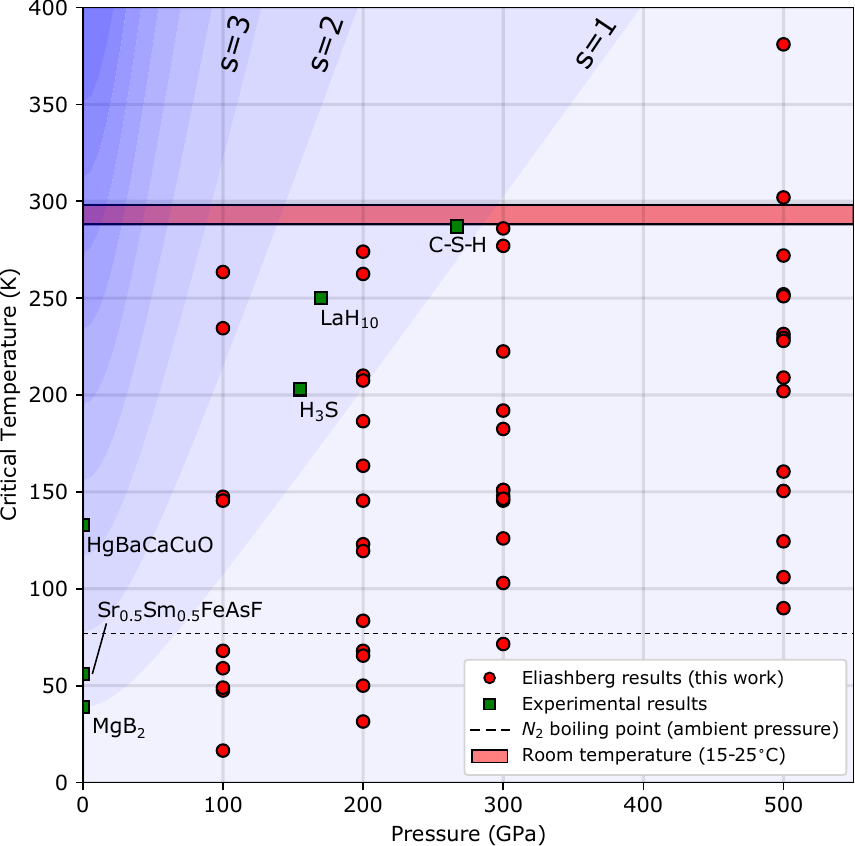}
    \caption{As Fig.\ \ref{fig:convergedresults}, but showing the Eliashberg results from Table\ \ref{tab:stable_superconductors} (dynamically stable structures only) alongside experimental results for specific superconductors (in order of increasing $T_c$: \cite{Nagamatsu2001}, \cite{Wu_2009}, \cite{Schilling1993}, \cite{drozdov2015conventional}, \cite{drozdov2019superconductivity} and \cite{Snider2020}).}
    \label{fig:convergedresultseliashberg}
\end{figure}

On completion of the focused searches, we again calculated $\eta_H$, $\eta_X$, $N(E_F)$, and $N_H(E_F)/N(E_F)$ for the stable and metastable structures found. Stoichiometries with $E_{stoic}\leq$ 25\ meV/formula unit were selected for further study. For the stoichiometries on the hull ($E_{stoic} = 0$), 2-5 of the most stable structures were chosen. For the selected off-hull stoichiometries, only the lowest energy structure ($E_{struc} = 0$) was chosen. The inputs were then fed into the final $T_c$ model trained in Sec.\ \ref{sec:Tcmodel} and fully converged electron-phonon calculations were performed for the structures with the highest predicted $T_c$ values at each pressure (as well as for the most promising candidates already identified during training). These results are shown in Figs.\ \ref{fig:convergedresults} and \ref{fig:convergedresultseliashberg}. Promising structures that remained dynamically stable after convergence of the $\mathbf{q}$-point grids are listed in Table\ \ref{tab:stable_superconductors}. We find near room-temperature superconductors at every pressure considered, which we elaborate on in the following sections. Unless otherwise stated, all critical temperatures reported in the following sections are from direct solution of the Eliashberg equations.

\newcommand{\hls}{\underline}

\begin{table*}
\centering
\footnotesize
\begin{tabular}{|@{\hspace{0.3cm}}l@{\hspace{0.3cm}}l@{\hspace{0.3cm}}l@{\hspace{0.3cm}}l@{\hspace{0.3cm}}l@{\hspace{0.3cm}}l@{\hspace{0.3cm}}l@{\hspace{0.3cm}}l@{\hspace{0.3cm}}|}
    \hline
    &&&&&&&\\[-0.25cm]
    Stoichiometry    & Space Group   &  Pressure (GPa) &  Allen-Dynes $T_c$ (K) & Eliashberg $T_c$ (K) & $\lambda$ & $E_{stoic}$ (meV/unit) & $E_{struc}$ (meV/unit) \\[0.1cm]
    \hline
    &&&&&&&\\
    \hls{NaH$_6$}          &  \hls{$Pm\bar{3}m$} &  100    &  228-267  & 248-279 & 2.54 & 28 & 0\\
    CaH$_6$          &  $Im\bar{3}m$ &  100    &  150-179  & 216-253 & 5.81 & 9 & 137\\
    \hls{Na$_2$H$_{11}$}   &  \hls{$Cmmm$}       &  100    &  127-156  & 134-161 & 1.28 & 0 & 0\\
    \hls{KH$_{10}$}        &  \hls{$C2/m$}       &  100    &  105-124  & 134-157 & 2.45 & 0 & 0\\
    NaH$_{16}$	     &  $Fmm2$       &	100	   &  47-60    & 61-75   & 1.10 & 0 & 0\\
    AcH$_5$          &	$P\bar{1}$	 &  100	   &  48-65    & 49-69   & 0.91 & 0 & 0\\
    LaH$_5$	         &  $P\bar{1}$	 &  100    &  37-55    & 40-58   & 0.83 & 0 & 0\\
    NaH$_{24}$	     &  $R\bar{3}$	 &  100	   &  40-57    & 40-55   & 0.82 & 0 & 0\\
    AcH$_{11}$	     &  $C2/m$	     &  100	   &  14-23    & 13-20   & 0.71 & 0 & 0\\
    &&&&&&&\\
    \hls{NaH$_6$}	         &  \hls{$Pm\bar{3}m$} &  200	   &  235-275  & 260-288 & 2.06 & 39 & 16\\
    \hls{AcH$_{12}$}	     &  \hls{$P6_3mc$}	 &  200	   &  197-231  & 245-280 & 3.92 & 11 & 0\\
    \hls{MgH$_{13}$}	     &  \hls{$Fm\bar{3}m$} &  200	   &  179-210  & 196-224 & 1.98 & 17 & 635\\
    SH$_{3}$	     &  $Im\bar{3}m$ &  200	   &  173-203  & 196-219 & 1.77 & 0 & 0\\
    \hls{AcH$_{6}$}	     &  \hls{$Fmmm$}       &	200    &  110-140  & 169-204 & 2.01 & 0 & 14\\
    \hls{NaH$_{8}$}	     &  \hls{$I4/mmm$}     &	200	   &  146-171  & 152-175 & 1.63 & 26 & 124\\
    \hls{Na$_2$H$_{11}$}   &  \hls{$Cmmm$}       &  200    &  120-156  & 129-162 & 1.11 & 0  & 0 \\
    \hls{MgH$_{14}$}       &  \hls{$P\bar{1}$}   &  200    &  106-132  & 112-134 & 1.35 & 23  & 0 \\
    LaH$_{7}$	     &  $C2/m$	     &  200	   &  98-120   & 105-134 & 1.23 & 3 & 0\\ 
    MgH$_4$	         &  $I4/mmm$	 &  200	   &  63-88    & 73-94   & 0.98 & 0 & 101\\
    SH$_7$	         &  $Fmmm$	     &  200	   &  57-78    & 58-78   & 0.91 & 29 & 0\\
    Mg$_2$H$_7$      &  $C2/m$       &  200    &  55-75    & 56-75   & 0.98 & 22  & 0 \\
    AcH$_4$	         &  $Cmcm$       &  200	   &  35-54    & 42-58   & 0.99 & 19 & 0\\
    Mg$_2$H$_5$      &	$R\bar{3}m$	 &  200	   &  22-39    & 24-39   & 0.74 & 139 & 21\\
    &&&&&&&\\
    MgH$_6$     	 &  $Im\bar{3}m$ &	300	   &  248-284  & 271-301 & 2.28 & 19 & 437\\
    \hls{YH$_9$}	         &  \hls{$F\bar{4}3m$} &	300	   &  220-255  & 261-293 & 2.58 & 2 & 0\\
    ScH$_8$	         &  $Immm$	     &  300	   &  185-217  & 212-233 & 2.06 & 3 & 0\\
    \hls{LiH$_2$}	         &  \hls{$P6/mmm$}     &	300	   &  162-193  & 177-207 & 1.45 & 40 & 75\\
    \hls{NaH$_7$}	         &  \hls{$C2/m$}	     &  300	   &  157-190  & 167-198 & 1.48 & 3 & 0\\
    \hls{ScH$_{12}$}   	 &  \hls{$P\bar{1}$}	 &  300	   &  127-157  & 137-165 & 1.28 & 0 & 103\\
    \hls{NaH$_{5}$}	     &  \hls{$P4/mmm$}	 &  300	   &  121-144  & 138-164 & 1.92 & 1 & 0\\
    \hls{LiH$_{6}$}	     &  \hls{$C2/m$}    	 &  300	   &  109-142  & 130-163 & 1.16 & 0 & 14\\
    LiH$_{6}$	     &  $R\bar{3}m$	 &  300	   &  121-152  & 130-161 & 1.30 & 0 & 0\\
    ScH$_{6}$	     &  $Im\bar{3}m$ &	300	   &  118-150  & 135-161 & 1.26 & 0 & 0\\
    \hls{LiH$_3$}	         &  \hls{$Cmcm$} 	     &  300	   &  104-137  & 112-140 & 1.06 & 1 & 0\\
    \hls{ScH$_{14}$}	     &  \hls{$P\bar{1}$}	 &  300	   &  87-109   & 91-115  & 1.20 & 6 & 0\\
    MgH$_4$	         &  $I4/mmm$	 &  300	   &  53-81    & 59-84   & 0.76 & 0 & 0\\
    &&&&&&&\\
    \hls{MgH$_{12}$}	     &  \hls{$Pm\bar{3}$}  &  500	   &  294-340  & 360-402 & 2.65 & 0 & 259\\
    \hls{SrH$_{10}$}	     &  \hls{$Fm\bar{3}m$} &  500	   &  239-275  & 285-319 & 2.22 & 8 & 120\\
    \hls{MgH$_{13}$}	     &  \hls{$P3m1$}	     &  500	   &  239-275  & 257-287 & 2.21 & 12 & 0\\
    \hls{MgH$_{10}$}	     &  \hls{$C2/m$}	     &  500	   &  209-250  & 232-270 & 1.63 & 9 & 0\\
    \hls{NaH$_9$}	         &  \hls{$P6_3/mmc$}	 &  500	   &  218-256  & 235-269 & 1.67 & 0 & 0\\
    \hls{YH$_{18}$}	     &  \hls{$P\bar{1}$}	 &  500	   &  179-212  & 213-246 & 1.99 & 25 & 0\\
    \hls{SrH$_{24}$}	     &  \hls{$R\bar{3}$}	 &  500	   &  195-227  & 218-245 & 1.88 & 9 & 0\\
    \hls{YH$_{20}$}	     &  \hls{$P\bar{1}$}	 &  500	   &  176-205  & 212-244 & 2.21 & 39 & 0\\
    SrH$_{10}$	     &  $R\bar{3}m$	 &  500	   &  165-199  & 190-228 & 1.31 & 8 & 0\\
    CaH$_{10}$	     &  $R\bar{3}m$	 &  500	   &  155-187  & 184-220 & 1.51 & 3 & 0\\
    \hls{Na$_2$H$_{11}$}	 &  \hls{$Cmmm$}	     &  500	   &  132-166  & 141-180 & 1.12 & 0 & 0\\
    \hls{CaH$_{15}$}	     &  \hls{$P\bar{6}2m$} &  500	   &  120-160  & 134-167 & 1.01 & 0 & 0\\
    \hls{SrH$_{15}$}	     &  \hls{$P\bar{6}2m$} &  500    &  100-136  & 110-139 & 0.93 & 0 & 0\\
    \hls{MgH$_8$}	         &  \hls{$C2/m$}	     &  500	   &  82-110   & 91-121  & 0.96 & 0 & 0\\
    Na$_2$H$_{11}$   &	$I4/mmm$	 &  500	   &  72-105   & 76-104  & 0.80 & 0 & 297\\
    &&&&&&&\\
    \hline
\end{tabular}
\caption{Allen-Dynes and Eliashberg $T_c$ values for dynamically stable superconductors found in this work, along with calculated $\lambda$ and stability measures. Structures that are \hls{underlined} have a calculated $T_c$ above 100\ K and have either not been reported in the literature before or have not had their superconducting properties studied before (to the best of our knowledge) - more details are given in the Supplementary Information \cite{supplement}. Structures available online \cite{table_1_structures}.}
\label{tab:stable_superconductors}
\end{table*}

\subsection{100GPa}
\begin{figure}
    \centering
    \subfigure[]{\includegraphics[width=0.5\columnwidth]{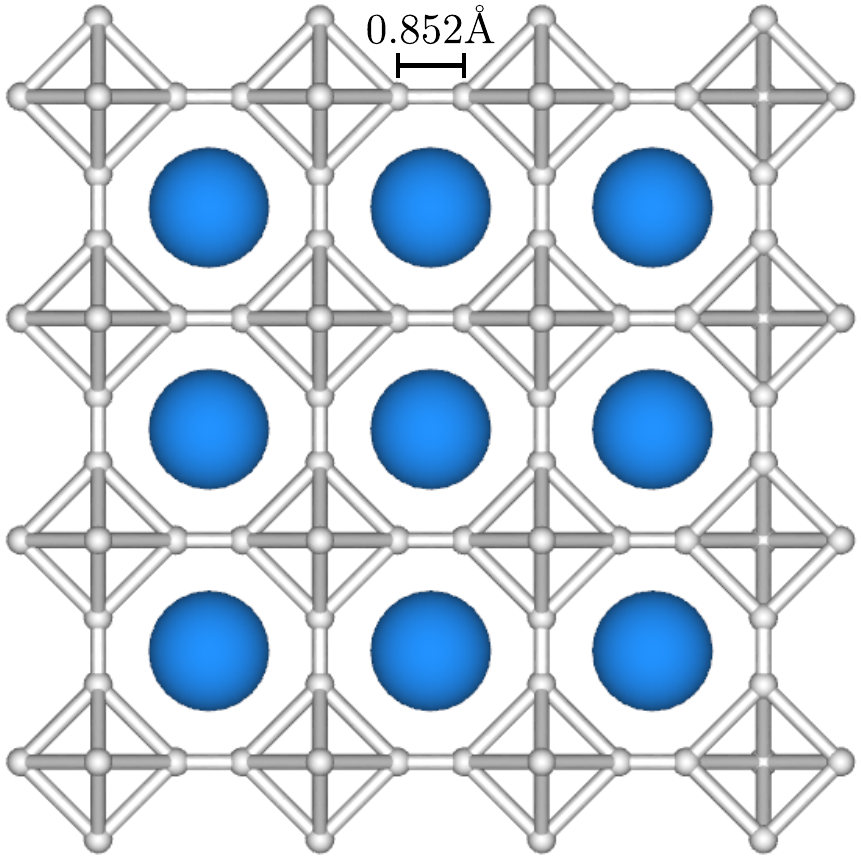}}
    \subfigure[]{\includegraphics[width=0.44\columnwidth]{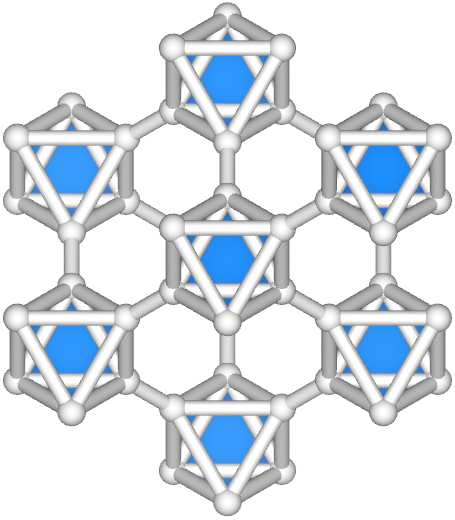}}
    \caption{The 100\ GPa structure of $Pm\bar{3}m$ NaH$_6$ as viewed along the (a) [100] and (b) [111] directions of the standardized cell.}
    \label{fig:NaH6_Pm-3m}
    
    \vspace{0.15cm}
    
    
    \vspace{0.25cm}
    
    \centering
    \includegraphics[width=\columnwidth]{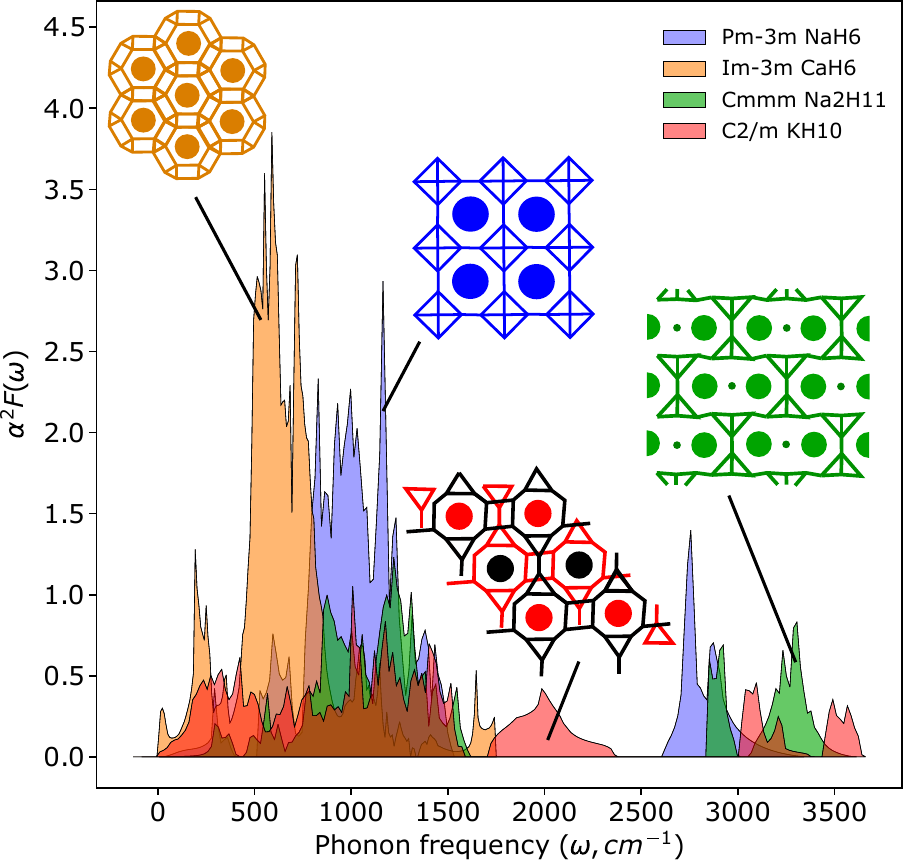}
    \caption{The highest-$T_c$ structures and corresponding Eliashberg functions at 100\ GPa.}
    \label{fig:a2f_100GPa}
\end{figure}

Of particular note at 100\ GPa is a $Pm\bar{3}m$ structure of NaH$_6$, a stoichiometry which we find to lie above the convex hull, in agreement with Ref.\ \cite{baettig2011} (although they find that the $Pm\bar{3}m$ structure is not the lowest energy structure until 150 GPa). This structure consists of a cubic lattice of H octahedra with Na at the body-centred positions (see Fig.\ \ref{fig:NaH6_Pm-3m}). Indeed, as we will see throughout this work, many of the best candidate systems contain hydrogen in such polyhedral clusters. While experimental synthesis of sodium polyhydrides has been demonstrated \cite{struzhkin2016}, superconductivity in the system seems under-studied given its promise here with a calculated $T_c$ of 248-279\ K. This places the structure at a crucial position in pressure-$T_c$ space, strongly influencing the apparent low-pressure trend of maximum $T_c$ (see Figs.\ \ref{fig:convergedresults} and \ref{fig:convergedresultseliashberg}) and hinting at the exciting possibility of other low-pressure high-$T_c$ superconductors. Such high-temperature superconductivity is only possible with both high average phonon frequencies and strong electron-phonon coupling. Indeed, the logarithmic average phonon frequency, $\omega_{\text{ln}} = 870$ cm$^{-1}$, and electron-phonon coupling constant, $\lambda = 2.54$, of $Pm\bar{3}m$-NaH$_6$ at 100\ GPa are comparable to the values for the $Fm\bar{3}m$-LaH$_{10}$ clathrate superconductor at 250\ GPa ($\omega_{\text{ln}} = 871$ cm$^{-1}$, $\lambda = 2.29$ \cite{kruglov2020}). The ability of $Pm\bar{3}m$-NaH$_6$ to maintain a high average phonon frequency at such low pressures is due to its compact structure, with a shortest hydrogen-hydrogen bond length of only $d_\text{HH} = 0.85$\r{A} (shown on Fig.\ \ref{fig:NaH6_Pm-3m}) - significantly shorter than in $Fm\bar{3}m$-LaH$_{10}$, where $d_\text{HH} > 1$\r{A} up to pressures as high as 300\ GPa \cite{1907.07820}.

\begin{figure}
    \centering
    \includegraphics[width=\columnwidth]{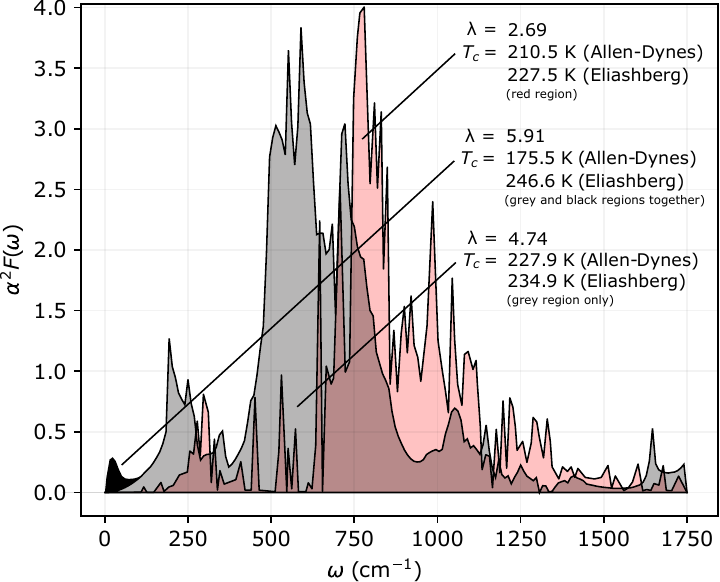}
    \caption{Grey and black: The Eliashberg function of $Im\bar{3}m$ CaH$_6$ at 100GPa, demonstrating the suppression of the Allen-Dynes $T_c$ relative to the one obtained by solving the Eliashberg equations. This suppression is due to electrons coupling to soft phonon modes. Red: The Eliashberg function of the same material at 150GPa from Ref.\ \cite{wang2012superconductive}.}
    \label{fig:CaH6_a2f}
\end{figure}

Also of interest at 100\ GPa is a cage-like $Im\bar{3}m$ structure of CaH$_6$ (see Fig.\ 3 of Ref.\ \cite{wang2012superconductive}). As was the case with NaH$_6$, this stoichiometry is found to be slightly above the static-lattice convex hull, in agreement with previous calculations \cite{wang2012superconductive}. However, despite strong electron-phonon coupling, its critical temperature (216-253\ K) is found to be slightly lower than that of $Pm\bar{3}m$-NaH$_6$ due to a lower average phonon frequency. Indeed, our CaH$_6$ phonons are significantly softer than those calculated at 150 GPa in Ref.\ \cite{wang2012superconductive}, resulting in a much higher $\lambda$ value (5.81 vs. 2.69 from \cite{wang2012superconductive}), suggesting that the structure is on the verge of a mechanical instability at 100 GPa. Fig.\ \ref{fig:CaH6_a2f} illustrates this, highlighting the strong dependence of both $\lambda$ and the Allen-Dynes $T_c$ on the low-frequency portion of $\alpha^2F(\omega)$ and, in particular, the appearance of a peak in $\alpha^2F(\omega)$ at very low frequencies. This is unsurprising as the Allen-Dynes equation is known to be sensitive to small changes in $\alpha^2F(\omega)$ at low frequencies (in particular, the functional derivative $\delta T_c(\text{AD}) / \delta\alpha^2F(\omega)$ diverges as $\omega\rightarrow0$ \cite{Bergmann1973}, in this case towards $-\infty$). However, the full Eliashberg calculation of $T_c$ is much less sensitive and the 100\ GPa result calculated here (216-253\ K) is comparable to the critical temperature of 220-235\ K calculated at 150\ GPa in Ref.\ \cite{wang2012superconductive}, also via solution of the Eliashberg equations. 

Fig.\ \ref{fig:a2f_100GPa} also demonstrates that while the next best structures, $Cmmm$-Na$_2$H$_{11}$ and $C2/m$-KH$_{10}$, have similar average phonon frequencies to $Pm\bar{3}m$-NaH$_6$, they do not exhibit such high coupling strengths. This leads to lower $T_c$ values of 134-161\ K and 134-157\ K, respectively. KH$_{10}$ adopts a cage like structure, whilst the Na$_2$H$_{11}$ structure contains broken octahedral hydrogen clusters (see Fig.\ \ref{fig:a2f_100GPa} inset). Superconductivity in the KH$_{10}$ stoichiometry has been studied previously \cite{semenok2018distribution}; it was found to be on the convex hull at 150\ GPa (and off-hull at 50\ GPa) where an Allen-Dynes $T_c$ of 148\ K was calculated for an $Immm$ structure. Ref.\ \cite{hooper2012potassium} found KH$_{10}$ to be above the convex hull at 100\ GPa and instead found metastable metallic structures of other stoichiometries, but superconductivity was not directly investigated. 

\subsection{200GPa}
\begin{figure}
    \centering
    \includegraphics[width=0.7\columnwidth]{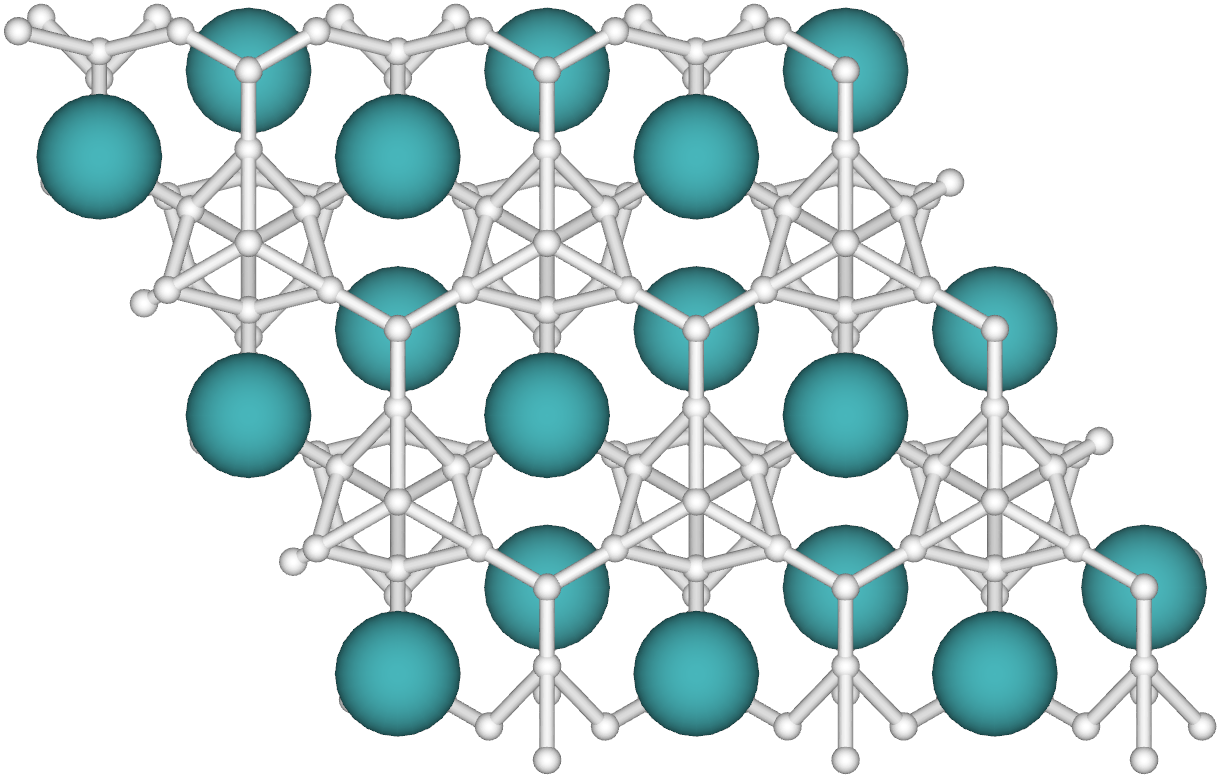}
    \caption{The 200\ GPa $P6_3mc$ structure of AcH$_{12}$ as viewed along the [001] direction of the standardized cell.}
    \label{fig:AcH12-P63mc}
\end{figure}

At 200\ GPa, the $Pm\bar{3}m$ structure of NaH$_6$ remains the highest $T_c$ structure found, rising slightly from its 100\ GPa value to 260-288\ K. However, as pressure increases we find this stoichiometry to be less stable with respect to decomposition. Similarly to Ref.\ \cite{semenok2018actinium}, we find several actinium hydrides to be high-temperature superconductors at this pressure, most notably a $P6_3mc$ structure of AcH$_{12}$ with a critical temperature of 245-280\ K. This structure is best described as cage-like, but with hydrogen atoms concentrated along specific channels along the $c$ axis (see Fig.\ \ref{fig:AcH12-P63mc}).

\begin{figure}
    \centering
    \includegraphics[width=\columnwidth]{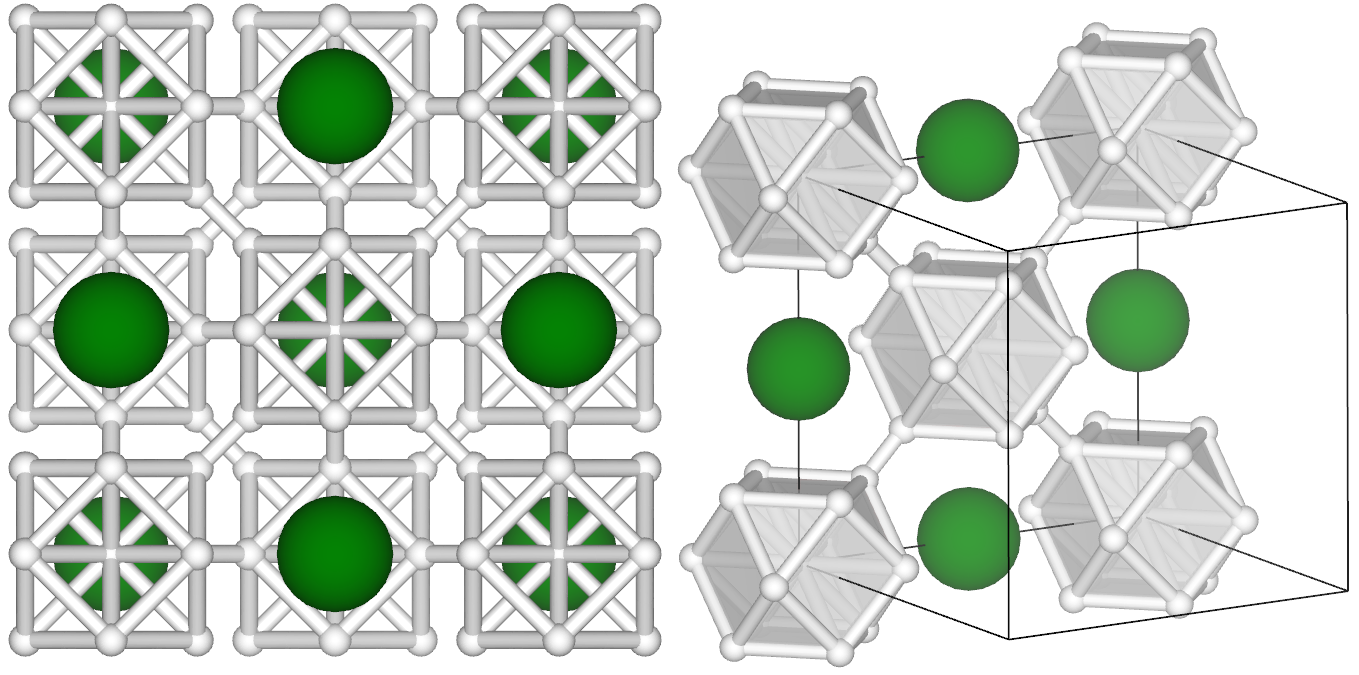}
    \caption{The 200\ GPa $Fm\bar{3}m$ structure of MgH$_{13}$. The structure consists of Mg atoms and axis-aligned cuboctahedra of hydrogen in a checkerboard pattern. Each cuboctahedra has an additional hydrogen atom at its centre to make up the necessary 13.}
    \label{fig:Fm3m_MgH13}
\end{figure}

200\ GPa also marks the appearance of magnesium hydrides, which become increasingly prevalent in our results with pressure. Of particular note is a cubic structure of MgH$_{13}$ with the space group $Fm\bar{3}m$, which possesses a slightly higher calculated $T_c$ (196-224\ K) than the experimentally-verified \cite{drozdov2015conventional} $Im\bar{3}m$-H$_3$S at the same pressure (196-219\ K). The MgH$_{13}$ structure consists of axis-aligned cuboctahedral hydrogen clusters (see Fig.\ \ref{fig:Fm3m_MgH13}). In agreement with Ref.\ \cite{lonie2013metallization}, we find MgH$_{13}$ to lie above the convex hull at 200\ GPa. They report significantly lower critical temperatures for on-hull structures.

A notable absence from the 200\ GPa results is LaH$_{10}$. Several structures of LaH$_{10}$ (including the experimentally-verified $Fm\bar{3}m$ structure \cite{drozdov2019superconductivity, somayazulu2019evidence}) were found in our structure searches and flagged as good candidates by our $T_c$ model, but were dynamically unstable at the harmonic level. This has been noted previously for the cubic structure \cite{geballe2018lanthanum, liu2018dynamics, shipley2020}. We elaborate on our treatment of structures with dynamical instabilities in Sec.\ \ref{sec:dynamical_instabilities}.

\subsection{300GPa}
At 300\ GPa, an $Im\bar{3}m$ structure of MgH$_6$ exhibits the highest $T_c$ value - 271-301\ K (in agreement with previous work \cite{feng2015compressed}). This structure can be obtained by substituting Ca for Mg in the CaH$_{6}$ structure investigated at 100\ GPa. By instead substituting only half of the Ca atoms at 200\ GPa, it has been reported that one obtains an even higher critical temperature ternary superconductor \cite{sukmas2020MgCa}. Hybridizing compatible binary crystal structures in this way could provide an efficient method to design future high-temperature ternary hydride superconductors.

\begin{figure}
    \centering
    \subfigure[]{\includegraphics[width=0.48\columnwidth]{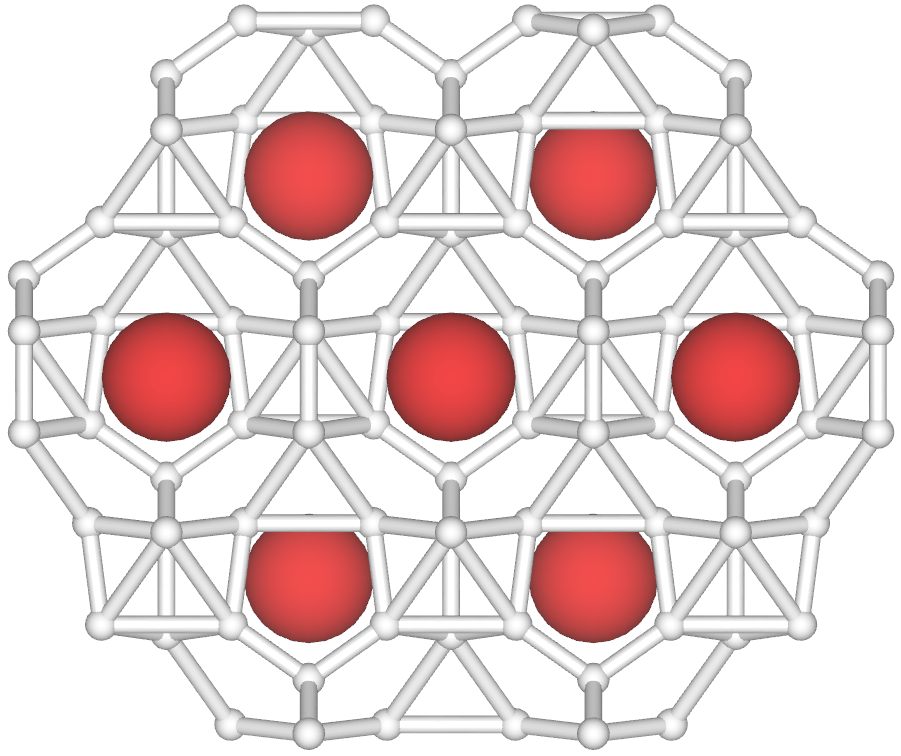}}
    \subfigure[]{\includegraphics[width=0.48\columnwidth]{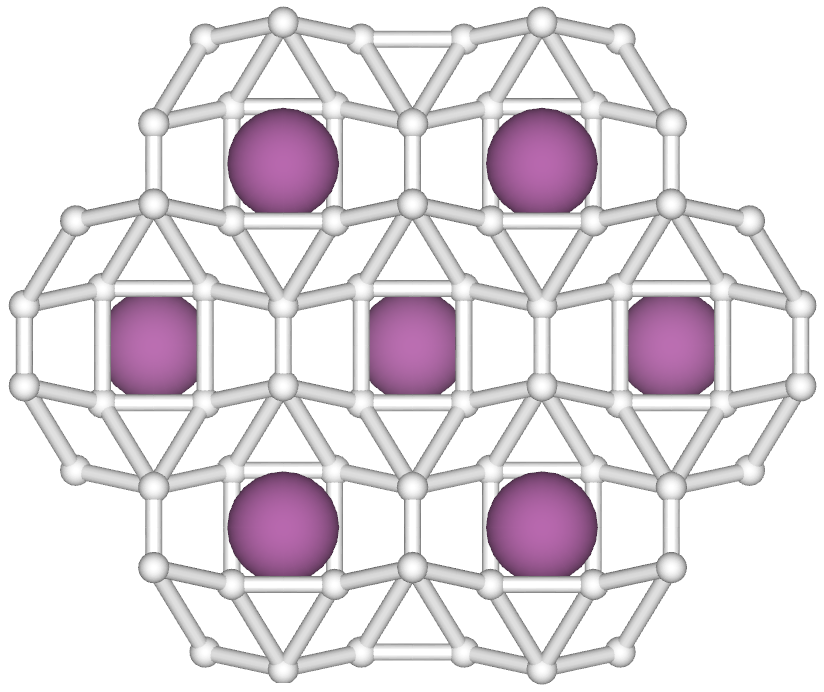}}
    \caption{(a) The 300\ GPa $F\bar{4}3m$ structure of YH$_9$ as viewed along the [110] direction of the standardized cell. (b) The 300\ GPa $Immm$ structure of ScH$_8$ as viewed along the [100] direction of the standardized cell.}
    \label{fig:YH9-ScH8}
\end{figure}

With a critical temperature of 261-293\ K, a cage-like $F\bar{4}3m$ structure of YH$_9$ (shown in Fig.\ \ref{fig:YH9-ScH8}(a)) is the next highest temperature superconductor found at 300\ GPa. This stoichiometry (although with $P6_3/mmc$ symmetry) has been synthesised experimentally, confirming theoretical predictions \cite{peng2017hydrogen}, and was found to exhibit a critical temperature of 243\ K at 201\ GPa \cite{kong2019superconductivity}. Y-H systems have been extensively studied \cite{heil2019superconductivity, troyan2019synthesis, li2015pressure, shipley2020} with critical temperatures in excess of 200\ K predicted over large pressure ranges.

We find that an $Immm$ structure of ScH$_8$ with a similar motif to $F\bar{4}3m$-YH$_9$ (see Fig.\ \ref{fig:YH9-ScH8}(b)) is also a high-temperature superconductor at this pressure with a $T_c$ of 212-233\ K. This is significantly higher than the (Allen-Dynes) value of $\sim115$\ K obtained in Ref.\ \cite{qian2017theoretical}; we elaborate on the source of this discrepancy in the Supplementary Information \cite{supplement}.

\begin{figure}
    \centering
    \includegraphics[width=0.68\columnwidth]{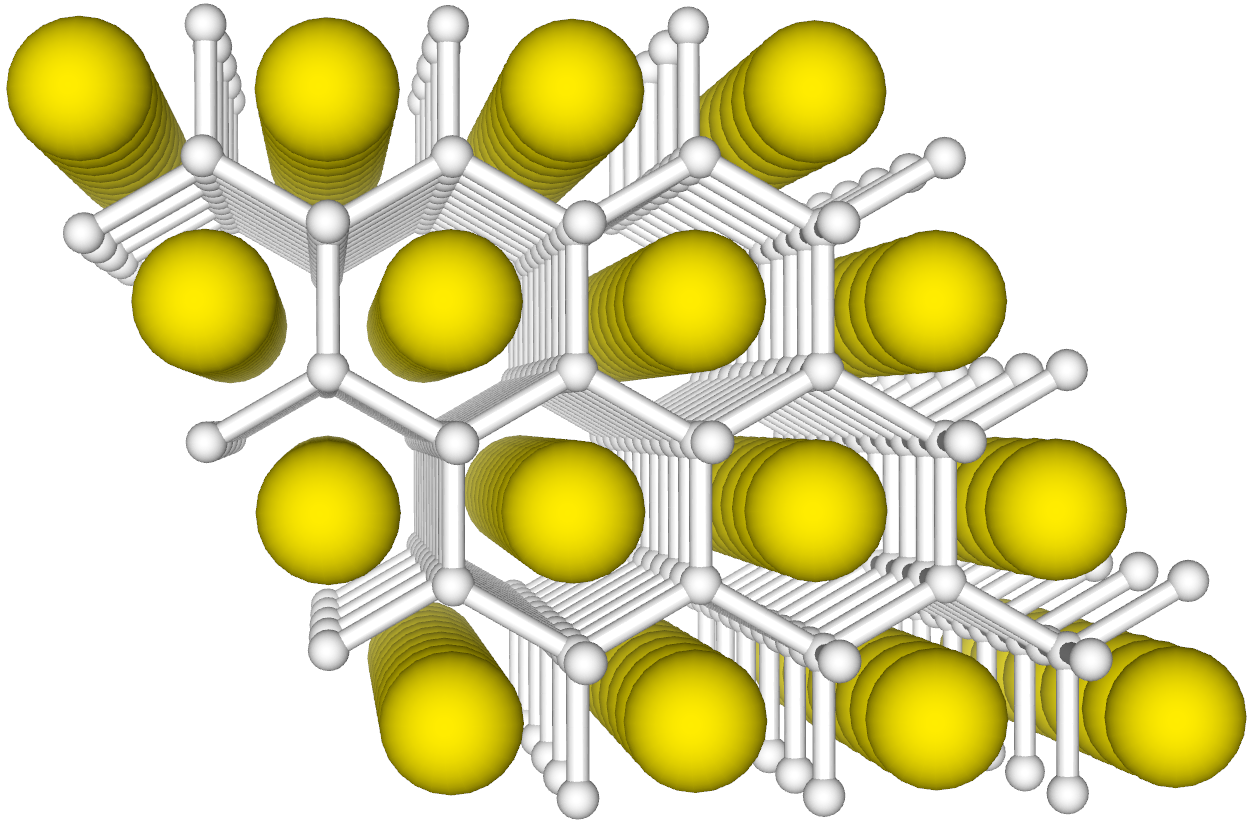}
    \caption{The 300\ GPa $P6/mmm$ structure of LiH$_2$ as viewed with perspective along the [001] direction of the standardized cell.}
    \label{fig:LiH2-P6_mmmm}
\end{figure}

The next structure of note is a metastable $P6/mmm$ structure of LiH$_2$, which is interesting both because of its relatively low hydrogen content and because its structure is analogous to the well-known ambient-pressure superconductor, MgB$_2$ \cite{Nagamatsu2001} (see Fig.\ \ref{fig:LiH2-P6_mmmm}). Superconductivity in Li-H systems has been investigated previously at lower pressures \cite{xie2014lithium}, where it was found that the LiH$_2$ stoichiometry did not exhibit superconductivity at 150\ GPa. We find that the LiH$_2$ stoichiometry lies above the static-lattice convex hull at 300\ GPa. However, we find this stoichiometry is stabilised (and lies on the hull, in agreement with Ref.\ \cite{zurek2009lithium}) when using harder pseudopotentials with denser $\textbf{k}$-point grids.

\subsection{500GPa}
\begin{figure}
    \centering
    \subfigure[]{\includegraphics[width=0.34\columnwidth]{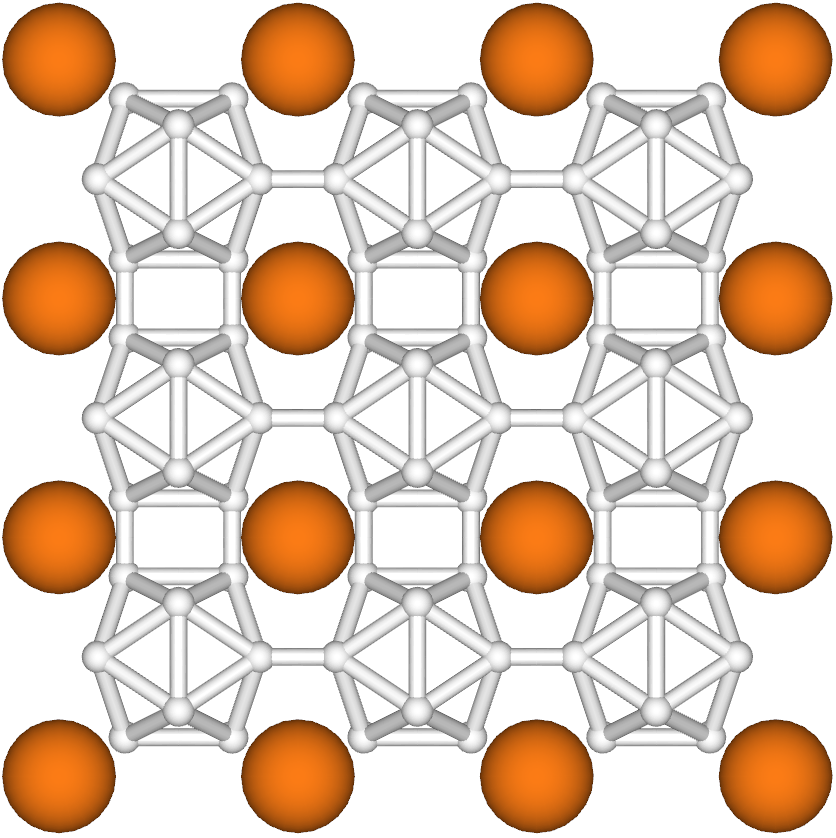}}
    \subfigure[]{\includegraphics[width=0.54\columnwidth]{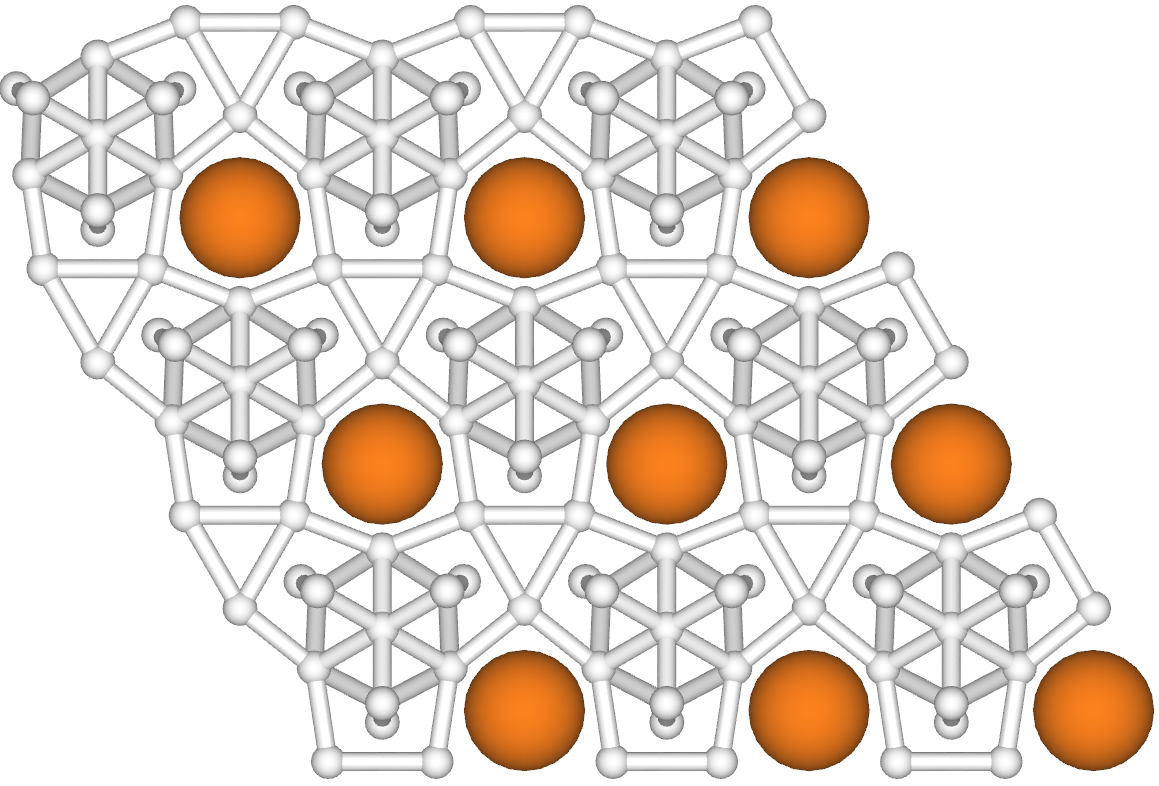}}
    \caption{(a) The 500\ GPa $Pm\bar{3}$ structure of MgH$_{12}$ as viewed along the [010] direction of the standardized cell. (b) The 500\ GPa $P3m1$ structure of MgH$_{13}$ as viewed along the [001] direction of the standardized cell.}
    \label{fig:MgH-500GPa}
\end{figure}

\begin{figure}
    \centering
    \includegraphics[width=\columnwidth]{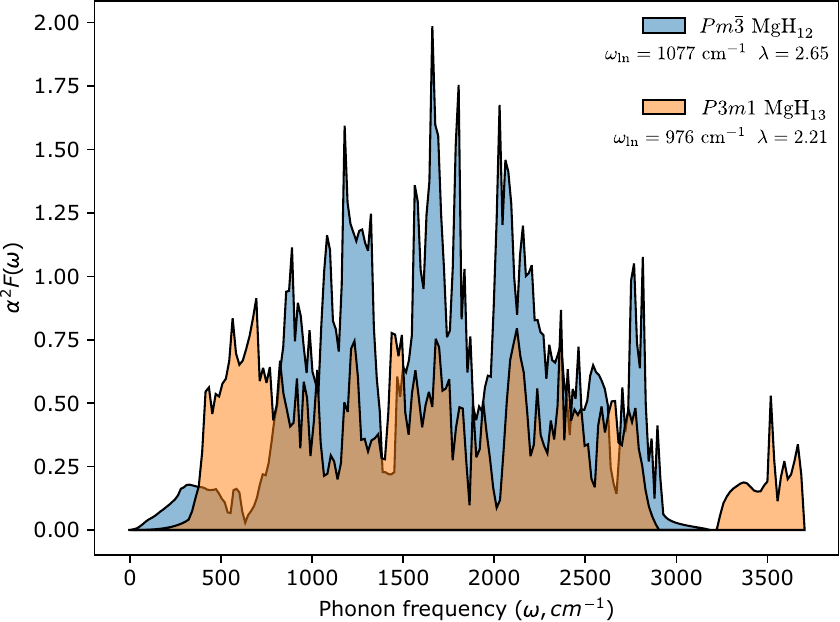}
    \caption{The Eliashberg functions of the $Pm\bar{3}$ structure of MgH$_{12}$ (a hot superconductor with a $T_c$ of 360-402\ K) and the $P3m1$ structure of MgH$_{13}$, both at 500\ GPa.}
    \label{fig:500GPa_Mg_eliashberg}
\end{figure}

As we increase pressure further, hydrides with higher hydrogen content can be metallised. In particular, at 500\ GPa we see the appearance of several MgH$_n$ structures with $n > 10$. The highest critical temperature of these belongs to MgH$_{12}$, which we find to be on the convex hull at this pressure, where a $Pm\bar{3}$ structure (see Fig.\ \ref{fig:MgH-500GPa}(a)) exhibits hot superconductivity with a critical temperature of 360-402\ K (87-129$^\circ$C). This extremely high critical temperature is due to strong electron-phonon coupling that persists over the entire phonon spectrum, extending to very high frequencies (see Fig.\ \ref{fig:500GPa_Mg_eliashberg}). Despite increased hydrogen content, $P3m1$-MgH$_{13}$ (see Fig.\ \ref{fig:MgH-500GPa}(b)) exhibits a lower $T_c$ of 257-287\ K due to reduced (but still extended) electron-phonon coupling (see Fig.\ \ref{fig:500GPa_Mg_eliashberg}). Superconductivity in the Mg-H system appears to be enhanced substantially with increased pressure; the MgH$_{12}$ stoichiometry has been previously investigated at lower pressures \cite{lonie2013metallization}, where it was also found to lie on the convex hull, but with a much reduced $T_c$ of 47-60\ K at 140\ GPa.

At 500\ GPa, we also find that a high-symmetry metastable $Fm\bar{3}m$ phase of SrH$_{10}$ exhibits room-temperature superconductivity with a $T_c$ of 285-319\ K (12-46$^\circ$C). This is significantly higher than the 190-228\ K we calculate for the ground-state structure of this stoichiometry, which we find to have $R\bar{3}m$ symmetry in agreement with previous calculations at 300\ GPa \cite{wang2015structural}. A $T_c$ of 259\ K had been calculated for the $R\bar{3}m$ structure previously at this lower pressure \cite{tanaka2017electron}.

\subsection{Dynamically unstable structures}
\label{sec:dynamical_instabilities}

Structures that are found to be dynamically unstable correspond to saddle-points, rather than minima, of the potential energy surface. Such structures can be stabilized by anharmonic or thermal effects, as is the case in $Fm\bar{3}m$-LaH$_{10}$ at 200\ GPa \cite{errea2019, shipley2020, drozdov2019superconductivity}, and locating them is the goal of sp-AIRSS \cite{monserrat2018v}. The calculations required to filter out, or correctly treat \cite{Errea2016, PhysRevB.89.064302, Hui_1974, PhysRevB.57.14453}, dynamically unstable superconductors are expensive; the solution of this problem will be important in the further development of high-throughput workflows. However, one can roughly establish the promise of a saddle-point superconductor by simply neglecting unstable modes in the calculation of the Eliashberg function \cite{PhysRevB.87.115124, shipley2020}. We provide the results of this procedure for dynamically unstable structures that were flagged by the final model in the Supplementary Information \cite{supplement} (see also Fig.\ \ref{fig:convergedresults}), along with a short discussion of the validity of this procedure.

\subsection{A modified Allen-Dynes equation}
\label{sec:allen_dynes_vs_eliashberg}
\begin{figure}
    \centering
    \includegraphics[width=0.75\columnwidth]{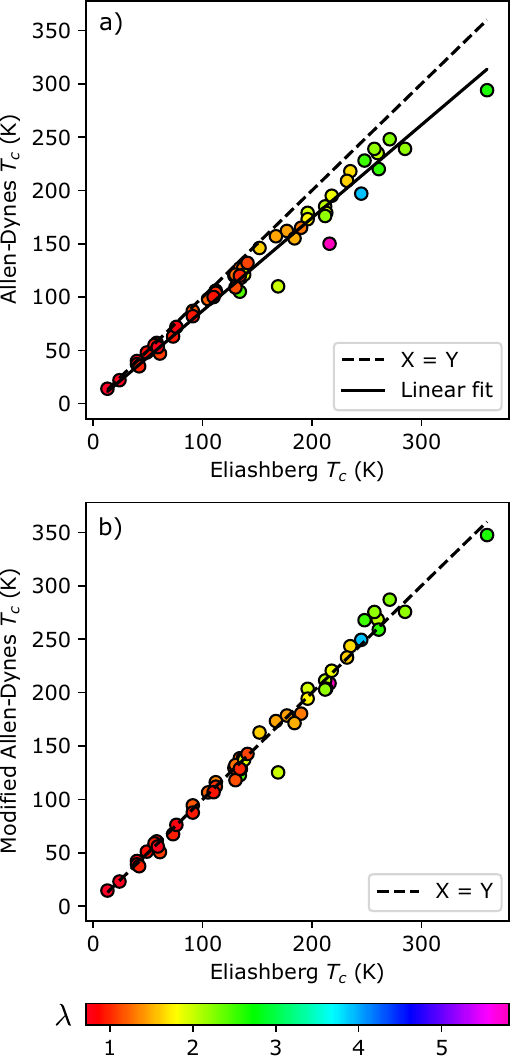}
    \caption{(a) Allen-Dynes critical temperatures, plotted against critical temperatures from solution of the Eliashberg equations (data from Table \ref{tab:stable_superconductors}). (b) The same as (a), but using the modified Allen-Dynes equation (Eq.\ \ref{eq:modified_allen_dynes}).}
    \label{fig:allen_dynes_vs_eliashberg}
\end{figure}

Having a large number of superconductors for which the Eliashberg equations have been solved directly (see Table \ref{tab:stable_superconductors}) provides a unique opportunity to test the Allen-Dynes equation. This comparison is made in Fig.\ \ref{fig:allen_dynes_vs_eliashberg} (a), where it is clear to see that, while the Allen-Dynes result correlates well with the Eliashberg result, it systematically underestimates its value (at least for the binary hydride superconductors studied in this work). We fit a modified version of the Allen-Dynes equation of the form
\begin{equation}
\label{eq:modified_allen_dynes}
    T_c = T_c^{(\text{AD})}(a + b\lambda)
\end{equation}
to the data given in Table \ref{tab:stable_superconductors}, which gives $a = 1.0061$ and $b = 0.0663$. As we can see in Fig.\ \ref{fig:allen_dynes_vs_eliashberg} (b) this removes the systematic underestimation and slightly reduces the variance of the prediction. However, given access to $\alpha^2F(\omega)$, we recommend that the Eliashberg equations be solved directly, to remove the need for approximate $T_c$ predictors entirely.

\section{Conclusions}
In this work, we demonstrate a high-throughput method to efficiently discover high-$T_c$ binary hydrides. We construct a $T_c$ model based on physically-motivated descriptors, trained initially on superconductivity data from the literature and iteratively updated using the results of our own DFPT calculations. Following extensive structure searching, a two-step screening process (based on stability criteria and the predictions of our model) allows us to identify energetically-competitive high-$T_c$ candidates from the large volume of search data. The best candidates include hydrides of sodium, calcium, actinium, lanthanum, magnesium, yttrium, scandium, lithium and strontium.

We have performed a total of 240 $T_c$ calculations using DFPT, split roughly equally between the training phase and final results. This was made possible by optimizing the \textsc{quantum espresso} electron-phonon code. In the final results, we identify 36 dynamically stable superconductors with $T_c > 100$\ K of which 18 have $T_c > 200$\ K (see Table\ \ref{tab:stable_superconductors}). To the best of our knowledge, superconductivity has not been investigated previously in 27 of the 36 (see Table \ref{tab:stable_superconductors} and the Supplementary Information \cite{supplement}). These findings add considerably to the known pressure-$T_c$ behaviour of the binary hydrides. Of particular note, we find a $Pm\bar{3}m$ structure of NaH$_6$ to have a $T_c$ of 248-279\ K at 100 GPa, suggesting the exciting possibility of other low-pressure high-$T_c$ hydride superconductors. We also identify $Pm\bar{3}$-MgH$_{12}$ and $Fm\bar{3}m$-SrH$_{10}$ as above-room-temperature superconductors at 500 GPa, as well as several near-room-temperature superconductors at lower pressures.

Throughout this work, our aim has been to consider as wide a range of binary compositions as possible; since our focus is on breadth, we make no claim that this study is exhaustive. Despite this, we identify a large number of high-$T_c$ candidates, suggesting the binaries have more to offer in future, more focused, studies. We also note that the highest critical temperature results at each pressure arise from metastable structures or off-hull stoichiometries. Many of these were introduced into the study during the training phase by focusing on high-symmetry structures using sp-AIRSS. This suggests that the additional freedom afforded by allowing some degree of metastability can reveal higher critical temperature superconductors. For example, in the case of SrH$_{10}$ at 500\ GPa we see that the ground state $R\bar{3}m$ structure has a critical temperature nearly 100\ K lower than a metastable $Fm\bar{3}m$ structure. Exploring avenues such as metastability will be important in future work in order to push the boundaries of high-temperature superconductivity.

Future work on superconducting hydrides is likely to focus on ternary hydrides and higher order systems. Given the increased complexity of these systems, high-throughput screening approaches, such as the one presented here, are likely to become increasingly important. Our high-throughput methodology could be extended to ternary hydrides, although it may be desirable to redefine these systems as effective binaries in order make use of the extensive binary hydride literature data during model training.

\nocite{xie2019high, ying2019ternary, szczesniak2016superconductivity, chen2017prediction, Camargo_Mart_nez_2020, wei2016pressure, smith2009high, shao2019unique, PhysRevLett.125.217001, hooper2014composition}

\section{Acknowledgements}
AMS is funded by an EPSRC studentship. MJH acknowledges the EPSRC Centre for Doctoral Training in Computational Methods for Materials Science for funding under grant number EP/L015552/1. RJN is supported by EPSRC under Critical Mass Grant EP/P034616/1 and the UKCP consortium grant EP/P022596/1. CJP was supported by the Royal Society through a Royal Society Wolfson Research Merit award. This work has been performed using resources provided by the Cambridge Tier-2 system operated by the University of Cambridge Research Computing Service (\url{http://www.hpc.cam.ac.uk}) funded by EPSRC Tier-2 capital grant EP/P020259/1.

\bibliography{references.bib}
\clearpage

\section*{Supplementary information}

\section{The critical temperature of \texorpdfstring{$Immm$-S\MakeLowercase{c}H$_8$}{Immm ScH8}}
We find that an $Immm$ structure of ScH$_8$ exhibits superconductivity at 300\ GPa with a $T_c$ of 212-233\ K. This is significantly higher than the (Allen-Dynes) value of $\sim115$\ K obtained in Ref.\ \cite{qian2017theoretical} using a $16\times16\times16$ $\mathbf{k}$-point grid and norm-conserving pseudopotentials with 3 valence electrons for Sc. In contrast, we use a $36\times36\times36$ $\mathbf{k}$-point grid and ultrasoft pseudopotentials with 11 valence electrons for Sc; a more substantial investigation into pseudopotentials is therefore needed to fully resolve this discrepancy. While they did not calculate $T_c$ for ScH$_8$, critical temperatures of 213\ K and 233\ K were obtained at 300\ GPa for ScH$_7$ and ScH$_9$, respectively, in Ref.\ \cite{ye2018high} (remarkably close to our range for ScH$_8$).

\section{Dynamically unstable critical temperatures}
For dynamically unstable structures, a large amount of spectral weight can be introduced to the Eliashberg function near to $\omega = 0$ due to the unstable modes. Similarly to what was seen for $Im\bar{3}m$ CaH6 at 100\ GPa in Fig.\ 7 of the main text, this strongly affects the Allen-Dynes critical temperature ($T_c^{(AD)}$), because the functional derivative $\delta T_c^{(AD)} / \delta \alpha^2F(\omega)$ diverges (towards $-\infty$) as $\omega \rightarrow 0$. However, the critical temperature derived from solution of the Eliashberg equations ($T_c^{(E)}$) is much less sensitive ($\delta T_c^{(E)} / \delta \alpha^2F(\omega) \rightarrow 0$ as $\omega \rightarrow 0$ \cite{Camargo_Mart_nez_2020}). Therefore, if one of these unstable superconductors is in reality stabilized by anharmonic effects (as is the case for $Fm\bar{3}m$-LaH$_{10}$ at 200\ GPa \cite{errea2019, shipley2020, drozdov2019superconductivity}), then a rough estimate of its promise as a high-temperature superconductor can be estimated by simply neglecting unstable modes in a harmonic calculation of $T_c^{(E)}$. The results of this procedure are shown in Table.\ \ref{tab:unstable_superconductors}, along with the same procedure for $T_c^{(AD)}$ for comparison. In carrying out a harmonic calculation, we are neglecting anharmonic renormalization of the phonon frequencies, which may affect the critical temperatures, especially if there is a large amount of spectral weight near to $\omega = 0$ to be renormalized. Therefore, an anharmonic treatment is needed to determine the stability and critical temperatures more accurately, especially when the Allen-Dynes and Eliashberg results strongly disagree (as they do in many cases). These results might therefore be used to identify interesting systems to investigate with anharmonic methods.

\begin{table}[H]
\centering
\footnotesize
\begin{tabular}{llllll}
    Stoichiometry & Space group &  P(GPa) & $T_c^{(AD)}$(K) & $T_c^{(E)}$(K) & $E_{stoic}$ (meV/unit)  \\
    \hline \hline
    LaH$_{6}$	   & $Im\bar{3}m$	 & 100	& 119-151 & 226-250 & 12$^*$ \\
    AcH$_{10}$	   & $Fm\bar{3}m$	 & 100	& 75-125  & 171-203 & 8$^*$ \\
    NaH$_{16}$	   & $P\bar{3}1m$ 	 & 100	& 81-99   & 138-163 & 0$^*$  \\
    CaH$_{11}$     & $R\bar{3}c$     & 100	& 30-50   & 128-161 & 28$^*$ \\
    LaH$_{14}$	   & $R\bar{3}m$ 	 & 100	& 80-98   & 99-123  & 33$^*$  \\
    LaH$_{24}$	   & $C2$      	     & 100	& 37-55   & 64-82 & 0 \\
    AcH$_{21}$	   & $P\bar{1}$	     & 100	& 28-40   & 28-41 & 0 \\
    CaH$_{24}$     & $Cmcm$	         & 100	& 19-29   & 19-32 & 0$^*$  \\
    \hline
    AcH$_6$	       & $I4/mmm$        & 200	& 129-166 & 279-315 & 0$^*$ \\
    LaH$_{10}$	   & $Fm\bar{3}m$	 & 200	& 235-260 & 269-279 & 0 \\
    LaH$_{10}$	   & $Immm$	         & 200	& 60-180  & 265-296 & 0$^*$  \\
    AcH$_{6}$	   & $Cmcm$	         & 200	& 171-220 & 248-304 & 0  \\
    NaH$_{10}$     & $Cmmm$          & 200  & 160-170 & 243-283 & 8 \\
    ScH$_8$	       & $R\bar{3}m$	 & 200	& 86-179  & 242-271 & - \\
    AcH$_6$ 	   & $Im\bar{3}m$	 & 200	& 63-78   & 241-265 & 0$^*$  \\
    LaH$_{10}$	   & $R\bar{3}m$	 & 200	& 45-54   & 225-270 & 0$^*$ \\
    LaH$_{10}$	   & $C222_1$        & 200	& 169-195 & 221-248 & 0$^*$  \\
    H$_3$S	       & $Fmmm$	         & 200	& 154-180 & 216-246 & 0$^*$  \\  
    AcH$_6$ 	   & $Pmmn$	         & 200	& 67-69   & 208-238 & 0$^*$  \\
    NaH$_4$	       & $Immm$	         & 200	& 106-126 & 191-218 & 61$^*$  \\
    MgH$_7$ 	   & $C2/m$	         & 200	& 85-102  & 188-220 & 29$^*$  \\
    AcH$_6$   	   & $C2/m$          & 200	& 100-112 & 184-214 & 0$^*$  \\
    AcH$_{11}$	   & $Imm2$	         & 200	& 88-108  & 165-184 & 0$^*$  \\
    NaH$_5$ 	   & $P4/mmm$        & 200	& 66-75   & 134-163 & 21$^*$  \\
    NaH$_{12}$     & $R\bar{3}m$     & 200  & 50-60   & 122-151 & 4 \\
    MgH$_6$ 	   & $P\bar{1}$	     & 200	& 71-104  & 98-121  & 19$^*$  \\
    MgH$_5$ 	   & $C2/m$	         & 200	& 58-75   & 71-95   & 30 \\
    Na$_2$H$_7$	   & $P\bar{1}$	     & 200	& 20-140  & 57-169  & 98$^*$  \\
    \hline
    ScH$_{12}$     & $P4/mcc$        & 300	& 120-136 & 302-332 & 0$^*$  \\
    NaH$_{10}$	   & $Cmmm$	         & 300	& 151-177 & 243-283 & 3  \\
    KH$_{11}$	   & $P4/nmm$        & 300	& 109-140 & 233-256 & 7$^*$  \\
    NaH$_{14}$     & $P4/mmm$        & 300	& 103-126 & 220-245 & 1$^*$  \\
    SrH$_{10}$	   & $Cmcm$	         & 300	& 133-156 & 215-250 & 0$^*$  \\
    Na$_2$H$_{11}$ & $P2/m$	         & 300	& 60-71   & 154-185 & 0$^*$  \\
    NaH$_9$        & $P2/m$	         & 300	& 67-80   & 151-186 & 0$^*$  \\
    LiH$_4$        & $C2/m$	         & 300	& 93-123  & 131-168 & 17  \\
    MgH$_{14}$     & $P\bar{1}$	     & 300	& 55-66   & 117-143 & 14  \\
    NaH$_{12}$	   & $R\bar{3}m$	 & 300	& 66-79   & 88-115  & 2  \\
    KH$_{6}$	   & $Immm$	         & 300	& 40-53   & 46-64   & 3  \\
    \hline
    LiH$_{12}$     & $Pm\bar{3}m$	 &  500 & 244-330 & 433-485 & 26$^*$ \\
    NaH$_6$	       & $Im\bar{3}m$	 &  500 & 215-267 & 357-406 & 30$^*$ \\
    YH$_{17}$	   & $R32$	         &  500 & 137-164 & 287-331 & 29 \\
    MgH$_9$	       & $Cmmm$	         &  500 & 133-168 & 284-336 & 7 \\
    CaH$_{10}$	   & $Fm\bar{3}m$    &  500 & 157-195 & 238-279 & 3$*$ \\
    SrH$_{22}$	   & $R32$	         &  500 & 167-194 & 227-266 & 0$^*$ \\
    Y$_2$H$_{19}$  & $C2$	         &  500 & 188-220 & 214-251 & 37 \\
    MgH$_{12}$     & $Fmmm$ 	     &  500 & 161-200 & 208-245 & 0  \\
    YH$_{22}$	   & $P\bar{1}$	     &  500 & 90-110  & 203-234 & 34 \\
    NaH$_4$	       & $I4_1/amd$      &  500 & 83-97   & 193-233 & 25 \\
    YbH$_{10}$	   & $R\bar{3}m$	 &  500 & 94-136  & 176-207 & 0$^*$ \\
    YbH$_{10}$	   &  $Fm\bar{3}m$   &	500	& 76-92   & 164-202 & 0  \\
    NaH$_5$	       & $P4/mmm$        &  500 & 86-116  & 85-118 & 26  \\
    MgH$_4$	       & $Immm$	         &  500 & 60-70   & 66-90 & 0$^*$  \\
    MgH$_4$	       & $I4/mmm$        &  500 & 40-63   & 48-77 & 0  \\
\end{tabular}
\caption{Dynamically unstable candidate superconductors. Those structures marked with $^*$ are not the lowest energy structure for the given stoichiometry (i.e. have non-zero $E_{struc}$). The Sc-H system was not in the focused searches at 200\ GPa.}
\label{tab:unstable_superconductors}
\end{table}

\clearpage
\section{100\ K+ dynamically stable superconductors}

\newcommand{\help}{\\&&&&}
\newcommand{\spacepls}{\vspace{0.1cm}}

\begin{table}[H]
\scriptsize
    \centering
    \begin{tabular}{lllll}
        Stoichiometry & Space group &  P(GPa) & $T_c$(K) & Comments \\
        \hline
        
        NaH6	& $Pm\bar{3}m$	   & 100 &	263.5 & 
        Superconductivity not previously studied to the best of our knowledge. \help
        Ref.\ \cite{struzhkin2016} found NaH$_6$ off hull at 50 GPa. \help
        Ref.\ \cite{baettig2011} found $Pm\bar{3}m$-NaH6 becomes stable 
        with respect to $P1$ above 150 GPa.\spacepls\\
        
        CaH6	& $Im\bar{3}m$	   & 100 &	234.5 & 
        Superconductivity studied previously for this structure in Ref.\ \cite{wang2012superconductive}. \help
        Ref.\ \cite{shao2019unique} found CaH6 slightly above the hull at 100 GPa.\spacepls\\
        
        NaH6	& $Pm\bar{3}m$	   & 200 &	274   & See above comments for NaH$_6$ at 100\ GPa.\spacepls\\
        
        AcH12	& $P63mc$	       & 200 &	262.5 & 
        Space group not studied elsewhere to the best of our knowledge. \help
        Ref.\ \cite{semenok2018actinium} studies $I4/mmm$ AcH12 ($T_c$=148-173 K at 150 GPa). \help
        They find AcH12 on hull at 150 GPa, slightly above at 250 GPa.\spacepls\\
        
        MgH13	& $Fm\bar{3}m$	   & 200 &	210   & 
        Space group not studied elsewhere to the best of our knowledge. \help
        Refs.\ \cite{lonie2013metallization} and \cite{ying2019ternary} predict MgH13 above hull at 200 and 300 GPa, respectively.\spacepls\\
        
        SH3	    & $Im\bar{3}m$	   & 200 &	207.5 & 
        Well-known structure from experiment \cite{drozdov2015conventional} and theory \cite{duan2014pressure, errea2015high}.\spacepls\\
        
        MgH6	& $Im\bar{3}m$	   & 300 &	286   & 
        Superconductivity studied previously for this structure. \help
        Ref.\ \cite{feng2015compressed} found MgH6 thermodynamically stable above \help
        263 GPa relative to MgH2 and H2, and $T_c$=260 K \help
        for the $Im\bar{3}m$ structure above 300 GPa. \help
        Ref.\ \cite{szczesniak2016superconductivity} calculated much higher $T_c$ for this structure - 420 K at 300 GPa.\spacepls\\
        
        YH9	    & $F\bar{4}3m$	   & 300 &	277   & 
        Space group not studied elsewhere to the best of our knowledge. \help
        Ref.\ \cite{peng2017hydrogen} predicted YH9 stoichiometry on hull 100-400 GPa, \help
        with $P63/mmc$ symmetry at 300 GPa. \help
        Ref.\ \cite{shipley2020} found YH9 on hull at 400 GPa with $P63/mmc$ lowest in energy. \help
        Ref.\ \cite{kong2019superconductivity} synthesised YH9 in $P63/mmc$ structure.\spacepls\\
        
        ScH8	& $Immm$	       & 300 &	222.5 & 
        Superconductivity studied previously for this structure. \help
        Ref.\ \cite{ye2018high} found ScH8 above static-lattice convex hull at \help
        150-350 GPa, but on hull at 350-400 GPa when ZPE included. \help 
        Their predicted phase behaviour is $Immm$ above 320 GPa. \help
        Ref.\ \cite{qian2017theoretical} found $Immm$ ScH8 stable above 300 GPa,
        with $T_c\approx$115 K at 300 GPa. \help
        Ref.\ \cite{peng2017hydrogen} found ScH8 above hull at 100, 200 and 300 GPa.\spacepls\\
        
        MgH12	& $Pm\bar{3}$	   & 500 &	381   & 
        Space group not studied elsewhere to the best of our knowledge. \help
        Ref.\ \cite{lonie2013metallization} looked at lower pressures, calculated $T_c$=47-60 K for $R3$ MgH12 at 140 GPa. \help
        Ref.\ \cite{ying2019ternary} found MgH12 on hull at 300 GPa.\spacepls\\
        
        MgH13	& $P3m1$	       & 500 &	272   & 
        Space group not studied elsewhere to the best of our knowledge. \help
        Ref.\ \cite{ying2019ternary} finds MgH13 above hull at 300 GPa.\spacepls\\
        
        SrH10	& $Fm\bar{3}m$	   & 500 &	302   &
        Space group not studied elsewhere to the best of our knowledge. \help
        Ref.\ \cite{wang2015structural} found a $P21/m$ structure at 50 GPa, \help
        a $P2/c$ structure at 150 GPa and transition to $R\bar{3}m$ at 300 GPa. \help
        Ref. \cite{semenok2018distribution} found $C2/m$ SrH10 at much lower pressures of ~100 GPa.\spacepls\\
        
        NaH9	& $P63/mmc$        & 500 &	252   & 
        Space group not studied elsewhere to the best of our knowledge. \help
        Ref.\ \cite{baettig2011} found a 25-300+ GPa stability range for NaH9 (though 500 GPa seems \help
        to be outside range of study) with $Cmc21$-NaH9 stable at 300 GPa.\spacepls\\
        
        MgH10	& $C2/m$	       & 500 &	251   & 
        Space group not studied elsewhere to the best of our knowledge. \help
        Ref.\ \cite{PhysRevLett.125.217001} found a $P63/mmc$ structure of MgH10 which was \help
        dynamically unstable at 300 GPa. \help
        Ref.\ \cite{ying2019ternary} predicts MgH10 to be above the hull at 300 GPa.\spacepls\\
        
        SrH24	& $R\bar{3}$	   & 500 &	231.5 & 
        Stoichiometry not studied elsewhere to the best of our knowledge. \help
        Refs.\ \cite{hooper2014composition, wang2015structural} both study Sr-H structures, \help
        but don't look at hydrogen content this high.\spacepls\\
        
        YH18	& $P\bar{1}$	   & 500 &	229.5 &
        Space group not studied elsewhere to the best of our knowledge. \help
        YH18 stoichiometry mentioned in Fig.\ 3b of Ref.\ \cite{kong2019superconductivity}.\spacepls\\
        
        YH20	& $P\bar{1}$	   & 500 &	228   & 
        Stoichiometry not studied elsewhere to the best of our knowledge. \help
        No high hydrogen content on/near hull up to 300 GPa in Ref.\ \cite{liu2017potential}. \help
        Higher H content YH24 on hull at 200 and 300 GPa in Ref.\ \cite{peng2017hydrogen}.\spacepls\\
        
        SrH10	& $R\bar{3}m$ 	   & 500 &	209   & 
        Superconductivity studied previously for this structure. \help
        Ref.\ \cite{wang2015structural} found a $P21/m$ structure at 50 GPa, a $P2/c$ structure  \help
        at 150 GPa and transition to $R\bar{3}m$ at 300 GPa. \help
        Ref.\ \cite{tanaka2017electron} calculated $T_c$=259 K for $R\bar{3}m$ at 300 GPa.\spacepls\\
        
        CaH10	& $R\bar{3}m$	   & 500 &	202   & 
        Superconductivity studied previously for this structure. \help
        Ref.\ \cite{shao2019unique} found $R\bar{3}m$-CaH10 metastable at 400 GPa with $T_c$=157-175 K. \help
        They find CaH10 lies above the hull at 50-400 GPa, \help
        but gets closer with increasing pressure.\spacepls\\
        
    \end{tabular}
    \caption{Dynamically stable superconductors with $T_c >$ 200\ K found in this work, along with their converged (average) Eliashberg $T_c$ values and notes on findings for these systems in previous work.}
    \label{tab:200K+supercondcutrs}
\end{table}

\clearpage
\begin{table}[H]
\scriptsize
    \centering
    \begin{tabular}{lllll}
        Stoichiometry & Space group &  P(GPa) & $T_c$(K) & Comments \\
        \hline
        
        Na2H11	& $Cmmm$	       & 100 &	147.5 & 
        Stoichiometry not studied elsewhere to the best of our knowledge.\spacepls\\
        
        KH10	& $C2/m$	       & 100 &	145.5 & 
        Space group not studied elsewhere to the best of our knowledge. \help 
        Ref.\ \cite{semenok2018distribution} finds KH10 not on hull at 50 GPa, but $Immm$-KH10 \help
        stable at 150 GPa with an Allen-Dynes $T_c$ of 148K.\spacepls\\
        
        NaH8	& $I4/mmm$         & 200 &	163.5 &
        Space group not studied elsewhere to the best of our knowledge. \help
        Ref.\ \cite{baettig2011} finds $P1$ structure below 180 GPa, and $Cmcm$ above.\spacepls\\
        
        AcH6	& $Fmmm$	       & 200 &	186.5 & 
        Space group not studied elsewhere to the best of our knowledge. \help
        Ref.\ \cite{semenok2018actinium} finds this stoichiometry is not on hull at 150 or 250 GPa.\spacepls\\
        
        Na2H11	& $Cmmm$	       & 200 &	145.5 & 
        Stoichiometry not studied elsewhere to the best of our knowledge.\spacepls\\
        
        MgH14	& $P\bar{1}$	   & 200 &	123   & 
        Stoichiometry not studied elsewhere to the best of our knowledge. \help
        Ref.\ \cite{ying2019ternary} did not find MgH14 on the hull at 300 GPa.\spacepls\\
        
        LaH7	& $C2/m$	       & 200 &	119.5 & 
        Superconductivity studied previously for this structure. \help
        Ref.\ \cite{kruglov2020} found LaH7 not on hull at 200 GPa. \help
        They report metastable $C2/m$ LaH7 (17meV/atom above the hull at 150 GPa) \help
        with Allen-Dynes $T_c$=158-185 K at 180 GPa.\spacepls\\
        
        LiH2	& $P6/mmm$         & 300 &	192   & 
        Space group not studied elsewhere to the best of our knowledge. \help
        Ref.\ \cite{zurek2009lithium} predicted 130-300+ GPa stability range for LiH2, \help
        with lowest energy structure having $P4/mbm$ symmetry. \help
        Ref.\ \cite{xie2014lithium} found that LiH2 does not exhibit superconductivity at 150 GPa. \help
        Ref.\ \cite{chen2017prediction} looked at lower pressures, found LiH2 to be \help
        stable at 130-200 GPa and also looked at $P4/mbm$.\spacepls\\
        
        NaH7	& $C2/m$	       & 300 &	182.5 & 
        Superconductivity not previously studied to the best of our knowledge. \help
        Stability range of NaH7 from Ref.\ \cite{baettig2011} is 25-100 GPa; \help
        they study a $Cc$ structure at low pressures and predict a transition to $C2/m$ at 245 GPa.\spacepls\\
        
        ScH12	& $P\bar{1}$	   & 300 &	151   & 
        Space group not studied elsewhere to the best of our knowledge. \help
        Ref.\ \cite{ye2018high} found $Immm$-ScH12 stable above 320 GPa with $T_c$=141-194 K at 350 GPa. \help
        They found the ScH12 stoichiometry to be on the static-lattice \help
        hull at 350 GPa and above, close at 300 GPa. \help
        Ref.\ \cite{peng2017hydrogen} found ScH12 on the hull at 300 GPa, identifying a $C2/c$ structure at this pressure.\spacepls\\
        
        LiH6	& $R\bar{3}m$	   & 300 &	145.5 & 
        Superconductivity studied previously for this structure. \help
        Ref.\ \cite{zurek2009lithium} found a 140-300+ GPa region of stability for LiH6 and found $R\bar{3}m$. \help
        Ref.\ \cite{xie2014lithium} calculated $T_c$=82 K at 300 GPa for $R\bar{3}m$ LiH6. \help
        Ref.\ \cite{chen2017prediction} also studied the $R\bar{3}m$ structure.\spacepls\\
        
        ScH6	& $Im\bar{3}m$	   & 300 &	148   & 
        Superconductivity studied previously for this structure. \help
        Ref.\ \cite{abe2017hydrogen} predicted $Im\bar{3}m$ ScH6 stable above 265 GPa with $T_c$=130 K at 285 GPa. \help
        Ref.\ \cite{peng2017hydrogen} found ScH6 on hull at 300 GPa and identified the $Im\bar{3}m$ structure. \help
        Ref.\ \cite{ye2018high} predicted a region of stability for $Im\bar{3}m$ ScH6 above 350 GPa.\spacepls\\
        
        NaH5	& $P4/mmm$         & 300 &	151   & 
        Space group not studied elsewhere to the best of our knowledge. \help
        Ref.\ \cite{struzhkin2016} studied lower pressures region, finding a structure with $P-1$ symmetry.\spacepls\\
        
        LiH6	& $C2/m$	       & 300 &	146.5 & 
        Space group not studied elsewhere to the best of our knowledge. \help
        See entry for $R\bar{3}m$ LiH6, $C2/m$ structure of LiH6 not mentioned in these references.\spacepls\\
        
        LiH3	& $Cmcm$	       & 300 &	126   & 
        Space group not studied elsewhere to the best of our knowledge. \help
        LiH3 found to be off-hull in Ref.\ \cite{zurek2009lithium} but no structure \help
        given, same as in Ref.\ \cite{chen2017prediction}.\spacepls\\
        
        ScH14	& $P\bar{1}$	   & 300 &	103   & 
        Stoichiometry not studied elsewhere to the best of our knowledge. \help
        Refs.\ \cite{ye2018high}, \cite{abe2017hydrogen} and \cite{wei2016pressure} do not study hydrogen content this high.\spacepls\\
        
        Na2H11	& $Cmmm$	       & 500 &	160.5 &
        Stoichiometry not studied elsewhere to the best of our knowledge.\spacepls\\
        
        CaH15	& $P\bar{6}2m$	   & 500 &	150.5 &
        Superconductivity not previously studied to the best of our knowledge. \help
        Structure found at 200 GPa in Ref.\ \cite{xie2019high}, but superconductivity not studied.\spacepls\\
        
        SrH15	& $P\bar{6}2m$	   & 500 &	124.5 &
        Stoichiometry not studied elsewhere to the best of our knowledge. \help
        Refs.\ \cite{tanaka2017electron, wang2015structural, hooper2014composition, smith2009high} do not study hydrogen content this high.\spacepls\\
        
        MgH8	& $C2/m$	       & 500 &	106   & 
        Space group not studied elsewhere to the best of our knowledge. \help
        MgH8 found to lie off of the convex hull at the lower pressures of 100/200\ GPa in Ref.\ \cite{lonie2013metallization}.\spacepls\\
        
    \end{tabular}
    \caption{Dynamically stable superconductors with $T_c$ between 100 and 200\ K found in this work, along with their converged (average) Eliashberg $T_c$ values and notes on findings for these systems in previous work.}
    \label{tab:100-200Ksuperconductors}
\end{table}

\clearpage

\section{Convex hulls}
On the following pages, we include the static-lattice convex hulls used to assess the stability of structures in the results stage of this work. These are produced from our focused searches as described in Section IV of the main text. To the right of each hull is a list of the on-hull stoichiometries and the space groups of the corresponding on-hull structures. The four columns in this list are stoichiometry, number of formula units in the cell, space group and composition label ($x$), respectively.

\clearpage

\newcommand{\convexwidth}{0.8\textwidth}

\begin{figure*}
    \centering
    \includegraphics[width=\convexwidth]{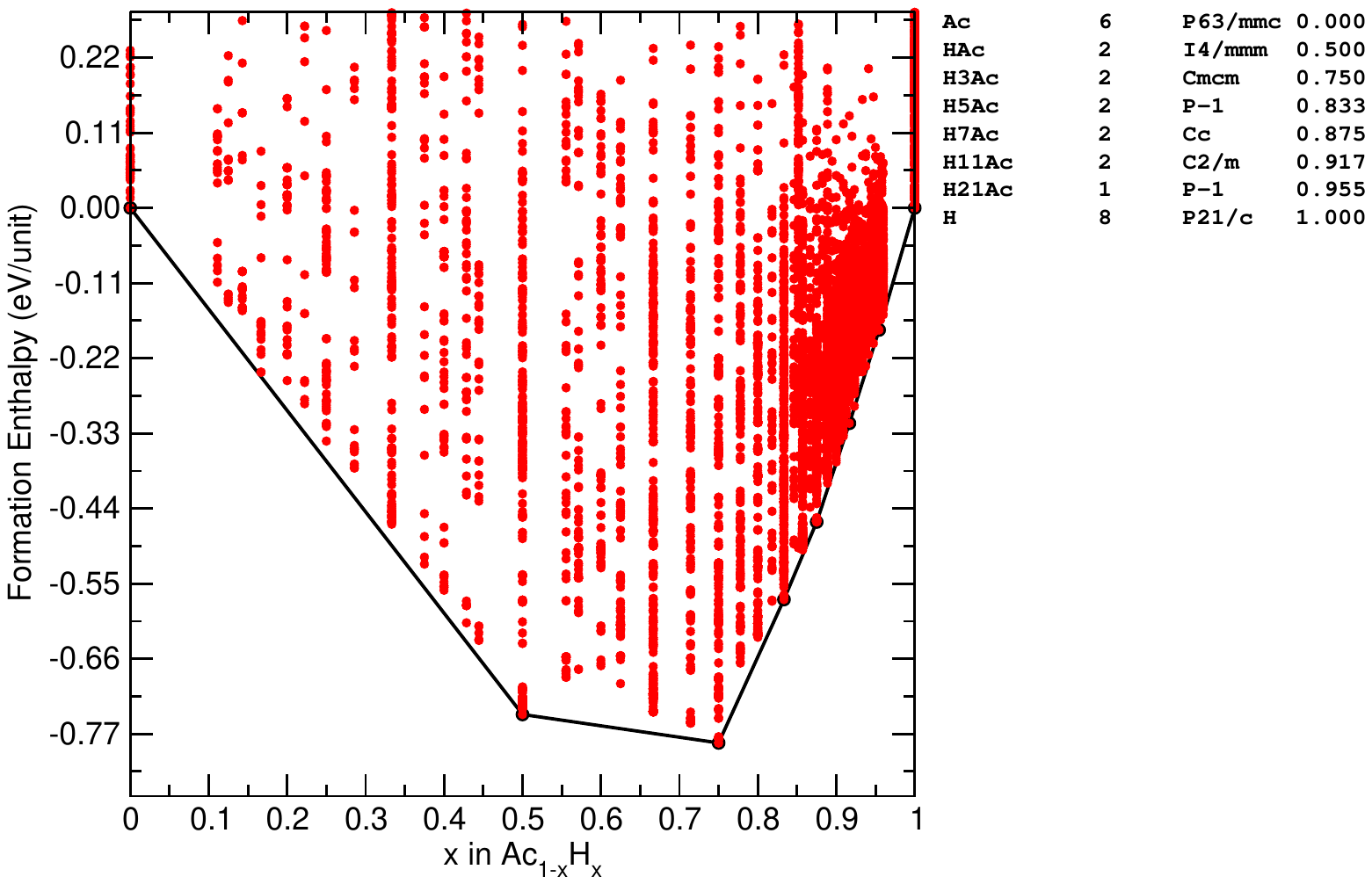}
    \caption{The convex hull of Ac-H at 100\ GPa.}
\end{figure*}

\begin{figure*}
    \centering
    \includegraphics[width=\convexwidth]{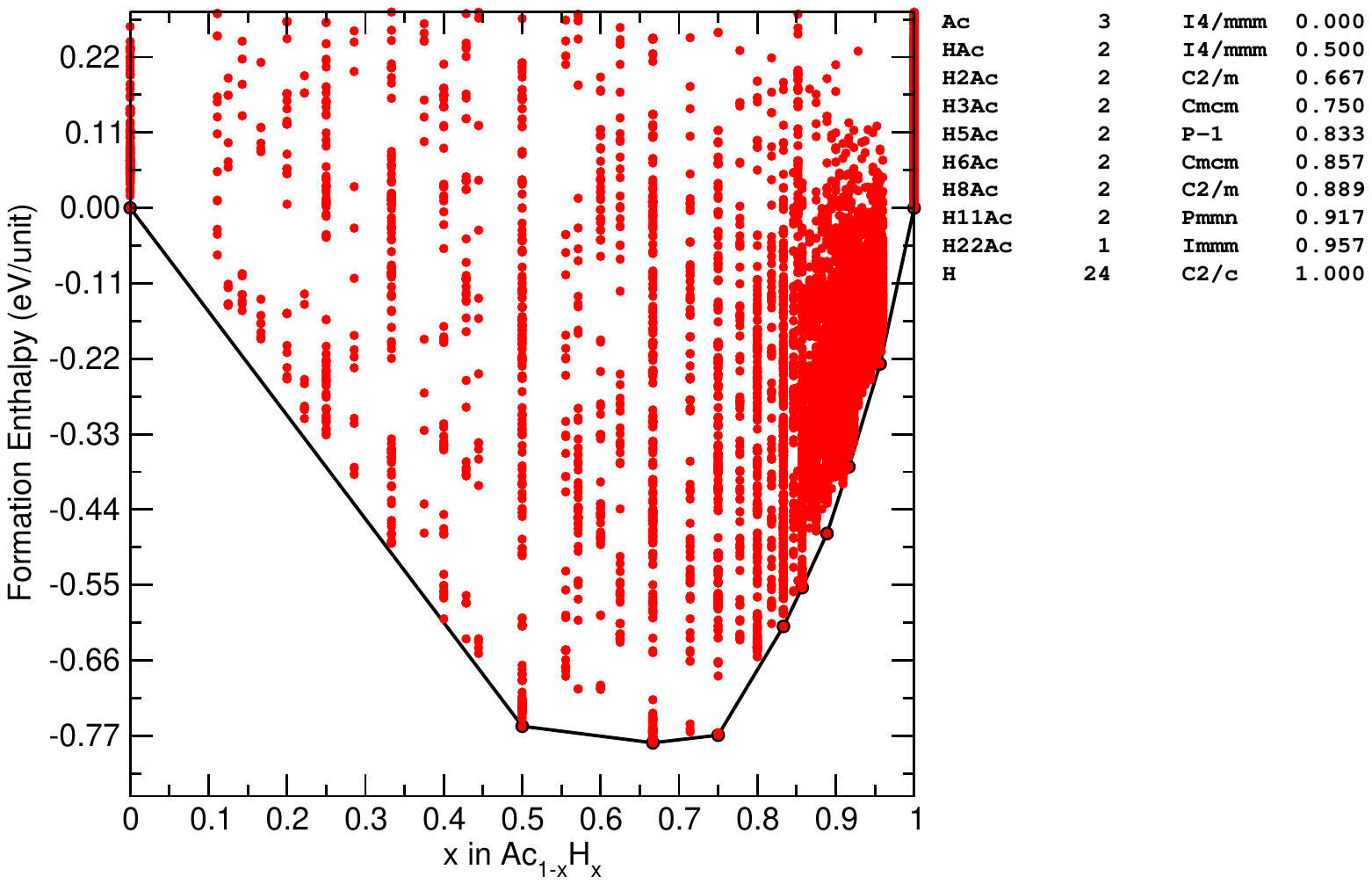}
    \caption{The convex hull of Ac-H at 200\ GPa.}
\end{figure*}

\begin{figure*}
    \centering
    \includegraphics[width=\convexwidth]{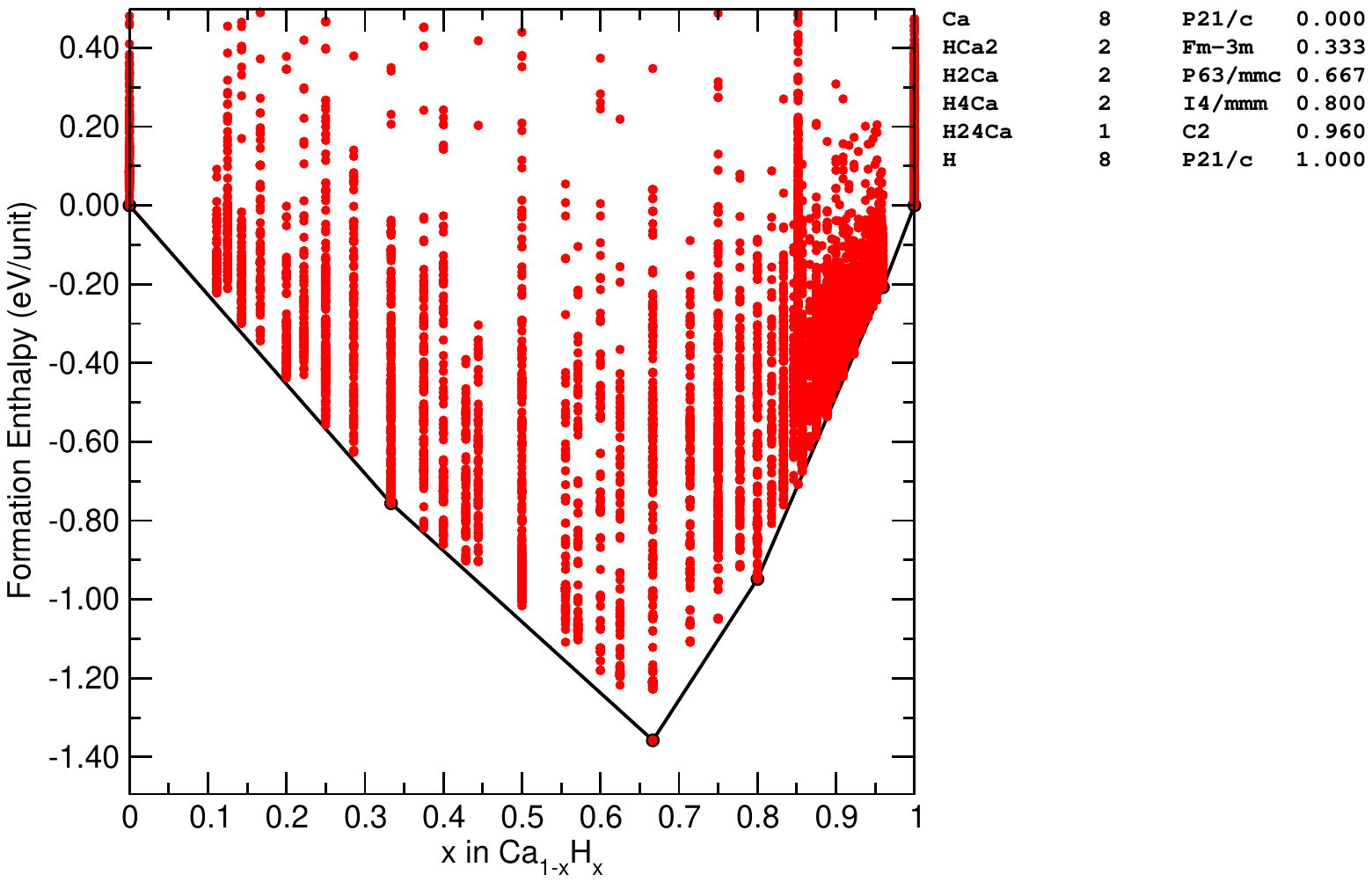}
    \caption{The convex hull of Ca-H at 100\ GPa.}
\end{figure*}

\begin{figure*}
    \centering
    \includegraphics[width=\convexwidth]{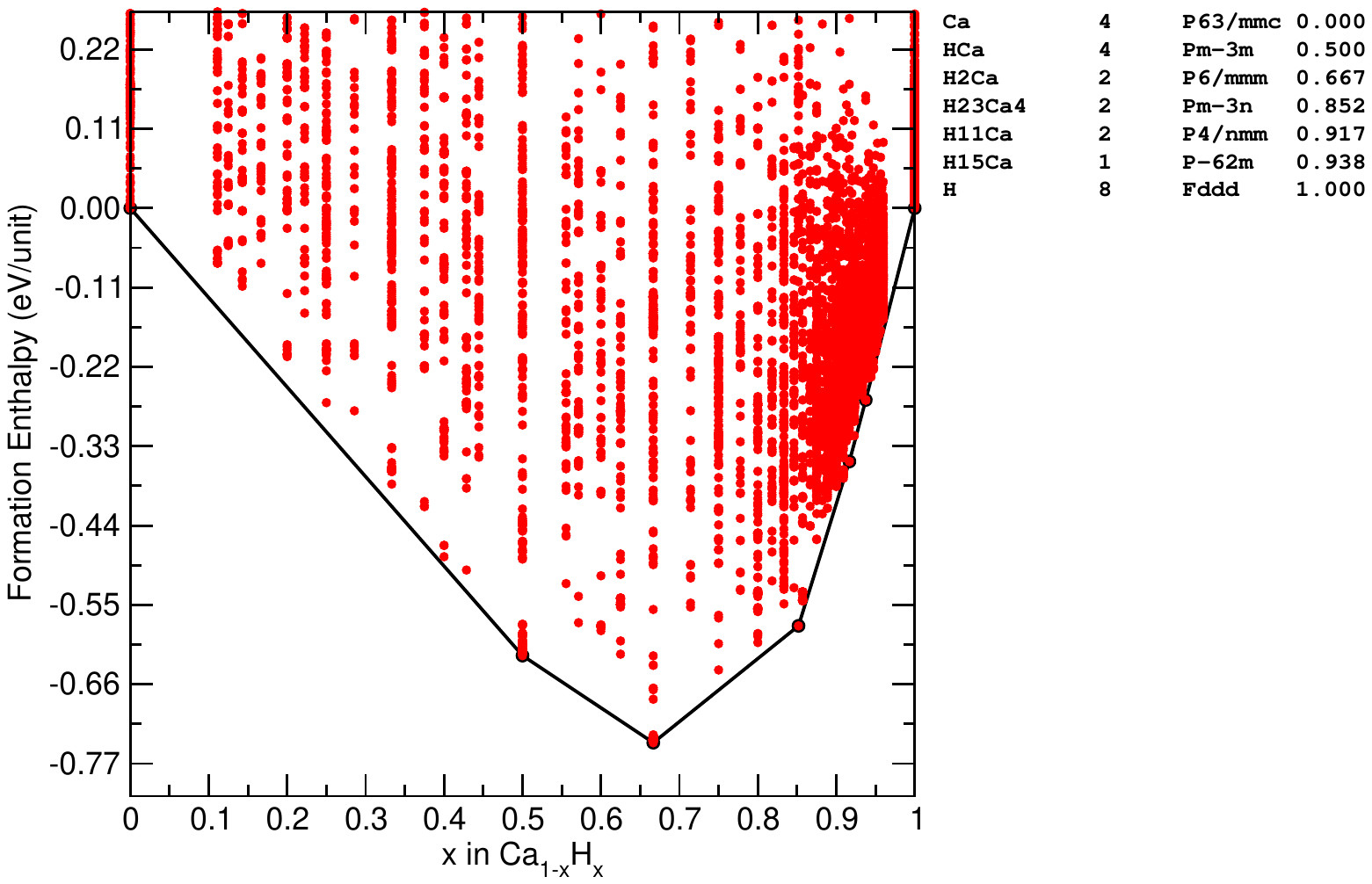}
    \caption{The convex hull of Ca-H at 500\ GPa.}
\end{figure*}

\begin{figure*}
    \centering
    \includegraphics[width=\convexwidth]{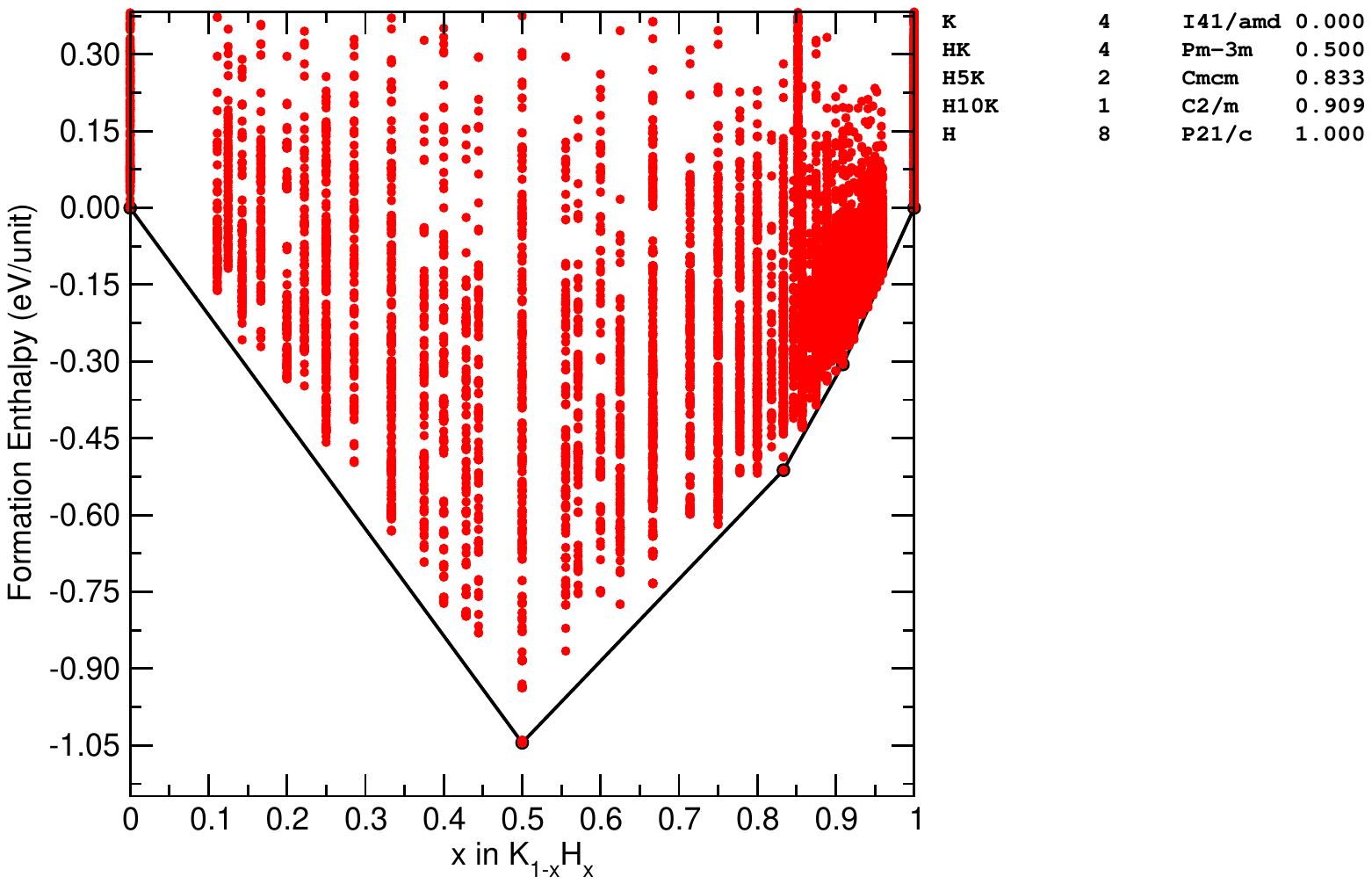}
    \caption{The convex hull of K-H at 100\ GPa.}
\end{figure*}

\begin{figure*}
    \centering
    \includegraphics[width=\convexwidth]{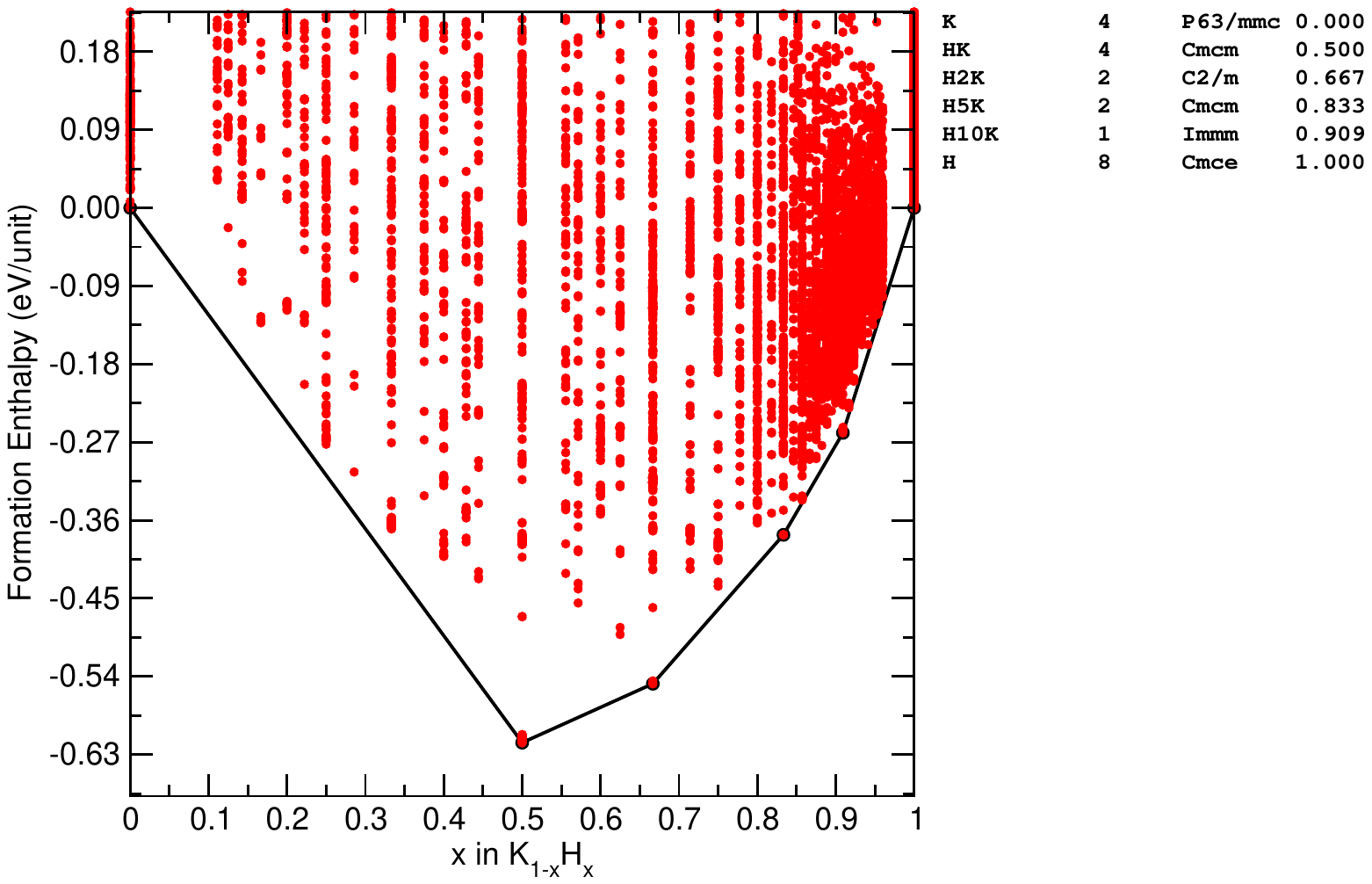}
    \caption{The convex hull of K-H at 300\ GPa.}
\end{figure*}

\begin{figure*}
    \centering
    \includegraphics[width=\convexwidth]{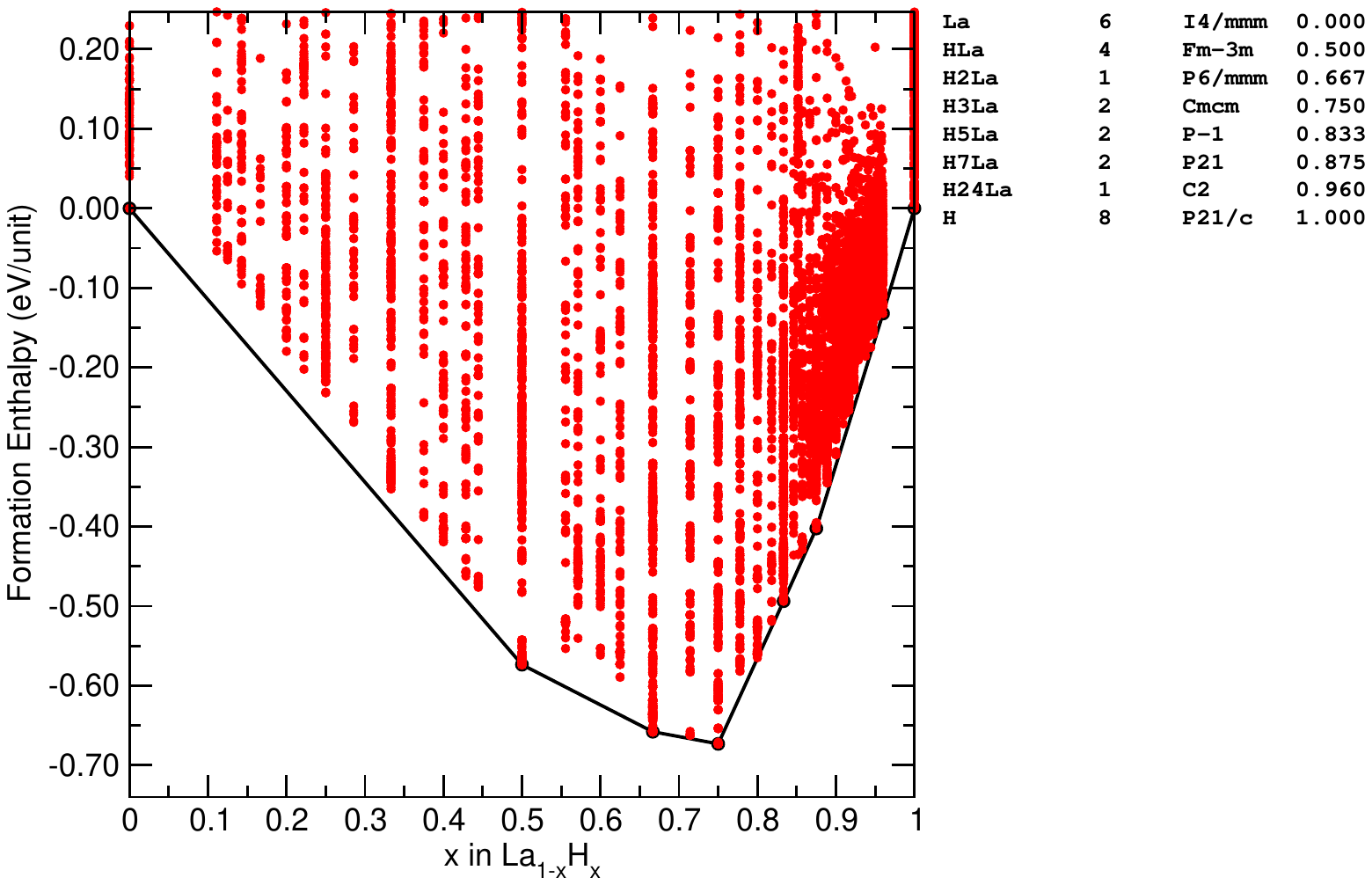}
    \caption{The convex hull of La-H at 100\ GPa.}
\end{figure*}

\begin{figure*}
    \centering
    \includegraphics[width=\convexwidth]{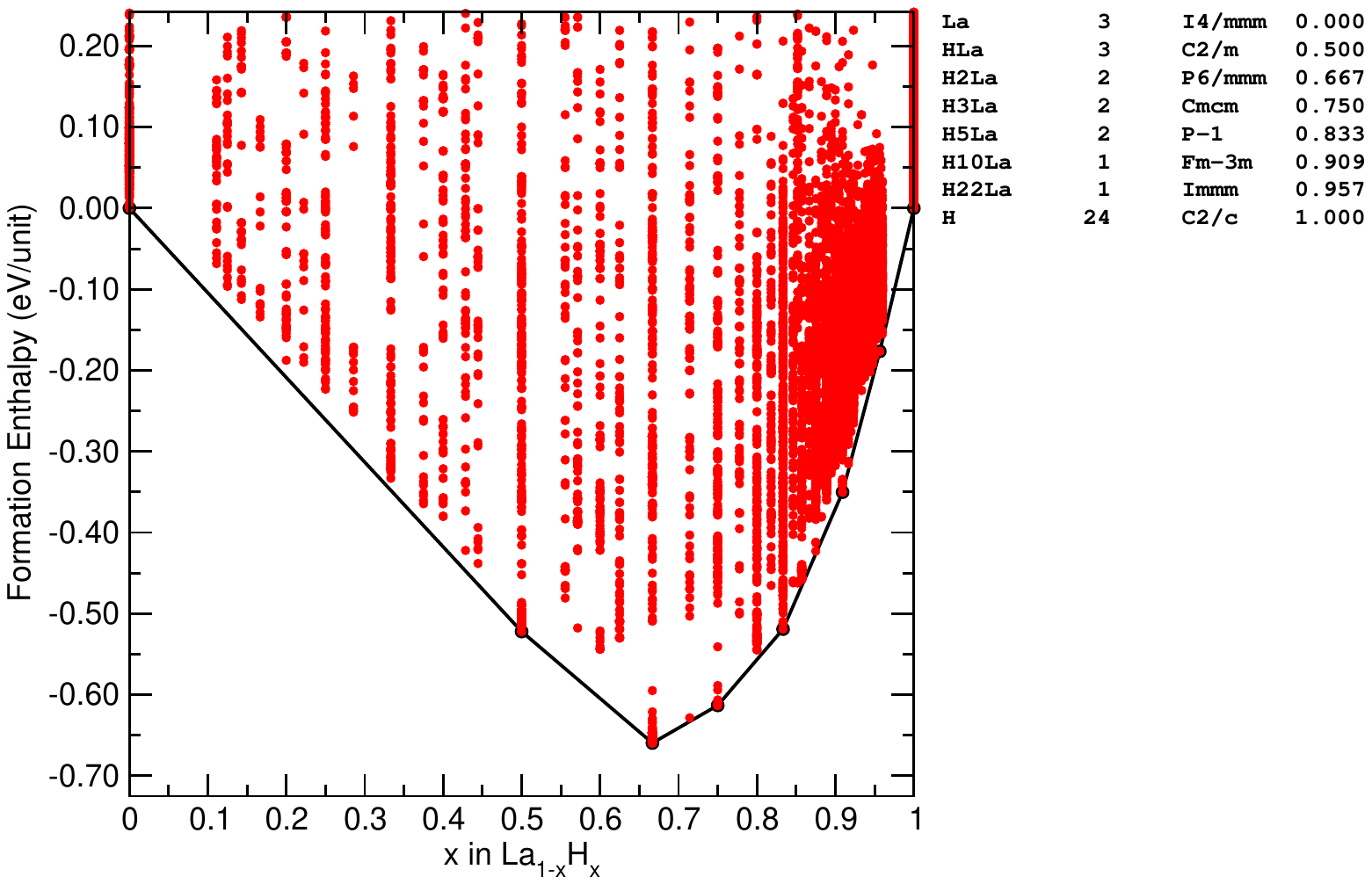}
    \caption{The convex hull of La-H at 200\ GPa.}
\end{figure*}

\begin{figure*}
    \centering
    \includegraphics[width=\convexwidth]{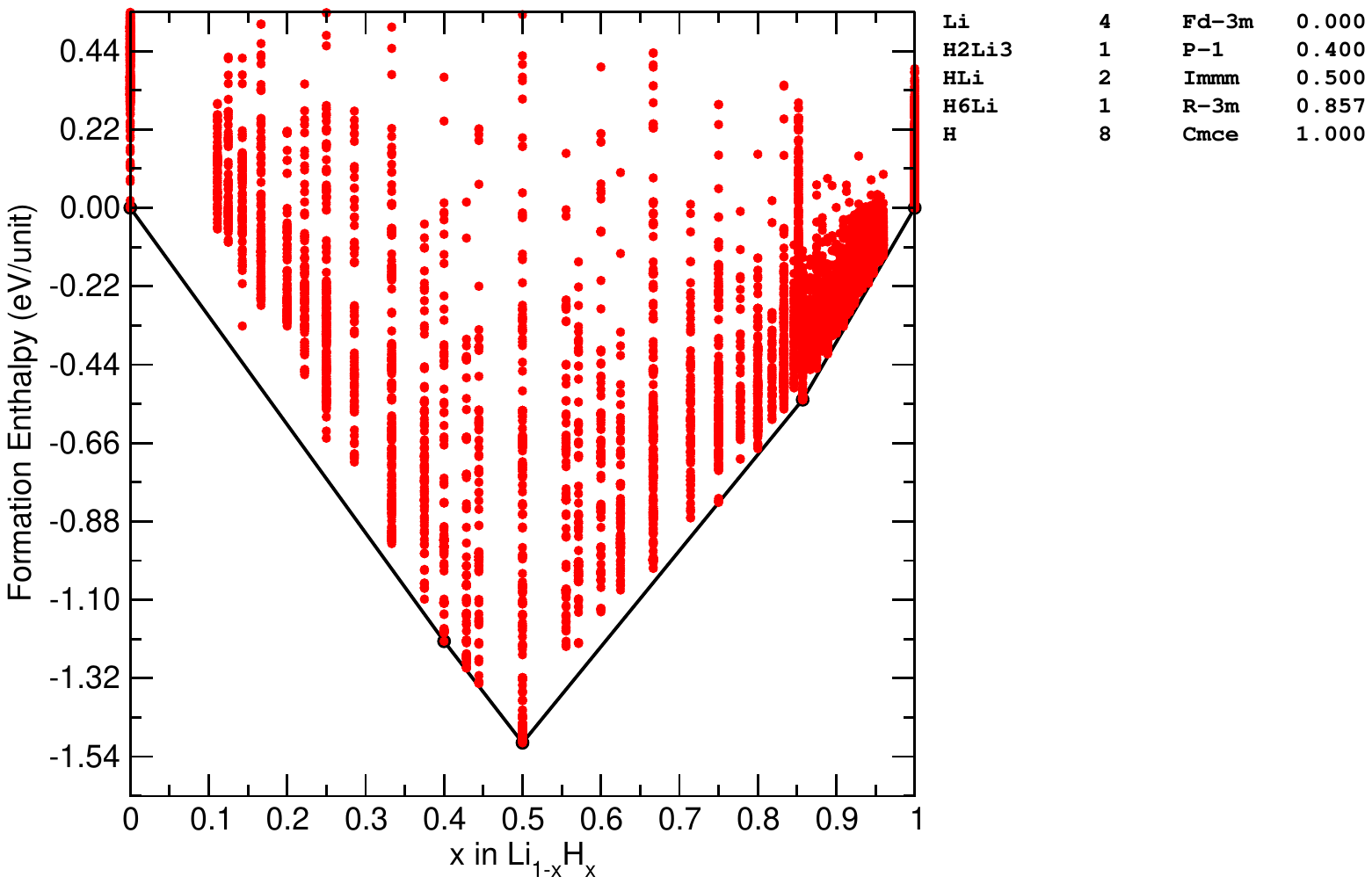}
    \caption{The convex hull of Li-H at 300\ GPa.}
\end{figure*}

\begin{figure*}
    \centering
    \includegraphics[width=\convexwidth]{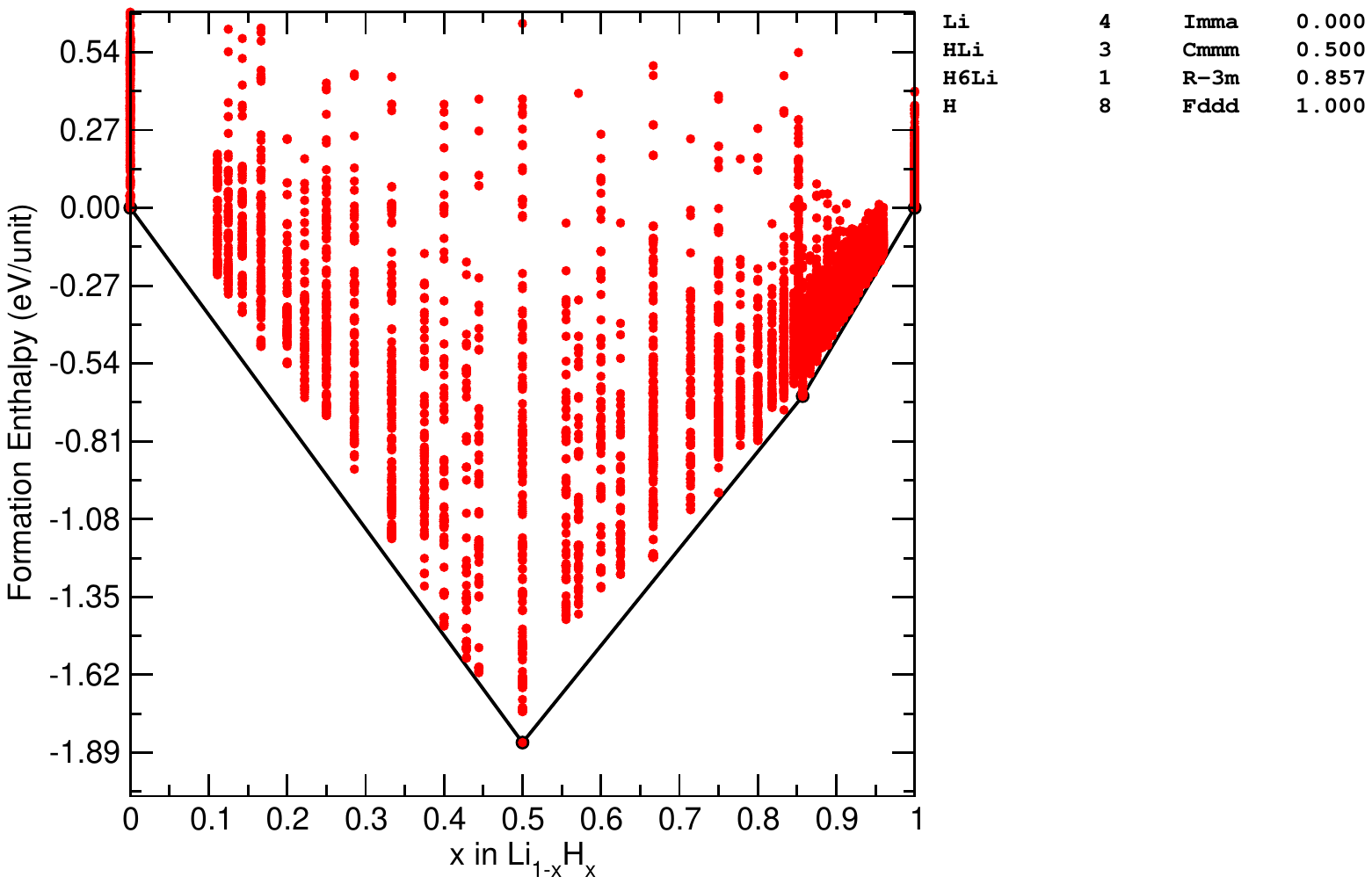}
    \caption{The convex hull of Li-H at 500\ GPa.}
\end{figure*}

\begin{figure*}
    \centering
    \includegraphics[width=\convexwidth]{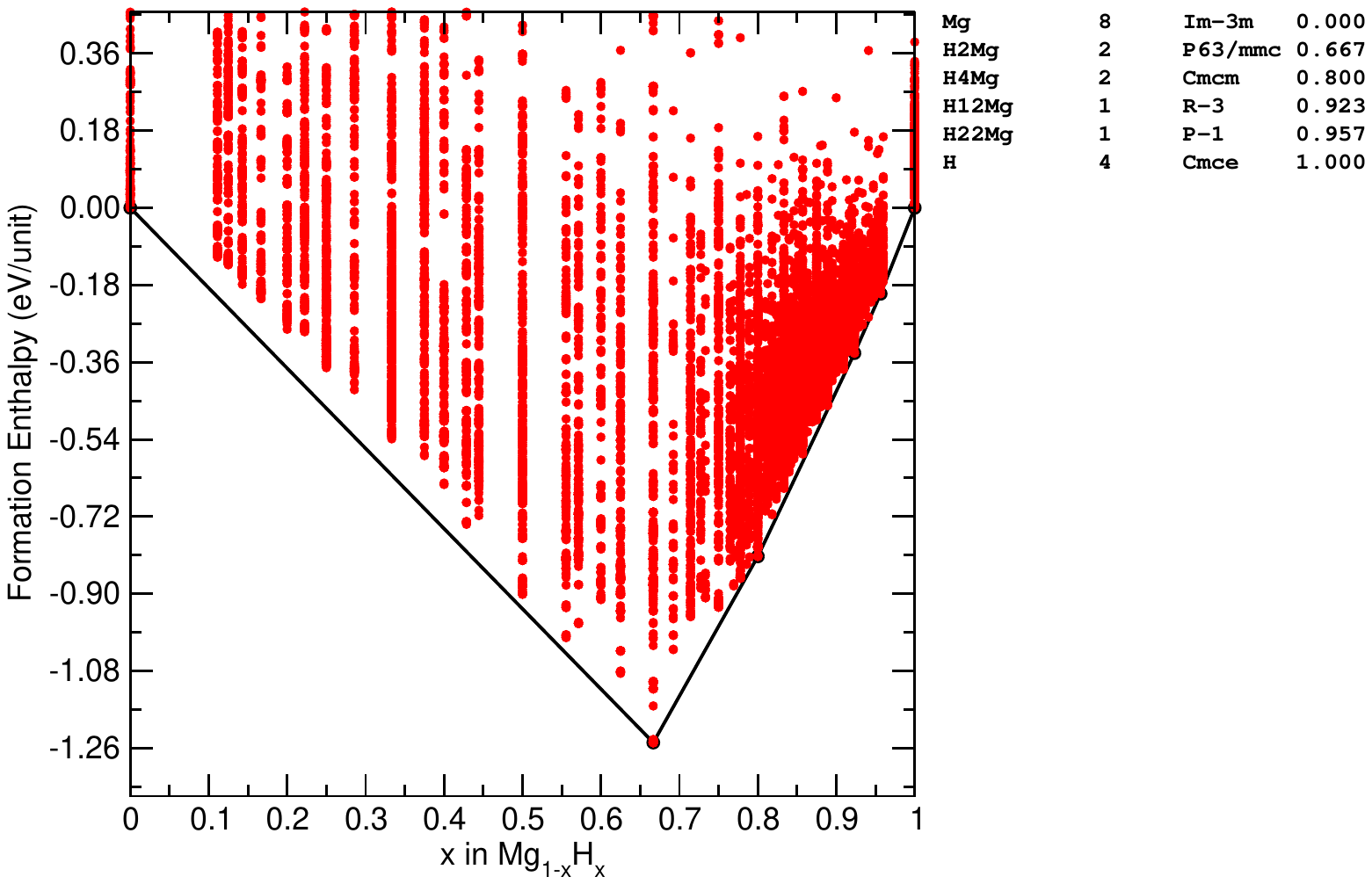}
    \caption{The convex hull of Mg-H at 200\ GPa.}
\end{figure*}

\begin{figure*}
    \centering
    \includegraphics[width=\convexwidth]{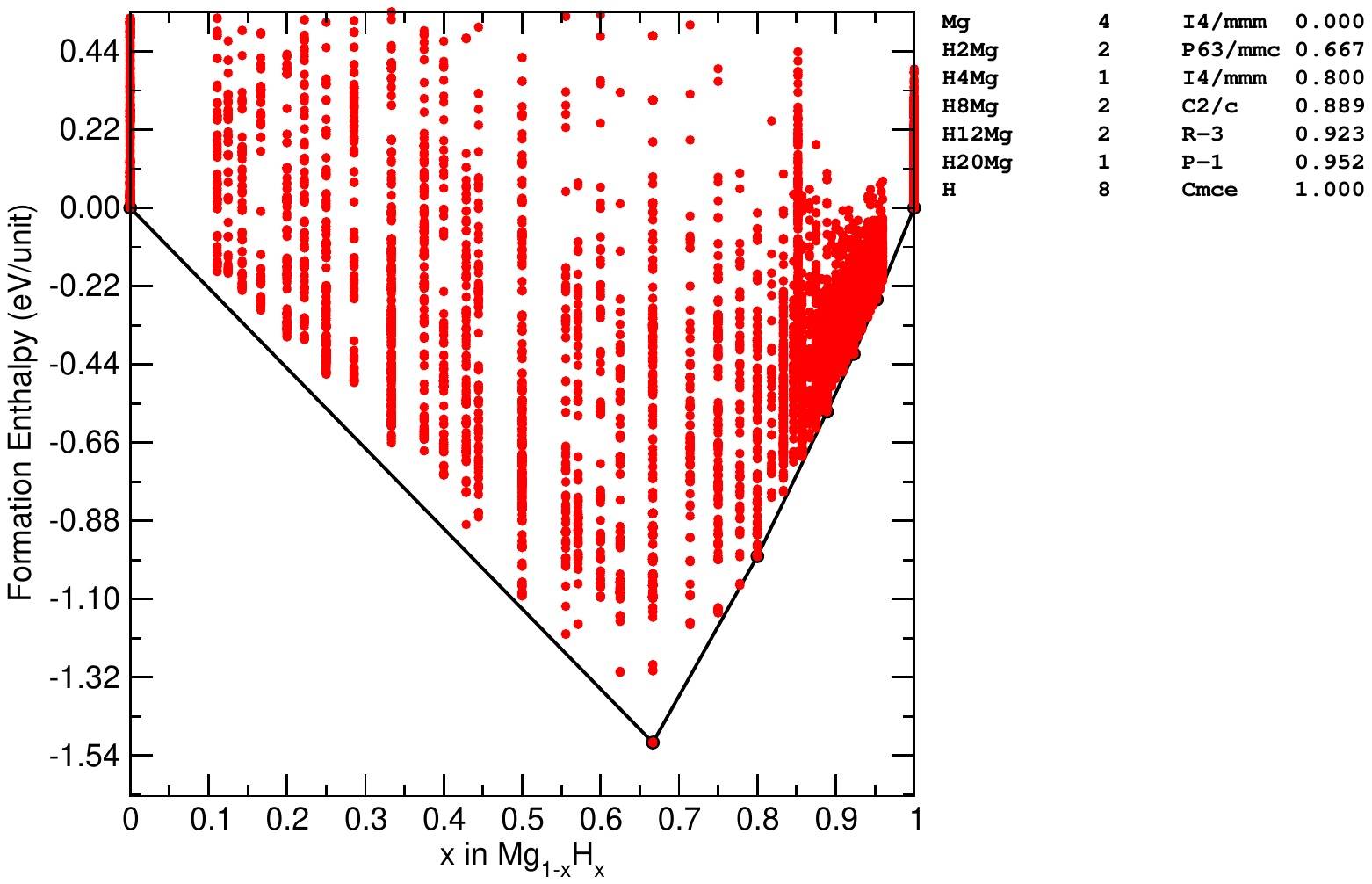}
    \caption{The convex hull of Mg-H at 300\ GPa.}
\end{figure*}

\begin{figure*}
    \centering
    \includegraphics[width=\convexwidth]{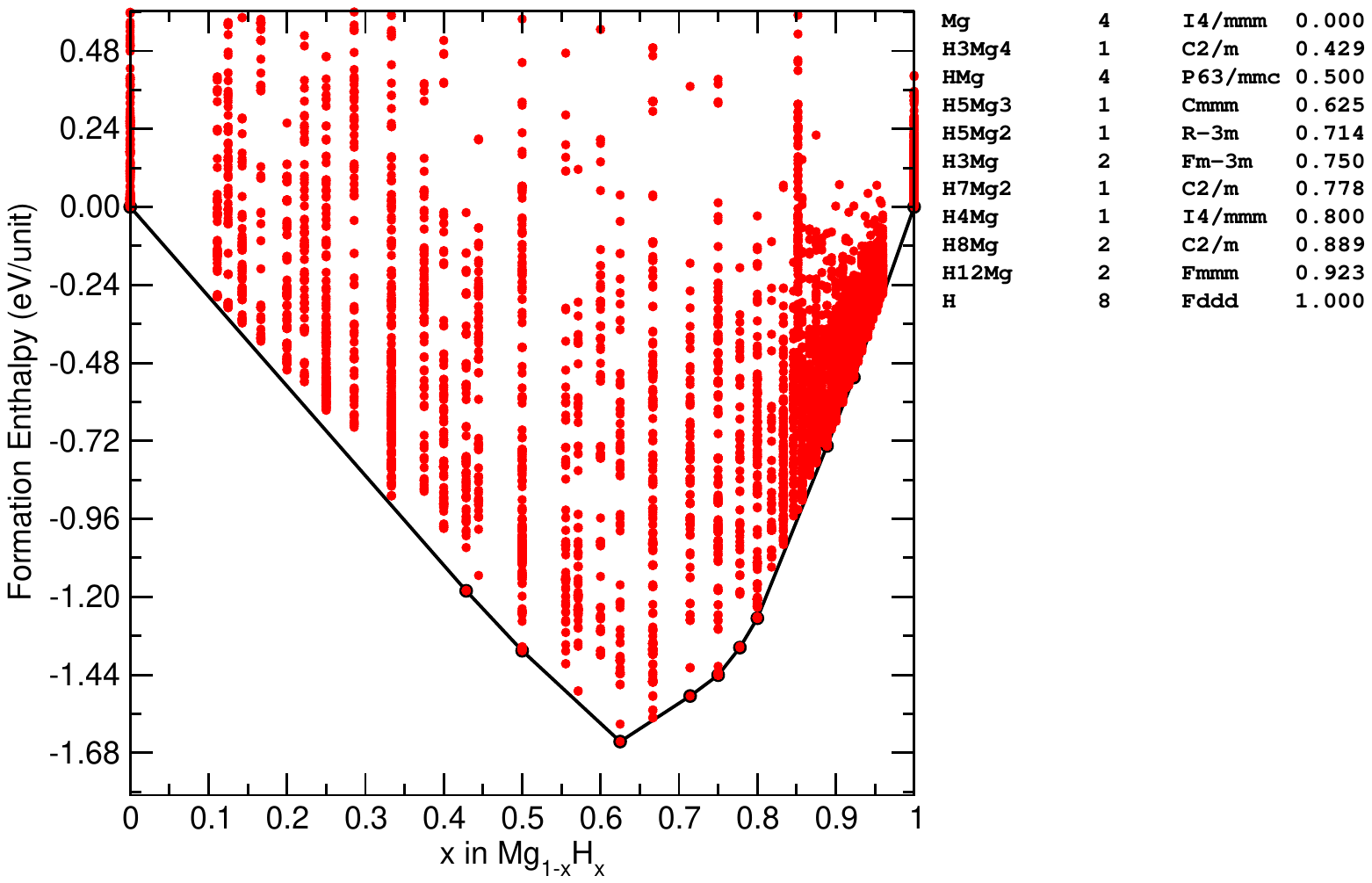}
    \caption{The convex hull of Mg-H at 500\ GPa.}
\end{figure*}

\begin{figure*}
    \centering
    \includegraphics[width=\convexwidth]{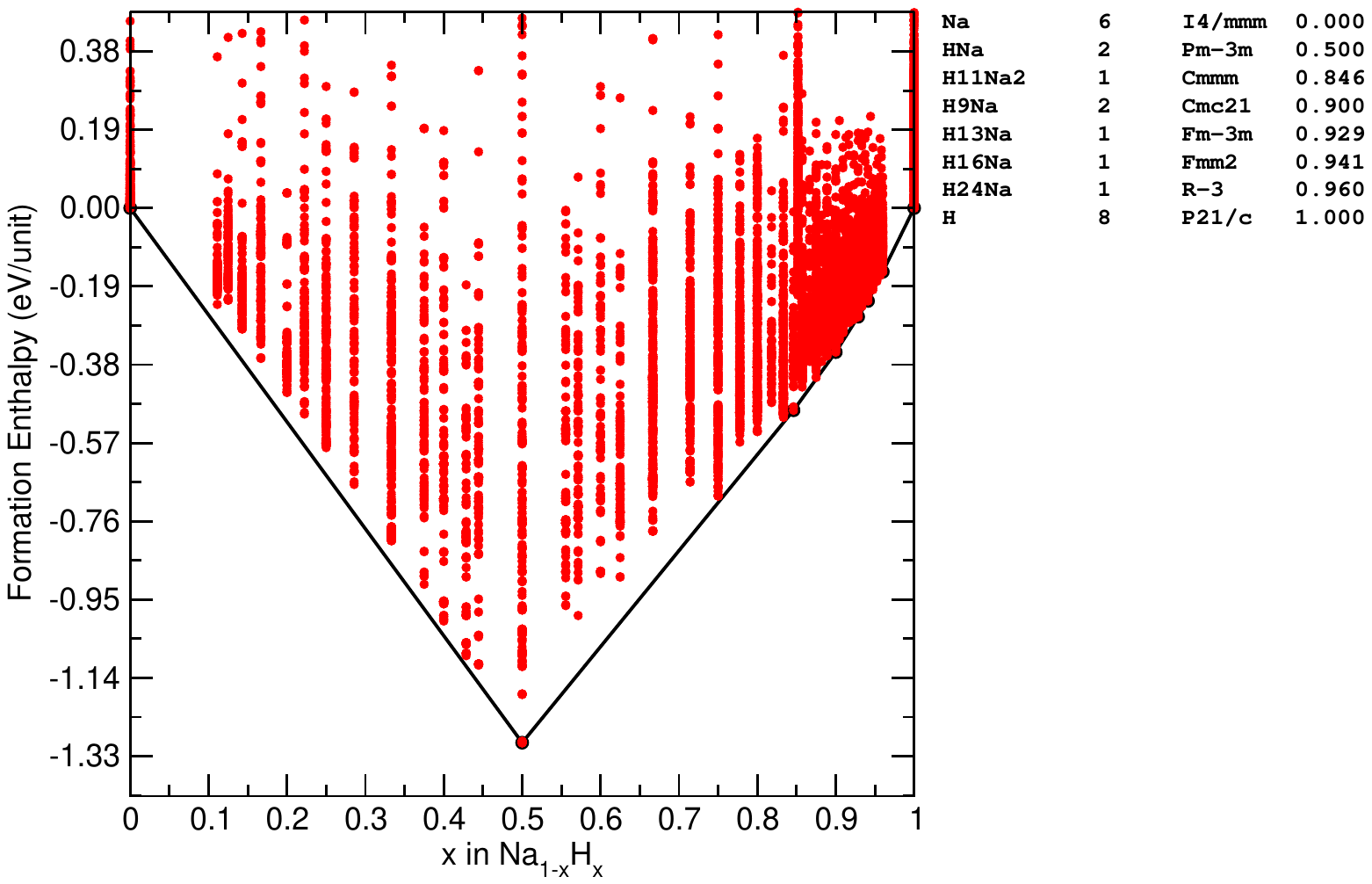}
    \caption{The convex hull of Na-H at 100\ GPa.}
\end{figure*}

\begin{figure*}
    \centering
    \includegraphics[width=\convexwidth]{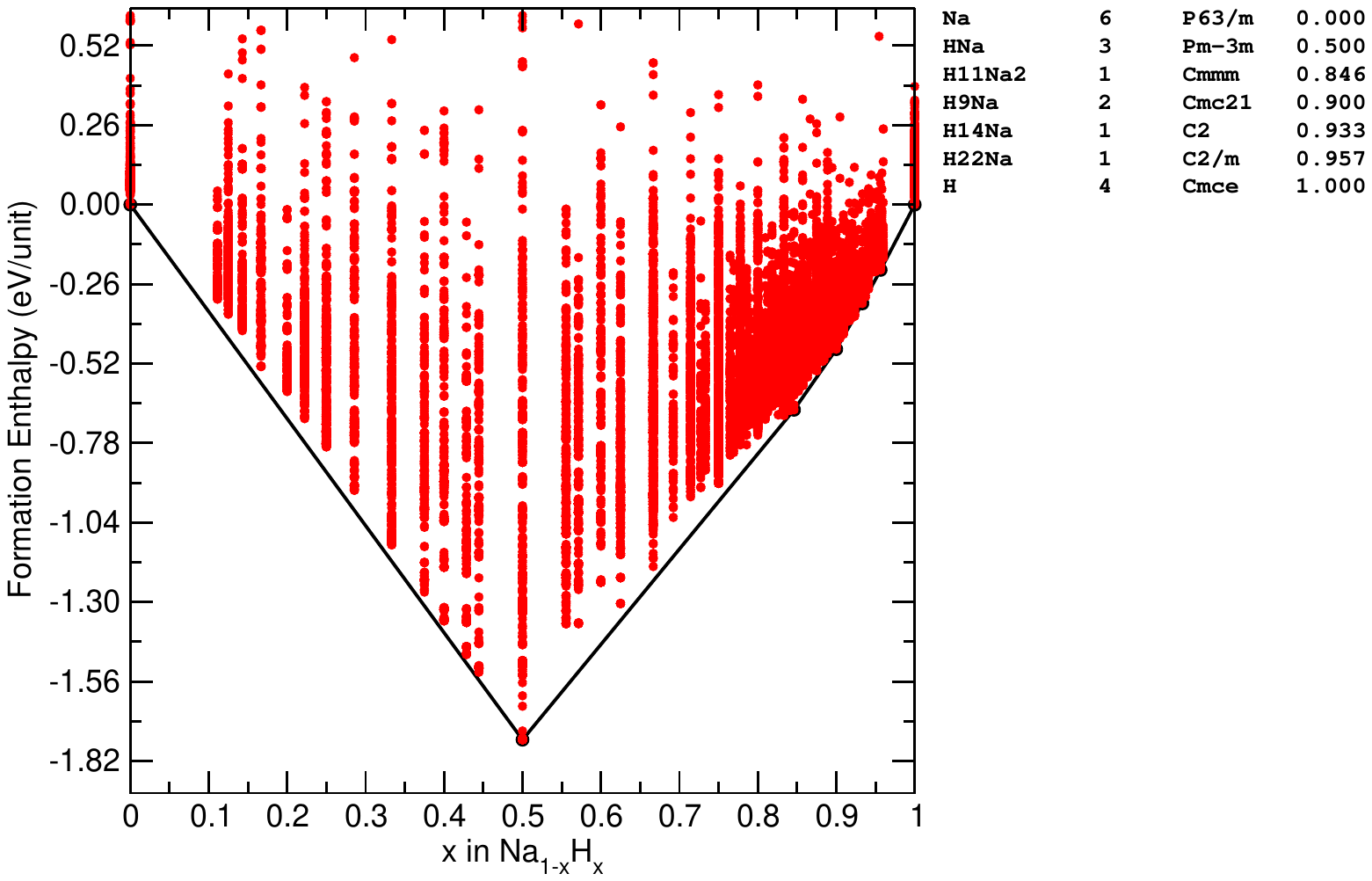}
    \caption{The convex hull of Na-H at 200\ GPa.}
\end{figure*}

\begin{figure*}
    \centering
    \includegraphics[width=\convexwidth]{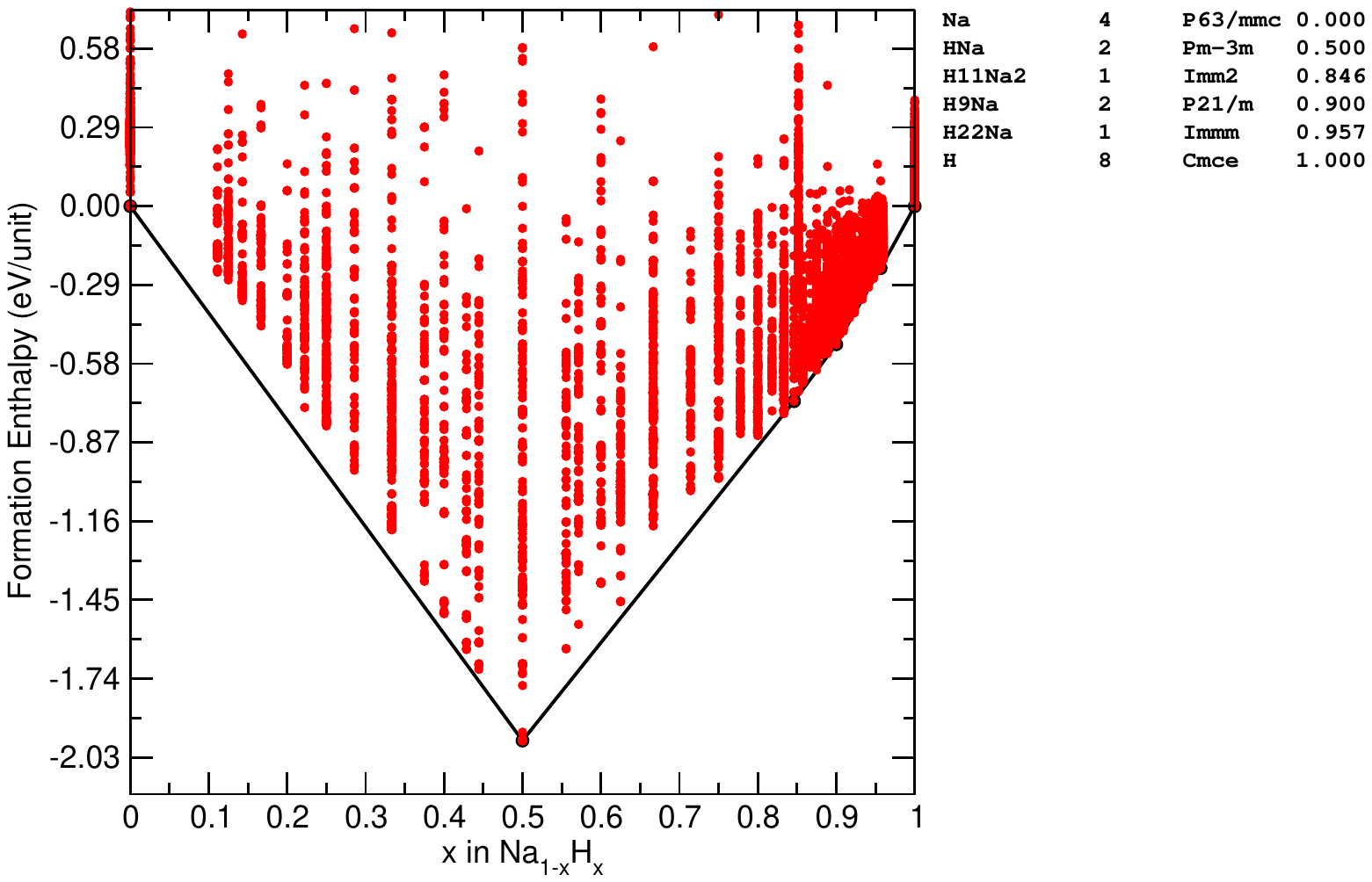}
    \caption{The convex hull of Na-H at 300\ GPa.}
\end{figure*}

\begin{figure*}
    \centering
    \includegraphics[width=\convexwidth]{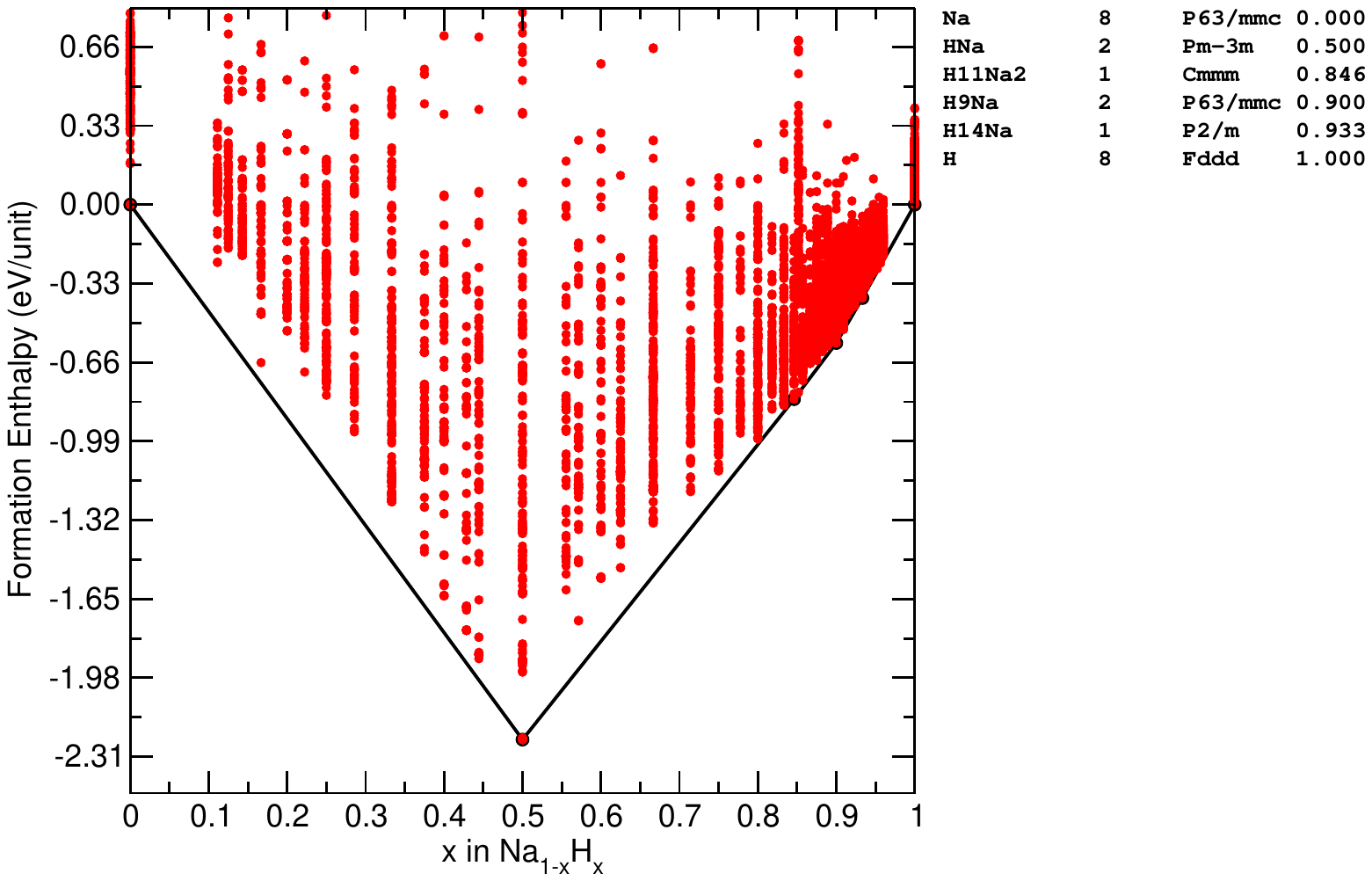}
    \caption{The convex hull of Na-H at 500\ GPa.}
\end{figure*}

\begin{figure*}
    \centering
    \includegraphics[width=\convexwidth]{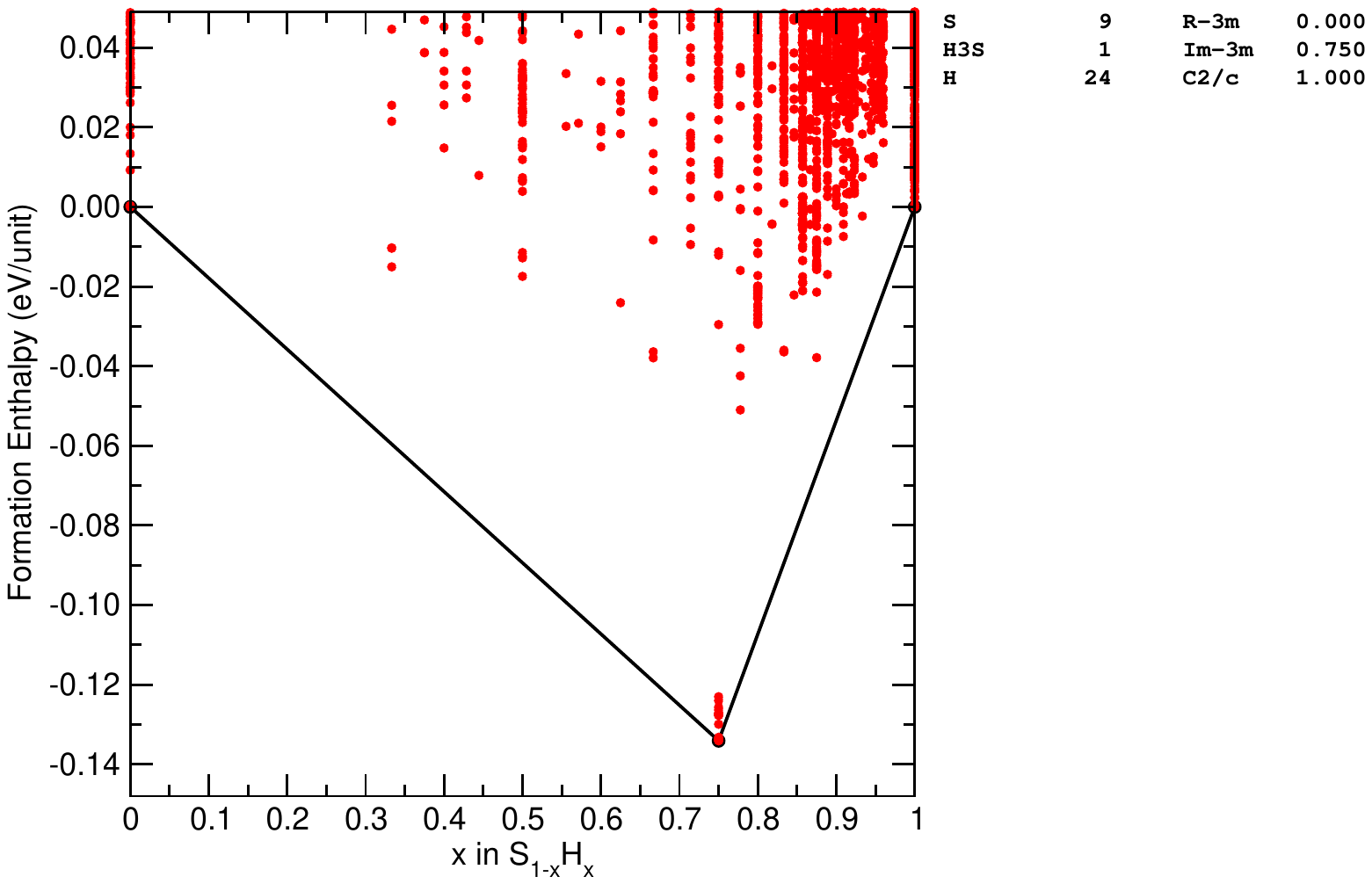}
    \caption{The convex hull of S-H at 200\ GPa.}
\end{figure*}

\begin{figure*}
    \centering
    \includegraphics[width=\convexwidth]{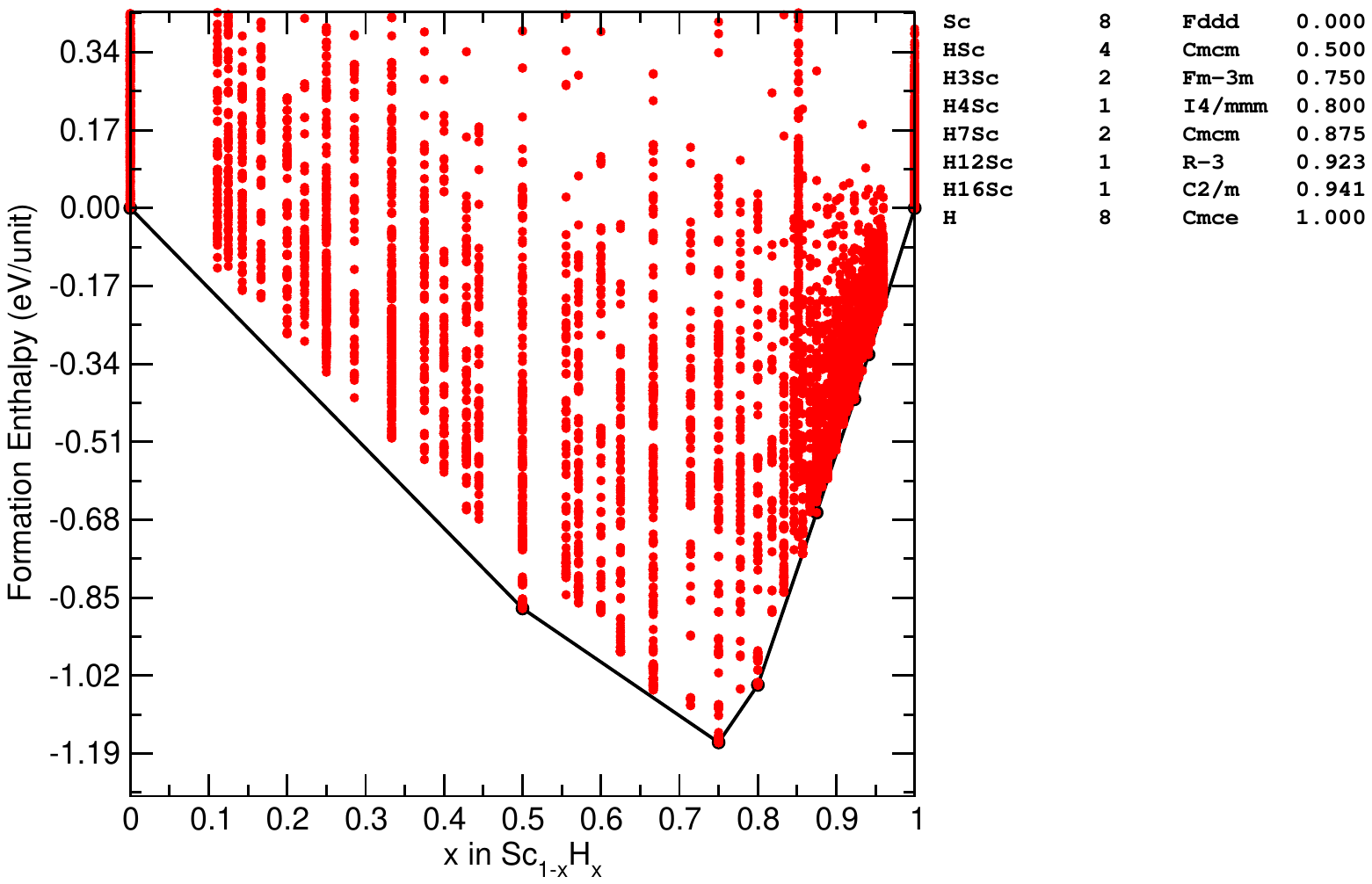}
    \caption{The convex hull of Sc-H at 300\ GPa.}
\end{figure*}

\begin{figure*}
    \centering
    \includegraphics[width=\convexwidth]{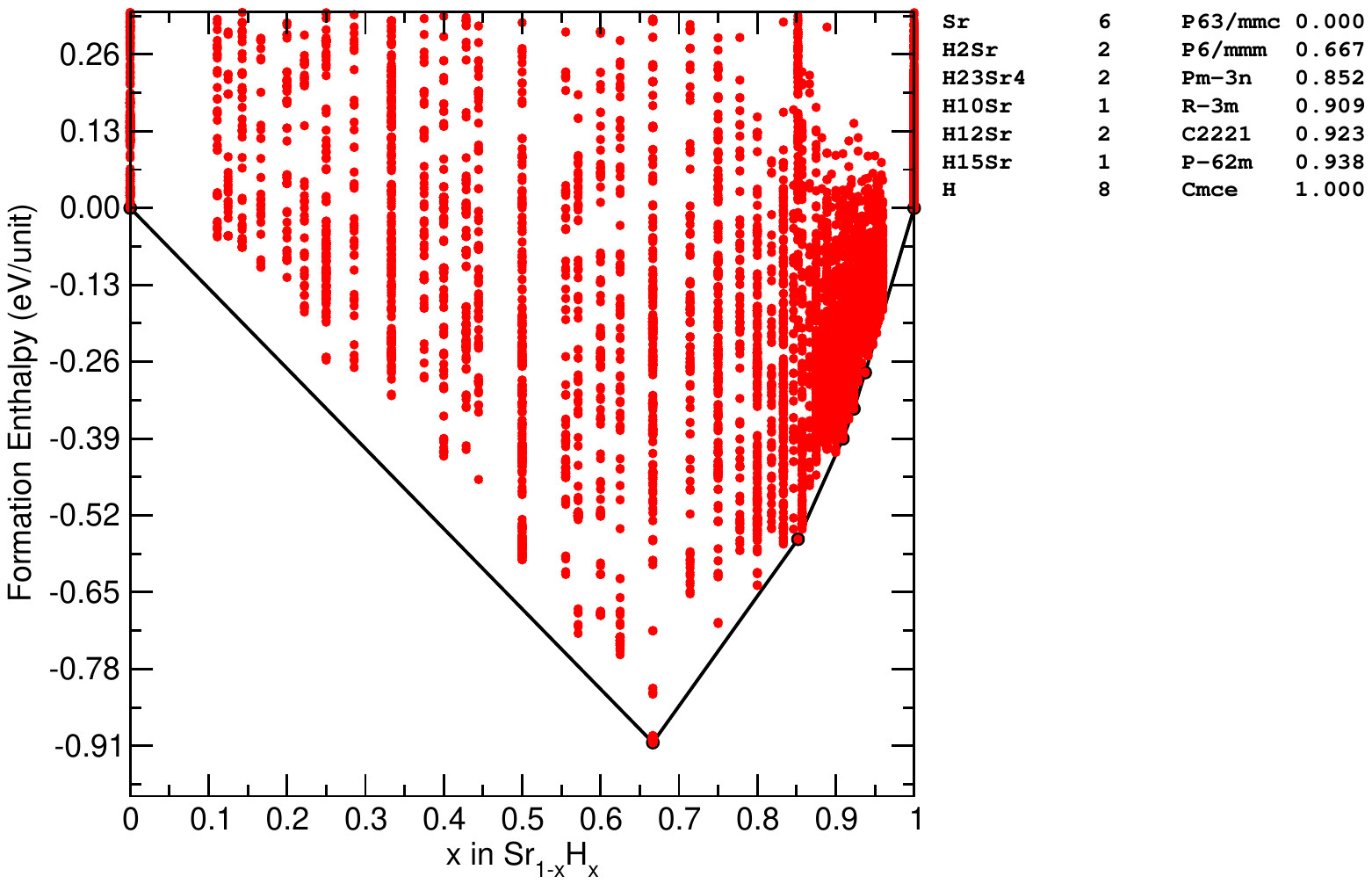}
    \caption{The convex hull of Sr-H at 300\ GPa.}
\end{figure*}

\begin{figure*}
    \centering
    \includegraphics[width=\convexwidth]{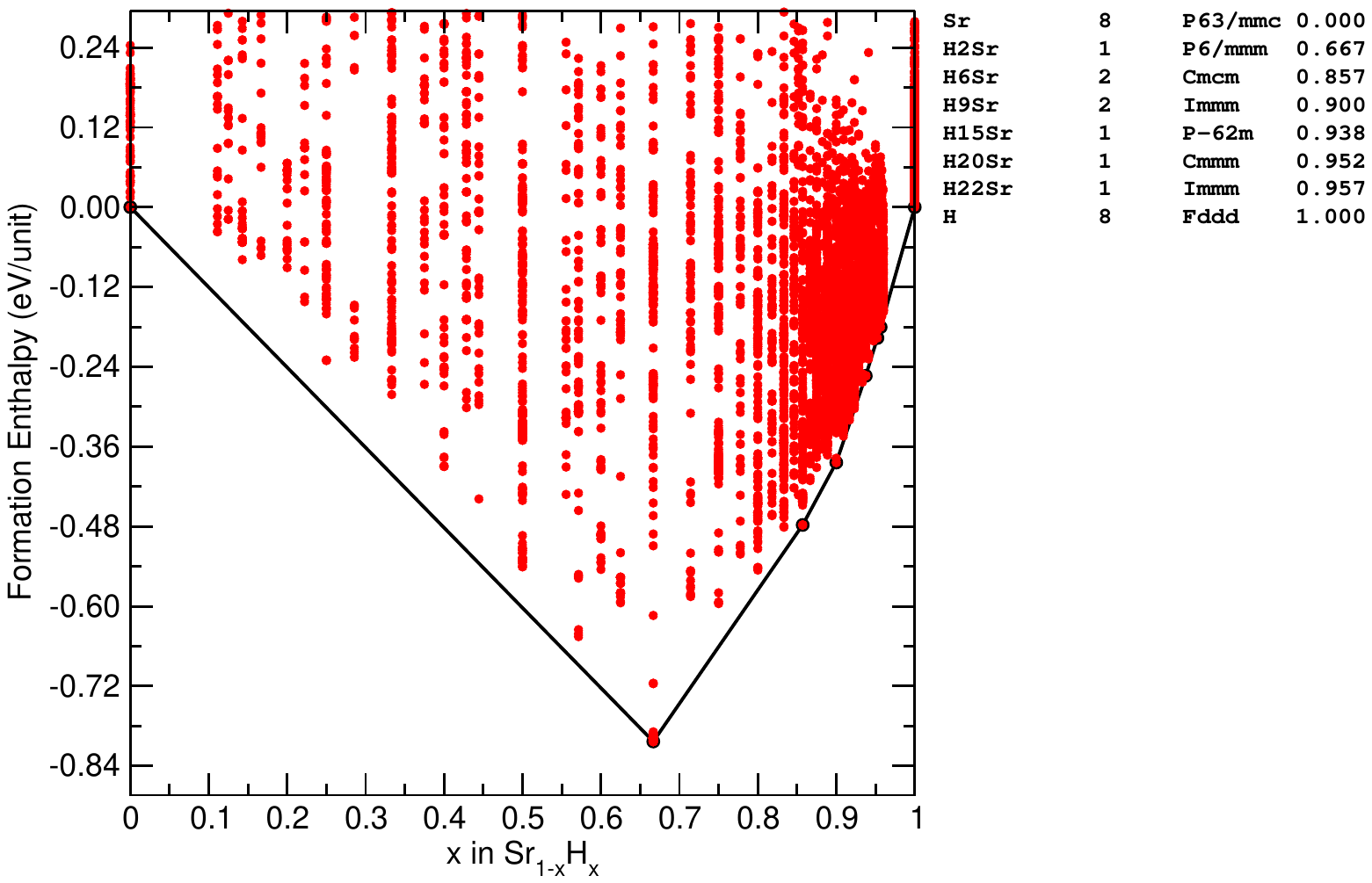}
    \caption{The convex hull of Sr-H at 500\ GPa.}
\end{figure*}

\begin{figure*}
    \centering
    \includegraphics[width=\convexwidth]{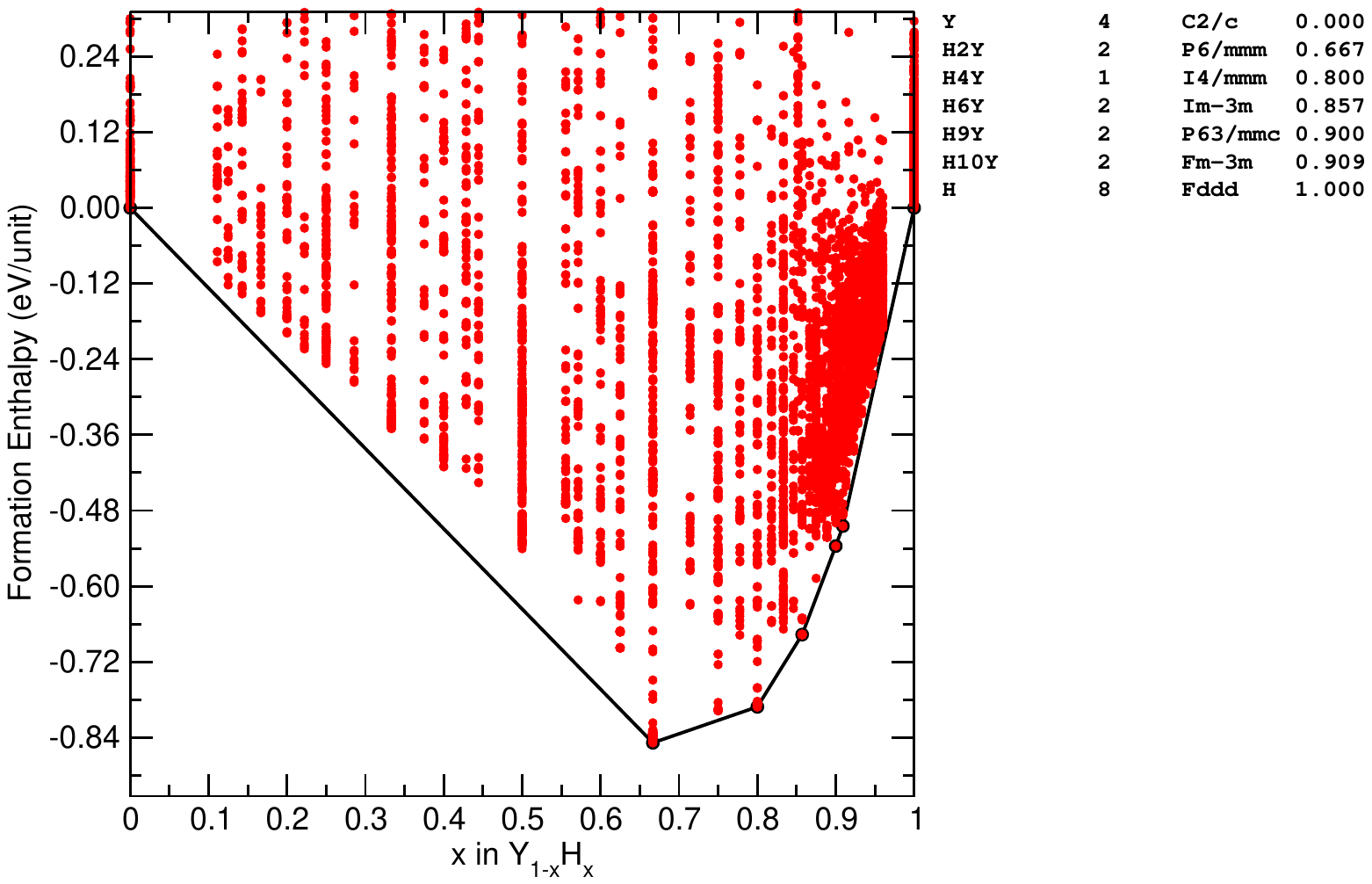}
    \caption{The convex hull of Y-H at 500\ GPa.}
\end{figure*}

\begin{figure*}
    \centering
    \includegraphics[width=\convexwidth]{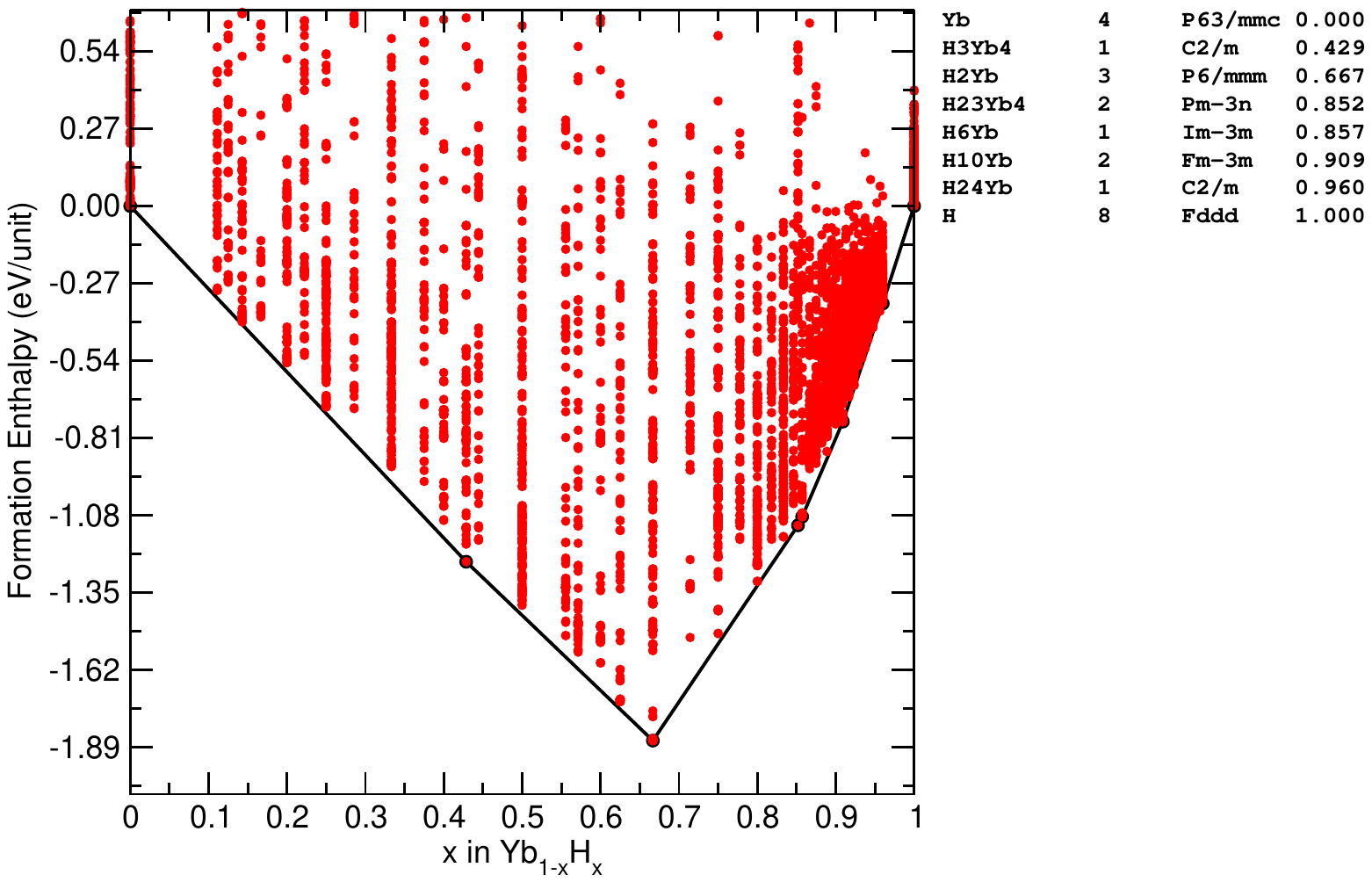}
    \caption{The convex hull of Yb-H at 500\ GPa.}
\end{figure*}

\clearpage

\bibliography{references.bib}

\begin{thebibliography}{121}%
\makeatletter
\providecommand \@ifxundefined [1]{%
 \@ifx{#1\undefined}
}%
\providecommand \@ifnum [1]{%
 \ifnum #1\expandafter \@firstoftwo
 \else \expandafter \@secondoftwo
 \fi
}%
\providecommand \@ifx [1]{%
 \ifx #1\expandafter \@firstoftwo
 \else \expandafter \@secondoftwo
 \fi
}%
\providecommand \natexlab [1]{#1}%
\providecommand \enquote  [1]{``#1''}%
\providecommand \bibnamefont  [1]{#1}%
\providecommand \bibfnamefont [1]{#1}%
\providecommand \citenamefont [1]{#1}%
\providecommand \href@noop [0]{\@secondoftwo}%
\providecommand \href [0]{\begingroup \@sanitize@url \@href}%
\providecommand \@href[1]{\@@startlink{#1}\@@href}%
\providecommand \@@href[1]{\endgroup#1\@@endlink}%
\providecommand \@sanitize@url [0]{\catcode `\\12\catcode `\$12\catcode
  `\&12\catcode `\#12\catcode `\^12\catcode `\_12\catcode `\%12\relax}%
\providecommand \@@startlink[1]{}%
\providecommand \@@endlink[0]{}%
\providecommand \url  [0]{\begingroup\@sanitize@url \@url }%
\providecommand \@url [1]{\endgroup\@href {#1}{\urlprefix }}%
\providecommand \urlprefix  [0]{URL }%
\providecommand \Eprint [0]{\href }%
\providecommand \doibase [0]{https://doi.org/}%
\providecommand \selectlanguage [0]{\@gobble}%
\providecommand \bibinfo  [0]{\@secondoftwo}%
\providecommand \bibfield  [0]{\@secondoftwo}%
\providecommand \translation [1]{[#1]}%
\providecommand \BibitemOpen [0]{}%
\providecommand \bibitemStop [0]{}%
\providecommand \bibitemNoStop [0]{.\EOS\space}%
\providecommand \EOS [0]{\spacefactor3000\relax}%
\providecommand \BibitemShut  [1]{\csname bibitem#1\endcsname}%
\let\auto@bib@innerbib\@empty
\bibitem [{\citenamefont {Duan}\ \emph {et~al.}(2017)\citenamefont {Duan},
  \citenamefont {Liu}, \citenamefont {Ma}, \citenamefont {Shao}, \citenamefont
  {Liu},\ and\ \citenamefont {Cui}}]{duan2017structure}%
  \BibitemOpen
  \bibfield  {author} {\bibinfo {author} {\bibfnamefont {D.}~\bibnamefont
  {Duan}}, \bibinfo {author} {\bibfnamefont {Y.}~\bibnamefont {Liu}}, \bibinfo
  {author} {\bibfnamefont {Y.}~\bibnamefont {Ma}}, \bibinfo {author}
  {\bibfnamefont {Z.}~\bibnamefont {Shao}}, \bibinfo {author} {\bibfnamefont
  {B.}~\bibnamefont {Liu}},\ and\ \bibinfo {author} {\bibfnamefont
  {T.}~\bibnamefont {Cui}},\ }\bibfield  {title} {\bibinfo {title} {Structure
  and superconductivity of hydrides at high pressures},\ }\href
  {https://doi.org/10.1093/nsr/nww029} {\bibfield  {journal} {\bibinfo
  {journal} {National Science Review}\ }\textbf {\bibinfo {volume} {4}},\
  \bibinfo {pages} {121} (\bibinfo {year} {2017})}\BibitemShut {NoStop}%
\bibitem [{\citenamefont {Zurek}\ and\ \citenamefont
  {Bi}(2019)}]{zurek2019high}%
  \BibitemOpen
  \bibfield  {author} {\bibinfo {author} {\bibfnamefont {E.}~\bibnamefont
  {Zurek}}\ and\ \bibinfo {author} {\bibfnamefont {T.}~\bibnamefont {Bi}},\
  }\bibfield  {title} {\bibinfo {title} {High-temperature superconductivity in
  alkaline and rare earth polyhydrides at high pressure: A theoretical
  perspective},\ }\href {https://doi.org/10.1093/nsr/nww029} {\bibfield
  {journal} {\bibinfo  {journal} {The Journal of chemical physics}\ }\textbf
  {\bibinfo {volume} {150}},\ \bibinfo {pages} {050901} (\bibinfo {year}
  {2019})}\BibitemShut {NoStop}%
\bibitem [{\citenamefont {Flores-Livas}\ \emph {et~al.}(2020)\citenamefont
  {Flores-Livas}, \citenamefont {Boeri}, \citenamefont {Sanna}, \citenamefont
  {Profeta}, \citenamefont {Arita},\ and\ \citenamefont
  {Eremets}}]{flores2019perspective}%
  \BibitemOpen
  \bibfield  {author} {\bibinfo {author} {\bibfnamefont {J.~A.}\ \bibnamefont
  {Flores-Livas}}, \bibinfo {author} {\bibfnamefont {L.}~\bibnamefont {Boeri}},
  \bibinfo {author} {\bibfnamefont {A.}~\bibnamefont {Sanna}}, \bibinfo
  {author} {\bibfnamefont {G.}~\bibnamefont {Profeta}}, \bibinfo {author}
  {\bibfnamefont {R.}~\bibnamefont {Arita}},\ and\ \bibinfo {author}
  {\bibfnamefont {M.}~\bibnamefont {Eremets}},\ }\bibfield  {title} {\bibinfo
  {title} {A perspective on conventional high-temperature superconductors at
  high pressure: Methods and materials},\ }\href
  {https://doi.org/10.1016/j.physrep.2020.02.003} {\bibfield  {journal}
  {\bibinfo  {journal} {Physics Reports}\ } (\bibinfo {year}
  {2020})}\BibitemShut {NoStop}%
\bibitem [{\citenamefont {Boeri}\ and\ \citenamefont
  {Bachelet}(2019)}]{boeri2019}%
  \BibitemOpen
  \bibfield  {author} {\bibinfo {author} {\bibfnamefont {L.}~\bibnamefont
  {Boeri}}\ and\ \bibinfo {author} {\bibfnamefont {G.~B.}\ \bibnamefont
  {Bachelet}},\ }\bibfield  {title} {\bibinfo {title} {Viewpoint: the road to
  room-temperature conventional superconductivity},\ }\href@noop {} {\bibfield
  {journal} {\bibinfo  {journal} {J. Phys: Condens. Matt.}\ }\textbf {\bibinfo
  {volume} {31}},\ \bibinfo {pages} {234002} (\bibinfo {year}
  {2019})}\BibitemShut {NoStop}%
\bibitem [{\citenamefont {Oganov}\ \emph {et~al.}(2019)\citenamefont {Oganov},
  \citenamefont {Pickard}, \citenamefont {Zhu},\ and\ \citenamefont
  {Needs}}]{oganov2019}%
  \BibitemOpen
  \bibfield  {author} {\bibinfo {author} {\bibfnamefont {A.~R.}\ \bibnamefont
  {Oganov}}, \bibinfo {author} {\bibfnamefont {C.~J.}\ \bibnamefont {Pickard}},
  \bibinfo {author} {\bibfnamefont {Q.}~\bibnamefont {Zhu}},\ and\ \bibinfo
  {author} {\bibfnamefont {R.~J.}\ \bibnamefont {Needs}},\ }\bibfield  {title}
  {\bibinfo {title} {Structure prediction drives materials discovery},\
  }\href@noop {} {\bibfield  {journal} {\bibinfo  {journal} {Nature Reviews
  Materials}\ }\textbf {\bibinfo {volume} {4}},\ \bibinfo {pages} {331}
  (\bibinfo {year} {2019})}\BibitemShut {NoStop}%
\bibitem [{\citenamefont {Pickard}\ \emph {et~al.}(2020)\citenamefont
  {Pickard}, \citenamefont {Errea},\ and\ \citenamefont
  {Eremets}}]{pickard2019}%
  \BibitemOpen
  \bibfield  {author} {\bibinfo {author} {\bibfnamefont {C.~J.}\ \bibnamefont
  {Pickard}}, \bibinfo {author} {\bibfnamefont {I.}~\bibnamefont {Errea}},\
  and\ \bibinfo {author} {\bibfnamefont {M.~I.}\ \bibnamefont {Eremets}},\
  }\bibfield  {title} {\bibinfo {title} {Superconducting hydrides under
  pressure},\ }\href {https://doi.org/10.1146/annurev-conmatphys-031218-013413}
  {\bibfield  {journal} {\bibinfo  {journal} {Annual Review of Condensed Matter
  Physics}\ }\textbf {\bibinfo {volume} {11}},\ \bibinfo {pages} {57} (\bibinfo
  {year} {2020})}\BibitemShut {NoStop}%
\bibitem [{\citenamefont {Gaspari}\ and\ \citenamefont
  {Gyorffy}(1972)}]{gaspari1972}%
  \BibitemOpen
  \bibfield  {author} {\bibinfo {author} {\bibfnamefont {G.~D.}\ \bibnamefont
  {Gaspari}}\ and\ \bibinfo {author} {\bibfnamefont {B.~L.}\ \bibnamefont
  {Gyorffy}},\ }\bibfield  {title} {\bibinfo {title} {Electron-phonon
  interactions, d resonances, and superconductivity in transition metals},\
  }\href@noop {} {\bibfield  {journal} {\bibinfo  {journal} {Phys. Rev. Lett.}\
  }\textbf {\bibinfo {volume} {28}},\ \bibinfo {pages} {801} (\bibinfo {year}
  {1972})}\BibitemShut {NoStop}%
\bibitem [{\citenamefont {Pickard}\ and\ \citenamefont
  {Needs}(2006)}]{pickard2006}%
  \BibitemOpen
  \bibfield  {author} {\bibinfo {author} {\bibfnamefont {C.~J.}\ \bibnamefont
  {Pickard}}\ and\ \bibinfo {author} {\bibfnamefont {R.~J.}\ \bibnamefont
  {Needs}},\ }\bibfield  {title} {\bibinfo {title} {High-pressure phases of
  silane},\ }\href@noop {} {\bibfield  {journal} {\bibinfo  {journal} {Phys.
  Rev. Lett.}\ }\textbf {\bibinfo {volume} {97}},\ \bibinfo {pages} {045504}
  (\bibinfo {year} {2006})}\BibitemShut {NoStop}%
\bibitem [{\citenamefont {Pickard}\ and\ \citenamefont
  {Needs}(2011)}]{pickard2011}%
  \BibitemOpen
  \bibfield  {author} {\bibinfo {author} {\bibfnamefont {C.~J.}\ \bibnamefont
  {Pickard}}\ and\ \bibinfo {author} {\bibfnamefont {R.~J.}\ \bibnamefont
  {Needs}},\ }\bibfield  {title} {\bibinfo {title} {Ab initio random structure
  searching},\ }\href@noop {} {\bibfield  {journal} {\bibinfo  {journal} {J.
  Phys: Condens. Matt.}\ }\textbf {\bibinfo {volume} {23}},\ \bibinfo {pages}
  {053201} (\bibinfo {year} {2011})}\BibitemShut {NoStop}%
\bibitem [{\citenamefont {Clark}\ \emph {et~al.}(2005)\citenamefont {Clark},
  \citenamefont {Segall}, \citenamefont {Pickard}, \citenamefont {Hasnip},
  \citenamefont {Probert}, \citenamefont {Refson},\ and\ \citenamefont
  {Payne}}]{castep2005}%
  \BibitemOpen
  \bibfield  {author} {\bibinfo {author} {\bibfnamefont {S.~J.}\ \bibnamefont
  {Clark}}, \bibinfo {author} {\bibfnamefont {M.~D.}\ \bibnamefont {Segall}},
  \bibinfo {author} {\bibfnamefont {C.~J.}\ \bibnamefont {Pickard}}, \bibinfo
  {author} {\bibfnamefont {P.~J.}\ \bibnamefont {Hasnip}}, \bibinfo {author}
  {\bibfnamefont {M.~I.~J.}\ \bibnamefont {Probert}}, \bibinfo {author}
  {\bibfnamefont {K.}~\bibnamefont {Refson}},\ and\ \bibinfo {author}
  {\bibfnamefont {M.~C.}\ \bibnamefont {Payne}},\ }\bibfield  {title} {\bibinfo
  {title} {First principles methods using {CASTEP}},\ }\href@noop {} {\bibfield
   {journal} {\bibinfo  {journal} {Zeitschrift f{\"u}r
  Kristallographie-Crystalline Materials}\ }\textbf {\bibinfo {volume} {220}},\
  \bibinfo {pages} {567} (\bibinfo {year} {2005})}\BibitemShut {NoStop}%
\bibitem [{\citenamefont {Perdew}\ \emph {et~al.}(1996)\citenamefont {Perdew},
  \citenamefont {Burke},\ and\ \citenamefont {Ernzerhof}}]{pbe1996}%
  \BibitemOpen
  \bibfield  {author} {\bibinfo {author} {\bibfnamefont {J.~P.}\ \bibnamefont
  {Perdew}}, \bibinfo {author} {\bibfnamefont {K.}~\bibnamefont {Burke}},\ and\
  \bibinfo {author} {\bibfnamefont {M.}~\bibnamefont {Ernzerhof}},\ }\bibfield
  {title} {\bibinfo {title} {Generalized gradient approximation made simple},\
  }\href@noop {} {\bibfield  {journal} {\bibinfo  {journal} {Phys. Rev. Lett.}\
  }\textbf {\bibinfo {volume} {77}},\ \bibinfo {pages} {3865} (\bibinfo {year}
  {1996})}\BibitemShut {NoStop}%
\bibitem [{\citenamefont {Monserrat}\ \emph {et~al.}(2018)\citenamefont
  {Monserrat}, \citenamefont {Drummond}, \citenamefont {Dalladay-Simpson},
  \citenamefont {Howie}, \citenamefont {L{\'o}pez~R{\'\i}os}, \citenamefont
  {Gregoryanz}, \citenamefont {Pickard},\ and\ \citenamefont
  {Needs}}]{monserrat2018v}%
  \BibitemOpen
  \bibfield  {author} {\bibinfo {author} {\bibfnamefont {B.}~\bibnamefont
  {Monserrat}}, \bibinfo {author} {\bibfnamefont {N.~D.}\ \bibnamefont
  {Drummond}}, \bibinfo {author} {\bibfnamefont {P.}~\bibnamefont
  {Dalladay-Simpson}}, \bibinfo {author} {\bibfnamefont {R.~T.}\ \bibnamefont
  {Howie}}, \bibinfo {author} {\bibfnamefont {P.}~\bibnamefont
  {L{\'o}pez~R{\'\i}os}}, \bibinfo {author} {\bibfnamefont {E.}~\bibnamefont
  {Gregoryanz}}, \bibinfo {author} {\bibfnamefont {C.~J.}\ \bibnamefont
  {Pickard}},\ and\ \bibinfo {author} {\bibfnamefont {R.~J.}\ \bibnamefont
  {Needs}},\ }\bibfield  {title} {\bibinfo {title} {Structure and metallicity
  of phase v of hydrogen},\ }\href@noop {} {\bibfield  {journal} {\bibinfo
  {journal} {Phys. Rev. Lett}\ }\textbf {\bibinfo {volume} {120}},\ \bibinfo
  {pages} {255701} (\bibinfo {year} {2018})}\BibitemShut {NoStop}%
\bibitem [{\citenamefont {McMillan}(1968)}]{mcmillan1968}%
  \BibitemOpen
  \bibfield  {author} {\bibinfo {author} {\bibfnamefont {W.~L.}\ \bibnamefont
  {McMillan}},\ }\bibfield  {title} {\bibinfo {title} {Transition temperature
  of strong-coupled superconductors},\ }\href@noop {} {\bibfield  {journal}
  {\bibinfo  {journal} {Physical Review}\ }\textbf {\bibinfo {volume} {167}},\
  \bibinfo {pages} {331} (\bibinfo {year} {1968})}\BibitemShut {NoStop}%
\bibitem [{\citenamefont {Hopfield}(1969)}]{hopfield1969}%
  \BibitemOpen
  \bibfield  {author} {\bibinfo {author} {\bibfnamefont {J.~J.}\ \bibnamefont
  {Hopfield}},\ }\bibfield  {title} {\bibinfo {title} {Angular momentum and
  transition-metal superconductivity},\ }\href@noop {} {\bibfield  {journal}
  {\bibinfo  {journal} {Physical Review}\ }\textbf {\bibinfo {volume} {186}},\
  \bibinfo {pages} {443} (\bibinfo {year} {1969})}\BibitemShut {NoStop}%
\bibitem [{\citenamefont {Slater}(1937)}]{slater1937wave}%
  \BibitemOpen
  \bibfield  {author} {\bibinfo {author} {\bibfnamefont {J.~C.}\ \bibnamefont
  {Slater}},\ }\bibfield  {title} {\bibinfo {title} {Wave functions in a
  periodic potential},\ }\href@noop {} {\bibfield  {journal} {\bibinfo
  {journal} {Physical Review}\ }\textbf {\bibinfo {volume} {51}},\ \bibinfo
  {pages} {846} (\bibinfo {year} {1937})}\BibitemShut {NoStop}%
\bibitem [{\citenamefont {Papaconstantopoulos}\ \emph
  {et~al.}(2015)\citenamefont {Papaconstantopoulos}, \citenamefont {Klein},
  \citenamefont {Mehl},\ and\ \citenamefont
  {Pickett}}]{papaconstantopoulos2015}%
  \BibitemOpen
  \bibfield  {author} {\bibinfo {author} {\bibfnamefont {D.~A.}\ \bibnamefont
  {Papaconstantopoulos}}, \bibinfo {author} {\bibfnamefont {B.}~\bibnamefont
  {Klein}}, \bibinfo {author} {\bibfnamefont {M.~J.}\ \bibnamefont {Mehl}},\
  and\ \bibinfo {author} {\bibfnamefont {W.~E.}\ \bibnamefont {Pickett}},\
  }\bibfield  {title} {\bibinfo {title} {Cubic {H}$_3${S} around 200{GP}a: An
  atomic hydrogen superconductor stabilized by sulfur},\ }\href@noop {}
  {\bibfield  {journal} {\bibinfo  {journal} {Phys. Rev. B}\ }\textbf {\bibinfo
  {volume} {91}},\ \bibinfo {pages} {184511} (\bibinfo {year}
  {2015})}\BibitemShut {NoStop}%
\bibitem [{\citenamefont {Chang}\ \emph {et~al.}(2020)\citenamefont {Chang},
  \citenamefont {Silayi}, \citenamefont {Papaconstantopoulos},\ and\
  \citenamefont {Mehl}}]{chang2019}%
  \BibitemOpen
  \bibfield  {author} {\bibinfo {author} {\bibfnamefont {P.-H.}\ \bibnamefont
  {Chang}}, \bibinfo {author} {\bibfnamefont {S.}~\bibnamefont {Silayi}},
  \bibinfo {author} {\bibfnamefont {D.}~\bibnamefont {Papaconstantopoulos}},\
  and\ \bibinfo {author} {\bibfnamefont {M.}~\bibnamefont {Mehl}},\ }\bibfield
  {title} {\bibinfo {title} {Pressure-induced high-temperature
  superconductivity in hypothetical h3x (x=as, se, br, sb, te and i) in the h3s
  structure with im3¯m symmetry},\ }\href
  {https://www.sciencedirect.com/science/article/pii/S0022369719321110}
  {\bibfield  {journal} {\bibinfo  {journal} {Journal of Physics and Chemistry
  of Solids}\ }\textbf {\bibinfo {volume} {139}},\ \bibinfo {pages} {109315}
  (\bibinfo {year} {2020})}\BibitemShut {NoStop}%
\bibitem [{\citenamefont {Sakurai}\ and\ \citenamefont
  {Napolitano}(2017)}]{sakurai2017}%
  \BibitemOpen
  \bibfield  {author} {\bibinfo {author} {\bibfnamefont {J.~J.}\ \bibnamefont
  {Sakurai}}\ and\ \bibinfo {author} {\bibfnamefont {J.}~\bibnamefont
  {Napolitano}},\ }\bibinfo {title} {Scattering theory},\ in\ \href
  {https://doi.org/10.1017/9781108499996.010} {\emph {\bibinfo {booktitle}
  {Modern Quantum Mechanics}}}\ (\bibinfo  {publisher} {Cambridge University
  Press},\ \bibinfo {year} {2017})\ p.\ \bibinfo {pages} {386–445},\ \bibinfo
  {edition} {2nd}\ ed.\BibitemShut {Stop}%
\bibitem [{elk()}]{elk_code}%
  \BibitemOpen
  \href@noop {} {}\bibinfo {howpublished} {http://elk.sourceforge.net/},\
  \bibinfo {note} {the \textsc{elk fp-lapw} code}\BibitemShut {NoStop}%
\bibitem [{\citenamefont {Hutcheon}\ \emph {et~al.}(2020)\citenamefont
  {Hutcheon}, \citenamefont {Shipley},\ and\ \citenamefont
  {Needs}}]{hutcheon2020}%
  \BibitemOpen
  \bibfield  {author} {\bibinfo {author} {\bibfnamefont {M.~J.}\ \bibnamefont
  {Hutcheon}}, \bibinfo {author} {\bibfnamefont {A.~M.}\ \bibnamefont
  {Shipley}},\ and\ \bibinfo {author} {\bibfnamefont {R.~J.}\ \bibnamefont
  {Needs}},\ }\bibfield  {title} {\bibinfo {title} {Predicting novel
  superconducting hydrides using machine learning approaches},\ }\href
  {https://doi.org/10.1103/PhysRevB.101.144505} {\bibfield  {journal} {\bibinfo
   {journal} {Phys. Rev. B}\ }\textbf {\bibinfo {volume} {101}},\ \bibinfo
  {pages} {144505} (\bibinfo {year} {2020})}\BibitemShut {NoStop}%
\bibitem [{\citenamefont {Allen}\ and\ \citenamefont
  {Dynes}(1975)}]{AD_formula}%
  \BibitemOpen
  \bibfield  {author} {\bibinfo {author} {\bibfnamefont {P.~B.}\ \bibnamefont
  {Allen}}\ and\ \bibinfo {author} {\bibfnamefont {R.~C.}\ \bibnamefont
  {Dynes}},\ }\bibfield  {title} {\bibinfo {title} {Transition temperature of
  strong-coupled superconductors reanalyzed},\ }\href
  {https://doi.org/10.1103/PhysRevB.12.905} {\bibfield  {journal} {\bibinfo
  {journal} {Phys. Rev. B}\ }\textbf {\bibinfo {volume} {12}},\ \bibinfo
  {pages} {905} (\bibinfo {year} {1975})}\BibitemShut {NoStop}%
\bibitem [{\citenamefont {Giannozzi}\ \emph {et~al.}(2009)\citenamefont
  {Giannozzi}, \citenamefont {Baroni}, \citenamefont {Bonini}, \citenamefont
  {Calandra}, \citenamefont {Car}, \citenamefont {Cavazzoni}, \citenamefont
  {Ceresoli}, \citenamefont {Chiarotti}, \citenamefont {Cococcioni},
  \citenamefont {Dabo}, \citenamefont {{Dal Corso}}, \citenamefont
  {de~Gironcoli}, \citenamefont {Fabris}, \citenamefont {Fratesi},
  \citenamefont {Gebauer}, \citenamefont {Gerstmann}, \citenamefont
  {Gougoussis}, \citenamefont {Kokalj}, \citenamefont {Lazzeri}, \citenamefont
  {Martin-Samos}, \citenamefont {Marzari}, \citenamefont {Mauri}, \citenamefont
  {Mazzarello}, \citenamefont {Paolini}, \citenamefont {Pasquarello},
  \citenamefont {Paulatto}, \citenamefont {Sbraccia}, \citenamefont {Scandolo},
  \citenamefont {Sclauzero}, \citenamefont {Seitsonen}, \citenamefont
  {Smogunov}, \citenamefont {Umari},\ and\ \citenamefont
  {Wentzcovitch}}]{QE-2009}%
  \BibitemOpen
  \bibfield  {author} {\bibinfo {author} {\bibfnamefont {P.}~\bibnamefont
  {Giannozzi}}, \bibinfo {author} {\bibfnamefont {S.}~\bibnamefont {Baroni}},
  \bibinfo {author} {\bibfnamefont {N.}~\bibnamefont {Bonini}}, \bibinfo
  {author} {\bibfnamefont {M.}~\bibnamefont {Calandra}}, \bibinfo {author}
  {\bibfnamefont {R.}~\bibnamefont {Car}}, \bibinfo {author} {\bibfnamefont
  {C.}~\bibnamefont {Cavazzoni}}, \bibinfo {author} {\bibfnamefont
  {D.}~\bibnamefont {Ceresoli}}, \bibinfo {author} {\bibfnamefont {G.~L.}\
  \bibnamefont {Chiarotti}}, \bibinfo {author} {\bibfnamefont {M.}~\bibnamefont
  {Cococcioni}}, \bibinfo {author} {\bibfnamefont {I.}~\bibnamefont {Dabo}},
  \bibinfo {author} {\bibfnamefont {A.}~\bibnamefont {{Dal Corso}}}, \bibinfo
  {author} {\bibfnamefont {S.}~\bibnamefont {de~Gironcoli}}, \bibinfo {author}
  {\bibfnamefont {S.}~\bibnamefont {Fabris}}, \bibinfo {author} {\bibfnamefont
  {G.}~\bibnamefont {Fratesi}}, \bibinfo {author} {\bibfnamefont
  {R.}~\bibnamefont {Gebauer}}, \bibinfo {author} {\bibfnamefont
  {U.}~\bibnamefont {Gerstmann}}, \bibinfo {author} {\bibfnamefont
  {C.}~\bibnamefont {Gougoussis}}, \bibinfo {author} {\bibfnamefont
  {A.}~\bibnamefont {Kokalj}}, \bibinfo {author} {\bibfnamefont
  {M.}~\bibnamefont {Lazzeri}}, \bibinfo {author} {\bibfnamefont
  {L.}~\bibnamefont {Martin-Samos}}, \bibinfo {author} {\bibfnamefont
  {N.}~\bibnamefont {Marzari}}, \bibinfo {author} {\bibfnamefont
  {F.}~\bibnamefont {Mauri}}, \bibinfo {author} {\bibfnamefont
  {R.}~\bibnamefont {Mazzarello}}, \bibinfo {author} {\bibfnamefont
  {S.}~\bibnamefont {Paolini}}, \bibinfo {author} {\bibfnamefont
  {A.}~\bibnamefont {Pasquarello}}, \bibinfo {author} {\bibfnamefont
  {L.}~\bibnamefont {Paulatto}}, \bibinfo {author} {\bibfnamefont
  {C.}~\bibnamefont {Sbraccia}}, \bibinfo {author} {\bibfnamefont
  {S.}~\bibnamefont {Scandolo}}, \bibinfo {author} {\bibfnamefont
  {G.}~\bibnamefont {Sclauzero}}, \bibinfo {author} {\bibfnamefont {A.~P.}\
  \bibnamefont {Seitsonen}}, \bibinfo {author} {\bibfnamefont {A.}~\bibnamefont
  {Smogunov}}, \bibinfo {author} {\bibfnamefont {P.}~\bibnamefont {Umari}},\
  and\ \bibinfo {author} {\bibfnamefont {R.~M.}\ \bibnamefont {Wentzcovitch}},\
  }\bibfield  {title} {\bibinfo {title} {{QUANTUM ESPRESSO}: a modular and
  open-source software project for quantum simulations of materials},\ }\href
  {http://www.quantum-espresso.org} {\bibfield  {journal} {\bibinfo  {journal}
  {Journal of Physics: Condensed Matter}\ }\textbf {\bibinfo {volume} {21}},\
  \bibinfo {pages} {395502 (19pp)} (\bibinfo {year} {2009})}\BibitemShut
  {NoStop}%
\bibitem [{\citenamefont {Giannozzi}\ \emph {et~al.}(2017)\citenamefont
  {Giannozzi}, \citenamefont {Andreussi}, \citenamefont {Brumme}, \citenamefont
  {Bunau}, \citenamefont {Nardelli}, \citenamefont {Calandra}, \citenamefont
  {Car}, \citenamefont {Cavazzoni}, \citenamefont {Ceresoli}, \citenamefont
  {Cococcioni}, \citenamefont {Colonna}, \citenamefont {Carnimeo},
  \citenamefont {Corso}, \citenamefont {de~Gironcoli}, \citenamefont {Delugas},
  \citenamefont {DiStasio}, \citenamefont {Ferretti}, \citenamefont {Floris},
  \citenamefont {Fratesi}, \citenamefont {Fugallo}, \citenamefont {Gebauer},
  \citenamefont {Gerstmann}, \citenamefont {Giustino}, \citenamefont {Gorni},
  \citenamefont {Jia}, \citenamefont {Kawamura}, \citenamefont {Ko},
  \citenamefont {Kokalj}, \citenamefont {Kü{\c{c}}ükbenli}, \citenamefont
  {Lazzeri}, \citenamefont {Marsili}, \citenamefont {Marzari}, \citenamefont
  {Mauri}, \citenamefont {Nguyen}, \citenamefont {Nguyen}, \citenamefont {de-la
  Roza}, \citenamefont {Paulatto}, \citenamefont {Ponc{\'{e}}}, \citenamefont
  {Rocca}, \citenamefont {Sabatini}, \citenamefont {Santra}, \citenamefont
  {Schlipf}, \citenamefont {Seitsonen}, \citenamefont {Smogunov}, \citenamefont
  {Timrov}, \citenamefont {Thonhauser}, \citenamefont {Umari}, \citenamefont
  {Vast}, \citenamefont {Wu},\ and\ \citenamefont {Baroni}}]{QE-2017}%
  \BibitemOpen
  \bibfield  {author} {\bibinfo {author} {\bibfnamefont {P.}~\bibnamefont
  {Giannozzi}}, \bibinfo {author} {\bibfnamefont {O.}~\bibnamefont
  {Andreussi}}, \bibinfo {author} {\bibfnamefont {T.}~\bibnamefont {Brumme}},
  \bibinfo {author} {\bibfnamefont {O.}~\bibnamefont {Bunau}}, \bibinfo
  {author} {\bibfnamefont {M.~B.}\ \bibnamefont {Nardelli}}, \bibinfo {author}
  {\bibfnamefont {M.}~\bibnamefont {Calandra}}, \bibinfo {author}
  {\bibfnamefont {R.}~\bibnamefont {Car}}, \bibinfo {author} {\bibfnamefont
  {C.}~\bibnamefont {Cavazzoni}}, \bibinfo {author} {\bibfnamefont
  {D.}~\bibnamefont {Ceresoli}}, \bibinfo {author} {\bibfnamefont
  {M.}~\bibnamefont {Cococcioni}}, \bibinfo {author} {\bibfnamefont
  {N.}~\bibnamefont {Colonna}}, \bibinfo {author} {\bibfnamefont
  {I.}~\bibnamefont {Carnimeo}}, \bibinfo {author} {\bibfnamefont {A.~D.}\
  \bibnamefont {Corso}}, \bibinfo {author} {\bibfnamefont {S.}~\bibnamefont
  {de~Gironcoli}}, \bibinfo {author} {\bibfnamefont {P.}~\bibnamefont
  {Delugas}}, \bibinfo {author} {\bibfnamefont {R.~A.}\ \bibnamefont
  {DiStasio}}, \bibinfo {author} {\bibfnamefont {A.}~\bibnamefont {Ferretti}},
  \bibinfo {author} {\bibfnamefont {A.}~\bibnamefont {Floris}}, \bibinfo
  {author} {\bibfnamefont {G.}~\bibnamefont {Fratesi}}, \bibinfo {author}
  {\bibfnamefont {G.}~\bibnamefont {Fugallo}}, \bibinfo {author} {\bibfnamefont
  {R.}~\bibnamefont {Gebauer}}, \bibinfo {author} {\bibfnamefont
  {U.}~\bibnamefont {Gerstmann}}, \bibinfo {author} {\bibfnamefont
  {F.}~\bibnamefont {Giustino}}, \bibinfo {author} {\bibfnamefont
  {T.}~\bibnamefont {Gorni}}, \bibinfo {author} {\bibfnamefont
  {J.}~\bibnamefont {Jia}}, \bibinfo {author} {\bibfnamefont {M.}~\bibnamefont
  {Kawamura}}, \bibinfo {author} {\bibfnamefont {H.-Y.}\ \bibnamefont {Ko}},
  \bibinfo {author} {\bibfnamefont {A.}~\bibnamefont {Kokalj}}, \bibinfo
  {author} {\bibfnamefont {E.}~\bibnamefont {Kü{\c{c}}ükbenli}}, \bibinfo
  {author} {\bibfnamefont {M.}~\bibnamefont {Lazzeri}}, \bibinfo {author}
  {\bibfnamefont {M.}~\bibnamefont {Marsili}}, \bibinfo {author} {\bibfnamefont
  {N.}~\bibnamefont {Marzari}}, \bibinfo {author} {\bibfnamefont
  {F.}~\bibnamefont {Mauri}}, \bibinfo {author} {\bibfnamefont {N.~L.}\
  \bibnamefont {Nguyen}}, \bibinfo {author} {\bibfnamefont {H.-V.}\
  \bibnamefont {Nguyen}}, \bibinfo {author} {\bibfnamefont {A.~O.}\
  \bibnamefont {de-la Roza}}, \bibinfo {author} {\bibfnamefont
  {L.}~\bibnamefont {Paulatto}}, \bibinfo {author} {\bibfnamefont
  {S.}~\bibnamefont {Ponc{\'{e}}}}, \bibinfo {author} {\bibfnamefont
  {D.}~\bibnamefont {Rocca}}, \bibinfo {author} {\bibfnamefont
  {R.}~\bibnamefont {Sabatini}}, \bibinfo {author} {\bibfnamefont
  {B.}~\bibnamefont {Santra}}, \bibinfo {author} {\bibfnamefont
  {M.}~\bibnamefont {Schlipf}}, \bibinfo {author} {\bibfnamefont {A.~P.}\
  \bibnamefont {Seitsonen}}, \bibinfo {author} {\bibfnamefont {A.}~\bibnamefont
  {Smogunov}}, \bibinfo {author} {\bibfnamefont {I.}~\bibnamefont {Timrov}},
  \bibinfo {author} {\bibfnamefont {T.}~\bibnamefont {Thonhauser}}, \bibinfo
  {author} {\bibfnamefont {P.}~\bibnamefont {Umari}}, \bibinfo {author}
  {\bibfnamefont {N.}~\bibnamefont {Vast}}, \bibinfo {author} {\bibfnamefont
  {X.}~\bibnamefont {Wu}},\ and\ \bibinfo {author} {\bibfnamefont
  {S.}~\bibnamefont {Baroni}},\ }\bibfield  {title} {\bibinfo {title} {Advanced
  capabilities for materials modelling with quantum {ESPRESSO}},\ }\href
  {https://doi.org/10.1088/1361-648x/aa8f79} {\bibfield  {journal} {\bibinfo
  {journal} {Journal of Physics: Condensed Matter}\ }\textbf {\bibinfo {volume}
  {29}},\ \bibinfo {pages} {465901} (\bibinfo {year} {2017})}\BibitemShut
  {NoStop}%
\bibitem [{\citenamefont {Eliashberg}(1960)}]{eliashberg1960}%
  \BibitemOpen
  \bibfield  {author} {\bibinfo {author} {\bibfnamefont {G.~M.}\ \bibnamefont
  {Eliashberg}},\ }\bibfield  {title} {\bibinfo {title} {Interactions between
  electrons and lattice vibrations in a superconductor},\ }\href@noop {}
  {\bibfield  {journal} {\bibinfo  {journal} {Sov. Phys. JETP}\ }\textbf
  {\bibinfo {volume} {11}},\ \bibinfo {pages} {696} (\bibinfo {year}
  {1960})}\BibitemShut {NoStop}%
\bibitem [{\citenamefont {Wierzbowska}\ \emph {et~al.}(2005)\citenamefont
  {Wierzbowska}, \citenamefont {de~Gironcoli},\ and\ \citenamefont
  {Giannozzi}}]{wierzbowska2005}%
  \BibitemOpen
  \bibfield  {author} {\bibinfo {author} {\bibfnamefont {M.}~\bibnamefont
  {Wierzbowska}}, \bibinfo {author} {\bibfnamefont {S.}~\bibnamefont
  {de~Gironcoli}},\ and\ \bibinfo {author} {\bibfnamefont {P.}~\bibnamefont
  {Giannozzi}},\ }\bibfield  {title} {\bibinfo {title} {Origins of low-and
  high-pressure discontinuities of $t_c$ in niobium},\ }\href@noop {}
  {\bibfield  {journal} {\bibinfo  {journal} {arXiv:cond-mat/0504077}\ }
  (\bibinfo {year} {2005})}\BibitemShut {NoStop}%
\bibitem [{\citenamefont {Shipley}\ \emph {et~al.}(2020)\citenamefont
  {Shipley}, \citenamefont {Hutcheon}, \citenamefont {Johnson}, \citenamefont
  {Needs},\ and\ \citenamefont {Pickard}}]{shipley2020}%
  \BibitemOpen
  \bibfield  {author} {\bibinfo {author} {\bibfnamefont {A.~M.}\ \bibnamefont
  {Shipley}}, \bibinfo {author} {\bibfnamefont {M.~J.}\ \bibnamefont
  {Hutcheon}}, \bibinfo {author} {\bibfnamefont {M.~S.}\ \bibnamefont
  {Johnson}}, \bibinfo {author} {\bibfnamefont {R.~J.}\ \bibnamefont {Needs}},\
  and\ \bibinfo {author} {\bibfnamefont {C.~J.}\ \bibnamefont {Pickard}},\
  }\bibfield  {title} {\bibinfo {title} {Stability and superconductivity of
  lanthanum and yttrium decahydrides},\ }\href
  {https://doi.org/10.1103/PhysRevB.101.224511} {\bibfield  {journal} {\bibinfo
   {journal} {Phys. Rev. B}\ }\textbf {\bibinfo {volume} {101}},\ \bibinfo
  {pages} {224511} (\bibinfo {year} {2020})}\BibitemShut {NoStop}%
\bibitem [{Note1()}]{Note1}%
  \BibitemOpen
  \bibinfo {note} {The optimizations to \protect \textsc {quantum espresso}
  resulting from this work have been submitted to the developers (see \protect
  \url {https://gitlab.com/miicck/q-e})}\BibitemShut {NoStop}%
\bibitem [{\citenamefont {Semenok}\ \emph {et~al.}(2018)\citenamefont
  {Semenok}, \citenamefont {Kvashnin}, \citenamefont {Kruglov},\ and\
  \citenamefont {Oganov}}]{semenok2018actinium}%
  \BibitemOpen
  \bibfield  {author} {\bibinfo {author} {\bibfnamefont {D.~V.}\ \bibnamefont
  {Semenok}}, \bibinfo {author} {\bibfnamefont {A.~G.}\ \bibnamefont
  {Kvashnin}}, \bibinfo {author} {\bibfnamefont {I.~A.}\ \bibnamefont
  {Kruglov}},\ and\ \bibinfo {author} {\bibfnamefont {A.~R.}\ \bibnamefont
  {Oganov}},\ }\bibfield  {title} {\bibinfo {title} {Actinium hydrides ach10,
  ach12, and ach16 as high-temperature conventional superconductors},\
  }\href@noop {} {\bibfield  {journal} {\bibinfo  {journal} {The journal of
  physical chemistry letters}\ }\textbf {\bibinfo {volume} {9}},\ \bibinfo
  {pages} {1920} (\bibinfo {year} {2018})}\BibitemShut {NoStop}%
\bibitem [{\citenamefont {Hou}\ \emph {et~al.}(2015)\citenamefont {Hou},
  \citenamefont {Zhao}, \citenamefont {Tian}, \citenamefont {Li}, \citenamefont
  {Duan}, \citenamefont {Zhao}, \citenamefont {Chu}, \citenamefont {Liu},\ and\
  \citenamefont {Cui}}]{hou2015high}%
  \BibitemOpen
  \bibfield  {author} {\bibinfo {author} {\bibfnamefont {P.}~\bibnamefont
  {Hou}}, \bibinfo {author} {\bibfnamefont {X.}~\bibnamefont {Zhao}}, \bibinfo
  {author} {\bibfnamefont {F.}~\bibnamefont {Tian}}, \bibinfo {author}
  {\bibfnamefont {D.}~\bibnamefont {Li}}, \bibinfo {author} {\bibfnamefont
  {D.}~\bibnamefont {Duan}}, \bibinfo {author} {\bibfnamefont {Z.}~\bibnamefont
  {Zhao}}, \bibinfo {author} {\bibfnamefont {B.}~\bibnamefont {Chu}}, \bibinfo
  {author} {\bibfnamefont {B.}~\bibnamefont {Liu}},\ and\ \bibinfo {author}
  {\bibfnamefont {T.}~\bibnamefont {Cui}},\ }\bibfield  {title} {\bibinfo
  {title} {High pressure structures and superconductivity of alh 3 (h 2)
  predicted by first principles},\ }\href@noop {} {\bibfield  {journal}
  {\bibinfo  {journal} {RSC Advances}\ }\textbf {\bibinfo {volume} {5}},\
  \bibinfo {pages} {5096} (\bibinfo {year} {2015})}\BibitemShut {NoStop}%
\bibitem [{\citenamefont {Fu}\ \emph {et~al.}(2016)\citenamefont {Fu},
  \citenamefont {Du}, \citenamefont {Zhang}, \citenamefont {Peng},
  \citenamefont {Zhang}, \citenamefont {Pickard}, \citenamefont {Needs},
  \citenamefont {Singh}, \citenamefont {Zheng},\ and\ \citenamefont
  {Ma}}]{fu2016high}%
  \BibitemOpen
  \bibfield  {author} {\bibinfo {author} {\bibfnamefont {Y.}~\bibnamefont
  {Fu}}, \bibinfo {author} {\bibfnamefont {X.}~\bibnamefont {Du}}, \bibinfo
  {author} {\bibfnamefont {L.}~\bibnamefont {Zhang}}, \bibinfo {author}
  {\bibfnamefont {F.}~\bibnamefont {Peng}}, \bibinfo {author} {\bibfnamefont
  {M.}~\bibnamefont {Zhang}}, \bibinfo {author} {\bibfnamefont {C.~J.}\
  \bibnamefont {Pickard}}, \bibinfo {author} {\bibfnamefont {R.~J.}\
  \bibnamefont {Needs}}, \bibinfo {author} {\bibfnamefont {D.~J.}\ \bibnamefont
  {Singh}}, \bibinfo {author} {\bibfnamefont {W.}~\bibnamefont {Zheng}},\ and\
  \bibinfo {author} {\bibfnamefont {Y.}~\bibnamefont {Ma}},\ }\bibfield
  {title} {\bibinfo {title} {High-pressure phase stability and
  superconductivity of pnictogen hydrides and chemical trends for compressed
  hydrides},\ }\href@noop {} {\bibfield  {journal} {\bibinfo  {journal}
  {Chemistry of Materials}\ }\textbf {\bibinfo {volume} {28}},\ \bibinfo
  {pages} {1746} (\bibinfo {year} {2016})}\BibitemShut {NoStop}%
\bibitem [{\citenamefont {Abe}\ and\ \citenamefont
  {Ashcroft}(2011)}]{abe2011crystalline}%
  \BibitemOpen
  \bibfield  {author} {\bibinfo {author} {\bibfnamefont {K.}~\bibnamefont
  {Abe}}\ and\ \bibinfo {author} {\bibfnamefont {N.}~\bibnamefont {Ashcroft}},\
  }\bibfield  {title} {\bibinfo {title} {Crystalline diborane at high
  pressures},\ }\href@noop {} {\bibfield  {journal} {\bibinfo  {journal}
  {Physical Review B}\ }\textbf {\bibinfo {volume} {84}},\ \bibinfo {pages}
  {104118} (\bibinfo {year} {2011})}\BibitemShut {NoStop}%
\bibitem [{\citenamefont {Yu}\ \emph {et~al.}(2014)\citenamefont {Yu},
  \citenamefont {Zeng}, \citenamefont {Oganov}, \citenamefont {Hu},
  \citenamefont {Frapper},\ and\ \citenamefont {Zhang}}]{yu2014exploration}%
  \BibitemOpen
  \bibfield  {author} {\bibinfo {author} {\bibfnamefont {S.}~\bibnamefont
  {Yu}}, \bibinfo {author} {\bibfnamefont {Q.}~\bibnamefont {Zeng}}, \bibinfo
  {author} {\bibfnamefont {A.~R.}\ \bibnamefont {Oganov}}, \bibinfo {author}
  {\bibfnamefont {C.}~\bibnamefont {Hu}}, \bibinfo {author} {\bibfnamefont
  {G.}~\bibnamefont {Frapper}},\ and\ \bibinfo {author} {\bibfnamefont
  {L.}~\bibnamefont {Zhang}},\ }\bibfield  {title} {\bibinfo {title}
  {Exploration of stable compounds, crystal structures, and superconductivity
  in the be-h system},\ }\href@noop {} {\bibfield  {journal} {\bibinfo
  {journal} {AIP Advances}\ }\textbf {\bibinfo {volume} {4}},\ \bibinfo {pages}
  {107118} (\bibinfo {year} {2014})}\BibitemShut {NoStop}%
\bibitem [{\citenamefont {Hu}\ \emph {et~al.}(2013)\citenamefont {Hu},
  \citenamefont {Oganov}, \citenamefont {Zhu}, \citenamefont {Qian},
  \citenamefont {Frapper}, \citenamefont {Lyakhov},\ and\ \citenamefont
  {Zhou}}]{hu2013pressure}%
  \BibitemOpen
  \bibfield  {author} {\bibinfo {author} {\bibfnamefont {C.-H.}\ \bibnamefont
  {Hu}}, \bibinfo {author} {\bibfnamefont {A.~R.}\ \bibnamefont {Oganov}},
  \bibinfo {author} {\bibfnamefont {Q.}~\bibnamefont {Zhu}}, \bibinfo {author}
  {\bibfnamefont {G.-R.}\ \bibnamefont {Qian}}, \bibinfo {author}
  {\bibfnamefont {G.}~\bibnamefont {Frapper}}, \bibinfo {author} {\bibfnamefont
  {A.~O.}\ \bibnamefont {Lyakhov}},\ and\ \bibinfo {author} {\bibfnamefont
  {H.-Y.}\ \bibnamefont {Zhou}},\ }\bibfield  {title} {\bibinfo {title}
  {Pressure-induced stabilization and insulator-superconductor transition of
  bh},\ }\href@noop {} {\bibfield  {journal} {\bibinfo  {journal} {Physical
  review letters}\ }\textbf {\bibinfo {volume} {110}},\ \bibinfo {pages}
  {165504} (\bibinfo {year} {2013})}\BibitemShut {NoStop}%
\bibitem [{\citenamefont {Ma}\ \emph {et~al.}(2015{\natexlab{a}})\citenamefont
  {Ma}, \citenamefont {Duan}, \citenamefont {Li}, \citenamefont {Liu},
  \citenamefont {Tian}, \citenamefont {Yu}, \citenamefont {Xu}, \citenamefont
  {Shao}, \citenamefont {Liu},\ and\ \citenamefont {Cui}}]{ma2015high}%
  \BibitemOpen
  \bibfield  {author} {\bibinfo {author} {\bibfnamefont {Y.}~\bibnamefont
  {Ma}}, \bibinfo {author} {\bibfnamefont {D.}~\bibnamefont {Duan}}, \bibinfo
  {author} {\bibfnamefont {D.}~\bibnamefont {Li}}, \bibinfo {author}
  {\bibfnamefont {Y.}~\bibnamefont {Liu}}, \bibinfo {author} {\bibfnamefont
  {F.}~\bibnamefont {Tian}}, \bibinfo {author} {\bibfnamefont {H.}~\bibnamefont
  {Yu}}, \bibinfo {author} {\bibfnamefont {C.}~\bibnamefont {Xu}}, \bibinfo
  {author} {\bibfnamefont {Z.}~\bibnamefont {Shao}}, \bibinfo {author}
  {\bibfnamefont {B.}~\bibnamefont {Liu}},\ and\ \bibinfo {author}
  {\bibfnamefont {T.}~\bibnamefont {Cui}},\ }\bibfield  {title} {\bibinfo
  {title} {High-pressure structures and superconductivity of bismuth
  hydrides},\ }\href@noop {} {\bibfield  {journal} {\bibinfo  {journal} {arXiv
  preprint arXiv:1511.05291}\ } (\bibinfo {year}
  {2015}{\natexlab{a}})}\BibitemShut {NoStop}%
\bibitem [{\citenamefont {Wang}\ \emph {et~al.}(2012)\citenamefont {Wang},
  \citenamefont {John}, \citenamefont {Tanaka}, \citenamefont {Iitaka},\ and\
  \citenamefont {Ma}}]{wang2012superconductive}%
  \BibitemOpen
  \bibfield  {author} {\bibinfo {author} {\bibfnamefont {H.}~\bibnamefont
  {Wang}}, \bibinfo {author} {\bibfnamefont {S.~T.}\ \bibnamefont {John}},
  \bibinfo {author} {\bibfnamefont {K.}~\bibnamefont {Tanaka}}, \bibinfo
  {author} {\bibfnamefont {T.}~\bibnamefont {Iitaka}},\ and\ \bibinfo {author}
  {\bibfnamefont {Y.}~\bibnamefont {Ma}},\ }\bibfield  {title} {\bibinfo
  {title} {Superconductive sodalite-like clathrate calcium hydride at high
  pressures},\ }\href {https://doi.org/10.1073/pnas.1118168109} {\bibfield
  {journal} {\bibinfo  {journal} {Proceedings of the National Academy of
  Sciences}\ }\textbf {\bibinfo {volume} {109}},\ \bibinfo {pages} {6463}
  (\bibinfo {year} {2012})}\BibitemShut {NoStop}%
\bibitem [{\citenamefont {Salke}\ \emph {et~al.}(2019)\citenamefont {Salke},
  \citenamefont {Esfahani}, \citenamefont {Zhang}, \citenamefont {Kruglov},
  \citenamefont {Zhou}, \citenamefont {Wang}, \citenamefont {Greenberg},
  \citenamefont {Prakapenka}, \citenamefont {Liu}, \citenamefont {Oganov},\
  and\ \citenamefont {Lin}}]{salke2019synthesis}%
  \BibitemOpen
  \bibfield  {author} {\bibinfo {author} {\bibfnamefont {N.~P.}\ \bibnamefont
  {Salke}}, \bibinfo {author} {\bibfnamefont {M.~M.~D.}\ \bibnamefont
  {Esfahani}}, \bibinfo {author} {\bibfnamefont {Y.}~\bibnamefont {Zhang}},
  \bibinfo {author} {\bibfnamefont {I.~A.}\ \bibnamefont {Kruglov}}, \bibinfo
  {author} {\bibfnamefont {J.}~\bibnamefont {Zhou}}, \bibinfo {author}
  {\bibfnamefont {Y.}~\bibnamefont {Wang}}, \bibinfo {author} {\bibfnamefont
  {E.}~\bibnamefont {Greenberg}}, \bibinfo {author} {\bibfnamefont {V.~B.}\
  \bibnamefont {Prakapenka}}, \bibinfo {author} {\bibfnamefont
  {J.}~\bibnamefont {Liu}}, \bibinfo {author} {\bibfnamefont {A.~R.}\
  \bibnamefont {Oganov}},\ and\ \bibinfo {author} {\bibfnamefont {J.-F.}\
  \bibnamefont {Lin}},\ }\bibfield  {title} {\bibinfo {title} {Synthesis of
  clathrate cerium superhydride ceh 9 at 80-100 gpa with atomic hydrogen
  sublattice},\ }\href@noop {} {\bibfield  {journal} {\bibinfo  {journal}
  {Nature communications}\ }\textbf {\bibinfo {volume} {10}},\ \bibinfo {pages}
  {1} (\bibinfo {year} {2019})}\BibitemShut {NoStop}%
\bibitem [{\citenamefont {Yu}\ \emph {et~al.}(2015)\citenamefont {Yu},
  \citenamefont {Jia}, \citenamefont {Frapper}, \citenamefont {Li},
  \citenamefont {Oganov}, \citenamefont {Zeng},\ and\ \citenamefont
  {Zhang}}]{yu2015pressure}%
  \BibitemOpen
  \bibfield  {author} {\bibinfo {author} {\bibfnamefont {S.}~\bibnamefont
  {Yu}}, \bibinfo {author} {\bibfnamefont {X.}~\bibnamefont {Jia}}, \bibinfo
  {author} {\bibfnamefont {G.}~\bibnamefont {Frapper}}, \bibinfo {author}
  {\bibfnamefont {D.}~\bibnamefont {Li}}, \bibinfo {author} {\bibfnamefont
  {A.~R.}\ \bibnamefont {Oganov}}, \bibinfo {author} {\bibfnamefont
  {Q.}~\bibnamefont {Zeng}},\ and\ \bibinfo {author} {\bibfnamefont
  {L.}~\bibnamefont {Zhang}},\ }\bibfield  {title} {\bibinfo {title}
  {Pressure-driven formation and stabilization of superconductive chromium
  hydrides},\ }\href@noop {} {\bibfield  {journal} {\bibinfo  {journal}
  {Scientific reports}\ }\textbf {\bibinfo {volume} {5}},\ \bibinfo {pages}
  {17764} (\bibinfo {year} {2015})}\BibitemShut {NoStop}%
\bibitem [{\citenamefont {Heil}\ \emph {et~al.}(2018)\citenamefont {Heil},
  \citenamefont {Bachelet},\ and\ \citenamefont {Boeri}}]{heil2018absence}%
  \BibitemOpen
  \bibfield  {author} {\bibinfo {author} {\bibfnamefont {C.}~\bibnamefont
  {Heil}}, \bibinfo {author} {\bibfnamefont {G.~B.}\ \bibnamefont {Bachelet}},\
  and\ \bibinfo {author} {\bibfnamefont {L.}~\bibnamefont {Boeri}},\ }\bibfield
   {title} {\bibinfo {title} {Absence of superconductivity in iron polyhydrides
  at high pressures},\ }\href@noop {} {\bibfield  {journal} {\bibinfo
  {journal} {Physical Review B}\ }\textbf {\bibinfo {volume} {97}},\ \bibinfo
  {pages} {214510} (\bibinfo {year} {2018})}\BibitemShut {NoStop}%
\bibitem [{\citenamefont {Esfahani}\ \emph {et~al.}(2017)\citenamefont
  {Esfahani}, \citenamefont {Oganov}, \citenamefont {Niu},\ and\ \citenamefont
  {Zhang}}]{esfahani2017superconductivity}%
  \BibitemOpen
  \bibfield  {author} {\bibinfo {author} {\bibfnamefont {M.~M.~D.}\
  \bibnamefont {Esfahani}}, \bibinfo {author} {\bibfnamefont {A.~R.}\
  \bibnamefont {Oganov}}, \bibinfo {author} {\bibfnamefont {H.}~\bibnamefont
  {Niu}},\ and\ \bibinfo {author} {\bibfnamefont {J.}~\bibnamefont {Zhang}},\
  }\bibfield  {title} {\bibinfo {title} {Superconductivity and unexpected
  chemistry of germanium hydrides under pressure},\ }\href@noop {} {\bibfield
  {journal} {\bibinfo  {journal} {Physical Review B}\ }\textbf {\bibinfo
  {volume} {95}},\ \bibinfo {pages} {134506} (\bibinfo {year}
  {2017})}\BibitemShut {NoStop}%
\bibitem [{\citenamefont {Gao}\ \emph {et~al.}(2008)\citenamefont {Gao},
  \citenamefont {Oganov}, \citenamefont {Bergara}, \citenamefont
  {Martinez-Canales}, \citenamefont {Cui}, \citenamefont {Iitaka},
  \citenamefont {Ma},\ and\ \citenamefont {Zou}}]{gao2008superconducting}%
  \BibitemOpen
  \bibfield  {author} {\bibinfo {author} {\bibfnamefont {G.}~\bibnamefont
  {Gao}}, \bibinfo {author} {\bibfnamefont {A.~R.}\ \bibnamefont {Oganov}},
  \bibinfo {author} {\bibfnamefont {A.}~\bibnamefont {Bergara}}, \bibinfo
  {author} {\bibfnamefont {M.}~\bibnamefont {Martinez-Canales}}, \bibinfo
  {author} {\bibfnamefont {T.}~\bibnamefont {Cui}}, \bibinfo {author}
  {\bibfnamefont {T.}~\bibnamefont {Iitaka}}, \bibinfo {author} {\bibfnamefont
  {Y.}~\bibnamefont {Ma}},\ and\ \bibinfo {author} {\bibfnamefont
  {G.}~\bibnamefont {Zou}},\ }\bibfield  {title} {\bibinfo {title}
  {Superconducting high pressure phase of germane},\ }\href@noop {} {\bibfield
  {journal} {\bibinfo  {journal} {Physical review letters}\ }\textbf {\bibinfo
  {volume} {101}},\ \bibinfo {pages} {107002} (\bibinfo {year}
  {2008})}\BibitemShut {NoStop}%
\bibitem [{\citenamefont {Zhong}\ \emph {et~al.}(2012)\citenamefont {Zhong},
  \citenamefont {Zhang}, \citenamefont {Chen}, \citenamefont {Li},
  \citenamefont {Zhang},\ and\ \citenamefont {Lin}}]{zhong2012structural}%
  \BibitemOpen
  \bibfield  {author} {\bibinfo {author} {\bibfnamefont {G.}~\bibnamefont
  {Zhong}}, \bibinfo {author} {\bibfnamefont {C.}~\bibnamefont {Zhang}},
  \bibinfo {author} {\bibfnamefont {X.}~\bibnamefont {Chen}}, \bibinfo {author}
  {\bibfnamefont {Y.}~\bibnamefont {Li}}, \bibinfo {author} {\bibfnamefont
  {R.}~\bibnamefont {Zhang}},\ and\ \bibinfo {author} {\bibfnamefont
  {H.}~\bibnamefont {Lin}},\ }\bibfield  {title} {\bibinfo {title} {Structural,
  electronic, dynamical, and superconducting properties in dense geh4 (h2) 2},\
  }\href@noop {} {\bibfield  {journal} {\bibinfo  {journal} {The Journal of
  Physical Chemistry C}\ }\textbf {\bibinfo {volume} {116}},\ \bibinfo {pages}
  {5225} (\bibinfo {year} {2012})}\BibitemShut {NoStop}%
\bibitem [{\citenamefont {Duan}\ \emph {et~al.}(2015)\citenamefont {Duan},
  \citenamefont {Tian}, \citenamefont {Liu}, \citenamefont {Huang},
  \citenamefont {Li}, \citenamefont {Yu}, \citenamefont {Ma}, \citenamefont
  {Liu},\ and\ \citenamefont {Cui}}]{duan2015enhancement}%
  \BibitemOpen
  \bibfield  {author} {\bibinfo {author} {\bibfnamefont {D.}~\bibnamefont
  {Duan}}, \bibinfo {author} {\bibfnamefont {F.}~\bibnamefont {Tian}}, \bibinfo
  {author} {\bibfnamefont {Y.}~\bibnamefont {Liu}}, \bibinfo {author}
  {\bibfnamefont {X.}~\bibnamefont {Huang}}, \bibinfo {author} {\bibfnamefont
  {D.}~\bibnamefont {Li}}, \bibinfo {author} {\bibfnamefont {H.}~\bibnamefont
  {Yu}}, \bibinfo {author} {\bibfnamefont {Y.}~\bibnamefont {Ma}}, \bibinfo
  {author} {\bibfnamefont {B.}~\bibnamefont {Liu}},\ and\ \bibinfo {author}
  {\bibfnamefont {T.}~\bibnamefont {Cui}},\ }\bibfield  {title} {\bibinfo
  {title} {Enhancement of t c in the atomic phase of iodine-doped hydrogen at
  high pressures},\ }\href@noop {} {\bibfield  {journal} {\bibinfo  {journal}
  {Physical Chemistry Chemical Physics}\ }\textbf {\bibinfo {volume} {17}},\
  \bibinfo {pages} {32335} (\bibinfo {year} {2015})}\BibitemShut {NoStop}%
\bibitem [{\citenamefont {Shamp}\ and\ \citenamefont
  {Zurek}(2015)}]{shamp2015superconducting}%
  \BibitemOpen
  \bibfield  {author} {\bibinfo {author} {\bibfnamefont {A.}~\bibnamefont
  {Shamp}}\ and\ \bibinfo {author} {\bibfnamefont {E.}~\bibnamefont {Zurek}},\
  }\bibfield  {title} {\bibinfo {title} {Superconducting high-pressure phases
  composed of hydrogen and iodine},\ }\href@noop {} {\bibfield  {journal}
  {\bibinfo  {journal} {The journal of physical chemistry letters}\ }\textbf
  {\bibinfo {volume} {6}},\ \bibinfo {pages} {4067} (\bibinfo {year}
  {2015})}\BibitemShut {NoStop}%
\bibitem [{\citenamefont {Duan}\ \emph {et~al.}(2014)\citenamefont {Duan},
  \citenamefont {Liu}, \citenamefont {Tian}, \citenamefont {Li}, \citenamefont
  {Huang}, \citenamefont {Zhao}, \citenamefont {Yu}, \citenamefont {Liu},
  \citenamefont {Tian},\ and\ \citenamefont {Cui}}]{duan2014pressure}%
  \BibitemOpen
  \bibfield  {author} {\bibinfo {author} {\bibfnamefont {D.}~\bibnamefont
  {Duan}}, \bibinfo {author} {\bibfnamefont {Y.}~\bibnamefont {Liu}}, \bibinfo
  {author} {\bibfnamefont {F.}~\bibnamefont {Tian}}, \bibinfo {author}
  {\bibfnamefont {D.}~\bibnamefont {Li}}, \bibinfo {author} {\bibfnamefont
  {X.}~\bibnamefont {Huang}}, \bibinfo {author} {\bibfnamefont
  {Z.}~\bibnamefont {Zhao}}, \bibinfo {author} {\bibfnamefont {H.}~\bibnamefont
  {Yu}}, \bibinfo {author} {\bibfnamefont {B.}~\bibnamefont {Liu}}, \bibinfo
  {author} {\bibfnamefont {W.}~\bibnamefont {Tian}},\ and\ \bibinfo {author}
  {\bibfnamefont {T.}~\bibnamefont {Cui}},\ }\bibfield  {title} {\bibinfo
  {title} {Pressure-induced metallization of dense (h 2 s) 2 h 2 with high-t c
  superconductivity},\ }\href@noop {} {\bibfield  {journal} {\bibinfo
  {journal} {Scientific reports}\ }\textbf {\bibinfo {volume} {4}},\ \bibinfo
  {pages} {6968} (\bibinfo {year} {2014})}\BibitemShut {NoStop}%
\bibitem [{\citenamefont {Errea}\ \emph {et~al.}(2015)\citenamefont {Errea},
  \citenamefont {Calandra}, \citenamefont {Pickard}, \citenamefont {Nelson},
  \citenamefont {Needs}, \citenamefont {Li}, \citenamefont {Liu}, \citenamefont
  {Zhang}, \citenamefont {Ma},\ and\ \citenamefont {Mauri}}]{errea2015high}%
  \BibitemOpen
  \bibfield  {author} {\bibinfo {author} {\bibfnamefont {I.}~\bibnamefont
  {Errea}}, \bibinfo {author} {\bibfnamefont {M.}~\bibnamefont {Calandra}},
  \bibinfo {author} {\bibfnamefont {C.~J.}\ \bibnamefont {Pickard}}, \bibinfo
  {author} {\bibfnamefont {J.}~\bibnamefont {Nelson}}, \bibinfo {author}
  {\bibfnamefont {R.~J.}\ \bibnamefont {Needs}}, \bibinfo {author}
  {\bibfnamefont {Y.}~\bibnamefont {Li}}, \bibinfo {author} {\bibfnamefont
  {H.}~\bibnamefont {Liu}}, \bibinfo {author} {\bibfnamefont {Y.}~\bibnamefont
  {Zhang}}, \bibinfo {author} {\bibfnamefont {Y.}~\bibnamefont {Ma}},\ and\
  \bibinfo {author} {\bibfnamefont {F.}~\bibnamefont {Mauri}},\ }\bibfield
  {title} {\bibinfo {title} {High-pressure hydrogen sulfide from first
  principles: A strongly anharmonic phonon-mediated superconductor},\
  }\href@noop {} {\bibfield  {journal} {\bibinfo  {journal} {Physical review
  letters}\ }\textbf {\bibinfo {volume} {114}},\ \bibinfo {pages} {157004}
  (\bibinfo {year} {2015})}\BibitemShut {NoStop}%
\bibitem [{\citenamefont {Zhang}\ \emph {et~al.}(2015)\citenamefont {Zhang},
  \citenamefont {Wang}, \citenamefont {Zhang}, \citenamefont {Liu},
  \citenamefont {Zhong}, \citenamefont {Song}, \citenamefont {Yang},
  \citenamefont {Zhang},\ and\ \citenamefont {Ma}}]{zhang2015phase}%
  \BibitemOpen
  \bibfield  {author} {\bibinfo {author} {\bibfnamefont {S.}~\bibnamefont
  {Zhang}}, \bibinfo {author} {\bibfnamefont {Y.}~\bibnamefont {Wang}},
  \bibinfo {author} {\bibfnamefont {J.}~\bibnamefont {Zhang}}, \bibinfo
  {author} {\bibfnamefont {H.}~\bibnamefont {Liu}}, \bibinfo {author}
  {\bibfnamefont {X.}~\bibnamefont {Zhong}}, \bibinfo {author} {\bibfnamefont
  {H.-F.}\ \bibnamefont {Song}}, \bibinfo {author} {\bibfnamefont
  {G.}~\bibnamefont {Yang}}, \bibinfo {author} {\bibfnamefont {L.}~\bibnamefont
  {Zhang}},\ and\ \bibinfo {author} {\bibfnamefont {Y.}~\bibnamefont {Ma}},\
  }\bibfield  {title} {\bibinfo {title} {Phase diagram and high-temperature
  superconductivity of compressed selenium hydrides},\ }\href@noop {}
  {\bibfield  {journal} {\bibinfo  {journal} {Scientific reports}\ }\textbf
  {\bibinfo {volume} {5}},\ \bibinfo {pages} {15433} (\bibinfo {year}
  {2015})}\BibitemShut {NoStop}%
\bibitem [{\citenamefont {Liu}\ \emph {et~al.}(2015{\natexlab{a}})\citenamefont
  {Liu}, \citenamefont {Duan}, \citenamefont {Tian}, \citenamefont {Wang},
  \citenamefont {Wu}, \citenamefont {Ma}, \citenamefont {Yu}, \citenamefont
  {Li}, \citenamefont {Liu},\ and\ \citenamefont {Cui}}]{liu2015prediction}%
  \BibitemOpen
  \bibfield  {author} {\bibinfo {author} {\bibfnamefont {Y.}~\bibnamefont
  {Liu}}, \bibinfo {author} {\bibfnamefont {D.}~\bibnamefont {Duan}}, \bibinfo
  {author} {\bibfnamefont {F.}~\bibnamefont {Tian}}, \bibinfo {author}
  {\bibfnamefont {C.}~\bibnamefont {Wang}}, \bibinfo {author} {\bibfnamefont
  {G.}~\bibnamefont {Wu}}, \bibinfo {author} {\bibfnamefont {Y.}~\bibnamefont
  {Ma}}, \bibinfo {author} {\bibfnamefont {H.}~\bibnamefont {Yu}}, \bibinfo
  {author} {\bibfnamefont {D.}~\bibnamefont {Li}}, \bibinfo {author}
  {\bibfnamefont {B.}~\bibnamefont {Liu}},\ and\ \bibinfo {author}
  {\bibfnamefont {T.}~\bibnamefont {Cui}},\ }\bibfield  {title} {\bibinfo
  {title} {Prediction of stoichiometric poh n compounds: crystal structures and
  properties},\ }\href@noop {} {\bibfield  {journal} {\bibinfo  {journal} {RSC
  Advances}\ }\textbf {\bibinfo {volume} {5}},\ \bibinfo {pages} {103445}
  (\bibinfo {year} {2015}{\natexlab{a}})}\BibitemShut {NoStop}%
\bibitem [{\citenamefont {Zhong}\ \emph {et~al.}(2016)\citenamefont {Zhong},
  \citenamefont {Wang}, \citenamefont {Zhang}, \citenamefont {Liu},
  \citenamefont {Zhang}, \citenamefont {Song}, \citenamefont {Yang},
  \citenamefont {Zhang},\ and\ \citenamefont {Ma}}]{zhong2016tellurium}%
  \BibitemOpen
  \bibfield  {author} {\bibinfo {author} {\bibfnamefont {X.}~\bibnamefont
  {Zhong}}, \bibinfo {author} {\bibfnamefont {H.}~\bibnamefont {Wang}},
  \bibinfo {author} {\bibfnamefont {J.}~\bibnamefont {Zhang}}, \bibinfo
  {author} {\bibfnamefont {H.}~\bibnamefont {Liu}}, \bibinfo {author}
  {\bibfnamefont {S.}~\bibnamefont {Zhang}}, \bibinfo {author} {\bibfnamefont
  {H.-F.}\ \bibnamefont {Song}}, \bibinfo {author} {\bibfnamefont
  {G.}~\bibnamefont {Yang}}, \bibinfo {author} {\bibfnamefont {L.}~\bibnamefont
  {Zhang}},\ and\ \bibinfo {author} {\bibfnamefont {Y.}~\bibnamefont {Ma}},\
  }\bibfield  {title} {\bibinfo {title} {Tellurium hydrides at high pressures:
  High-temperature superconductors},\ }\href@noop {} {\bibfield  {journal}
  {\bibinfo  {journal} {Physical review letters}\ }\textbf {\bibinfo {volume}
  {116}},\ \bibinfo {pages} {057002} (\bibinfo {year} {2016})}\BibitemShut
  {NoStop}%
\bibitem [{\citenamefont {Zhou}\ \emph {et~al.}(2012)\citenamefont {Zhou},
  \citenamefont {Jin}, \citenamefont {Meng}, \citenamefont {Bao}, \citenamefont
  {Ma}, \citenamefont {Liu},\ and\ \citenamefont {Cui}}]{zhou2012ab}%
  \BibitemOpen
  \bibfield  {author} {\bibinfo {author} {\bibfnamefont {D.}~\bibnamefont
  {Zhou}}, \bibinfo {author} {\bibfnamefont {X.}~\bibnamefont {Jin}}, \bibinfo
  {author} {\bibfnamefont {X.}~\bibnamefont {Meng}}, \bibinfo {author}
  {\bibfnamefont {G.}~\bibnamefont {Bao}}, \bibinfo {author} {\bibfnamefont
  {Y.}~\bibnamefont {Ma}}, \bibinfo {author} {\bibfnamefont {B.}~\bibnamefont
  {Liu}},\ and\ \bibinfo {author} {\bibfnamefont {T.}~\bibnamefont {Cui}},\
  }\bibfield  {title} {\bibinfo {title} {Ab initio study revealing a layered
  structure in hydrogen-rich kh 6 under high pressure},\ }\href@noop {}
  {\bibfield  {journal} {\bibinfo  {journal} {Physical Review B}\ }\textbf
  {\bibinfo {volume} {86}},\ \bibinfo {pages} {014118} (\bibinfo {year}
  {2012})}\BibitemShut {NoStop}%
\bibitem [{\citenamefont {Kruglov}\ \emph {et~al.}(2020)\citenamefont
  {Kruglov}, \citenamefont {Semenok}, \citenamefont {Song}, \citenamefont
  {Szczesniak}, \citenamefont {Wrona}, \citenamefont {Akashi}, \citenamefont
  {Esfahani}, \citenamefont {Duan}, \citenamefont {Cui}, \citenamefont
  {Kvashnin},\ and\ \citenamefont {Oganov}}]{kruglov2020}%
  \BibitemOpen
  \bibfield  {author} {\bibinfo {author} {\bibfnamefont {I.~A.}\ \bibnamefont
  {Kruglov}}, \bibinfo {author} {\bibfnamefont {D.~V.}\ \bibnamefont
  {Semenok}}, \bibinfo {author} {\bibfnamefont {H.}~\bibnamefont {Song}},
  \bibinfo {author} {\bibfnamefont {R.~L.}\ \bibnamefont {Szczesniak}},
  \bibinfo {author} {\bibfnamefont {I.~A.}\ \bibnamefont {Wrona}}, \bibinfo
  {author} {\bibfnamefont {R.}~\bibnamefont {Akashi}}, \bibinfo {author}
  {\bibfnamefont {M.~M.~D.}\ \bibnamefont {Esfahani}}, \bibinfo {author}
  {\bibfnamefont {D.}~\bibnamefont {Duan}}, \bibinfo {author} {\bibfnamefont
  {T.}~\bibnamefont {Cui}}, \bibinfo {author} {\bibfnamefont {A.~G.}\
  \bibnamefont {Kvashnin}},\ and\ \bibinfo {author} {\bibfnamefont {A.~R.}\
  \bibnamefont {Oganov}},\ }\bibfield  {title} {\bibinfo {title}
  {Superconductivity of lah 10 and lah 16 polyhydrides},\ }\href@noop {}
  {\bibfield  {journal} {\bibinfo  {journal} {Phys. Rev. B}\ }\textbf {\bibinfo
  {volume} {101}},\ \bibinfo {pages} {024508} (\bibinfo {year}
  {2020})}\BibitemShut {NoStop}%
\bibitem [{\citenamefont {Gao}\ \emph {et~al.}(2013)\citenamefont {Gao},
  \citenamefont {Hoffmann}, \citenamefont {Ashcroft}, \citenamefont {Liu},
  \citenamefont {Bergara},\ and\ \citenamefont {Ma}}]{gao2013theoretical}%
  \BibitemOpen
  \bibfield  {author} {\bibinfo {author} {\bibfnamefont {G.}~\bibnamefont
  {Gao}}, \bibinfo {author} {\bibfnamefont {R.}~\bibnamefont {Hoffmann}},
  \bibinfo {author} {\bibfnamefont {N.~W.}\ \bibnamefont {Ashcroft}}, \bibinfo
  {author} {\bibfnamefont {H.}~\bibnamefont {Liu}}, \bibinfo {author}
  {\bibfnamefont {A.}~\bibnamefont {Bergara}},\ and\ \bibinfo {author}
  {\bibfnamefont {Y.}~\bibnamefont {Ma}},\ }\bibfield  {title} {\bibinfo
  {title} {Theoretical study of the ground-state structures and properties of
  niobium hydrides under pressure},\ }\href@noop {} {\bibfield  {journal}
  {\bibinfo  {journal} {Physical Review B}\ }\textbf {\bibinfo {volume} {88}},\
  \bibinfo {pages} {184104} (\bibinfo {year} {2013})}\BibitemShut {NoStop}%
\bibitem [{\citenamefont {Zhou}\ \emph
  {et~al.}(2020{\natexlab{a}})\citenamefont {Zhou}, \citenamefont {Semenok},
  \citenamefont {Xie}, \citenamefont {Huang}, \citenamefont {Duan},
  \citenamefont {Aperis}, \citenamefont {Oppeneer}, \citenamefont {Galasso},
  \citenamefont {Kartsev}, \citenamefont {Kvashnin}, \citenamefont {Oganov},\
  and\ \citenamefont {Cui}}]{zhou2019high}%
  \BibitemOpen
  \bibfield  {author} {\bibinfo {author} {\bibfnamefont {D.}~\bibnamefont
  {Zhou}}, \bibinfo {author} {\bibfnamefont {D.~V.}\ \bibnamefont {Semenok}},
  \bibinfo {author} {\bibfnamefont {H.}~\bibnamefont {Xie}}, \bibinfo {author}
  {\bibfnamefont {X.}~\bibnamefont {Huang}}, \bibinfo {author} {\bibfnamefont
  {D.}~\bibnamefont {Duan}}, \bibinfo {author} {\bibfnamefont {A.}~\bibnamefont
  {Aperis}}, \bibinfo {author} {\bibfnamefont {P.~M.}\ \bibnamefont
  {Oppeneer}}, \bibinfo {author} {\bibfnamefont {M.}~\bibnamefont {Galasso}},
  \bibinfo {author} {\bibfnamefont {A.~I.}\ \bibnamefont {Kartsev}}, \bibinfo
  {author} {\bibfnamefont {A.~G.}\ \bibnamefont {Kvashnin}}, \bibinfo {author}
  {\bibfnamefont {A.~R.}\ \bibnamefont {Oganov}},\ and\ \bibinfo {author}
  {\bibfnamefont {T.}~\bibnamefont {Cui}},\ }\bibfield  {title} {\bibinfo
  {title} {High-pressure synthesis of magnetic neodymium polyhydrides},\ }\href
  {https://doi.org/10.1021/jacs.9b10439} {\bibfield  {journal} {\bibinfo
  {journal} {Journal of the American Chemical Society}\ }\textbf {\bibinfo
  {volume} {142}},\ \bibinfo {pages} {2803} (\bibinfo {year}
  {2020}{\natexlab{a}})}\BibitemShut {NoStop}%
\bibitem [{\citenamefont {Liu}\ \emph {et~al.}(2015{\natexlab{b}})\citenamefont
  {Liu}, \citenamefont {Duan}, \citenamefont {Huang}, \citenamefont {Tian},
  \citenamefont {Li}, \citenamefont {Sha}, \citenamefont {Wang}, \citenamefont
  {Zhang}, \citenamefont {Yang}, \citenamefont {Liu},\ and\ \citenamefont
  {Cui}}]{liu2015structures}%
  \BibitemOpen
  \bibfield  {author} {\bibinfo {author} {\bibfnamefont {Y.}~\bibnamefont
  {Liu}}, \bibinfo {author} {\bibfnamefont {D.}~\bibnamefont {Duan}}, \bibinfo
  {author} {\bibfnamefont {X.}~\bibnamefont {Huang}}, \bibinfo {author}
  {\bibfnamefont {F.}~\bibnamefont {Tian}}, \bibinfo {author} {\bibfnamefont
  {D.}~\bibnamefont {Li}}, \bibinfo {author} {\bibfnamefont {X.}~\bibnamefont
  {Sha}}, \bibinfo {author} {\bibfnamefont {C.}~\bibnamefont {Wang}}, \bibinfo
  {author} {\bibfnamefont {H.}~\bibnamefont {Zhang}}, \bibinfo {author}
  {\bibfnamefont {T.}~\bibnamefont {Yang}}, \bibinfo {author} {\bibfnamefont
  {B.}~\bibnamefont {Liu}},\ and\ \bibinfo {author} {\bibfnamefont
  {T.}~\bibnamefont {Cui}},\ }\bibfield  {title} {\bibinfo {title} {Structures
  and properties of osmium hydrides under pressure from first principle
  calculation},\ }\href@noop {} {\bibfield  {journal} {\bibinfo  {journal} {The
  Journal of Physical Chemistry C}\ }\textbf {\bibinfo {volume} {119}},\
  \bibinfo {pages} {15905} (\bibinfo {year} {2015}{\natexlab{b}})}\BibitemShut
  {NoStop}%
\bibitem [{\citenamefont {Zhou}\ \emph
  {et~al.}(2020{\natexlab{b}})\citenamefont {Zhou}, \citenamefont {Semenok},
  \citenamefont {Duan}, \citenamefont {Xie}, \citenamefont {Chen},
  \citenamefont {Huang}, \citenamefont {Li}, \citenamefont {Liu}, \citenamefont
  {Oganov},\ and\ \citenamefont {Cui}}]{zhou2019superconducting}%
  \BibitemOpen
  \bibfield  {author} {\bibinfo {author} {\bibfnamefont {D.}~\bibnamefont
  {Zhou}}, \bibinfo {author} {\bibfnamefont {D.~V.}\ \bibnamefont {Semenok}},
  \bibinfo {author} {\bibfnamefont {D.}~\bibnamefont {Duan}}, \bibinfo {author}
  {\bibfnamefont {H.}~\bibnamefont {Xie}}, \bibinfo {author} {\bibfnamefont
  {W.}~\bibnamefont {Chen}}, \bibinfo {author} {\bibfnamefont {X.}~\bibnamefont
  {Huang}}, \bibinfo {author} {\bibfnamefont {X.}~\bibnamefont {Li}}, \bibinfo
  {author} {\bibfnamefont {B.}~\bibnamefont {Liu}}, \bibinfo {author}
  {\bibfnamefont {A.~R.}\ \bibnamefont {Oganov}},\ and\ \bibinfo {author}
  {\bibfnamefont {T.}~\bibnamefont {Cui}},\ }\bibfield  {title} {\bibinfo
  {title} {Superconducting praseodymium superhydrides},\ }\href
  {https://advances.sciencemag.org/content/6/9/eaax6849} {\bibfield  {journal}
  {\bibinfo  {journal} {Science Advances}\ }\textbf {\bibinfo {volume} {6}}
  (\bibinfo {year} {2020}{\natexlab{b}})}\BibitemShut {NoStop}%
\bibitem [{\citenamefont {Liu}\ \emph {et~al.}(2016)\citenamefont {Liu},
  \citenamefont {Duan}, \citenamefont {Tian}, \citenamefont {Wang},
  \citenamefont {Ma}, \citenamefont {Li}, \citenamefont {Huang}, \citenamefont
  {Liu},\ and\ \citenamefont {Cui}}]{liu2016stability}%
  \BibitemOpen
  \bibfield  {author} {\bibinfo {author} {\bibfnamefont {Y.}~\bibnamefont
  {Liu}}, \bibinfo {author} {\bibfnamefont {D.}~\bibnamefont {Duan}}, \bibinfo
  {author} {\bibfnamefont {F.}~\bibnamefont {Tian}}, \bibinfo {author}
  {\bibfnamefont {C.}~\bibnamefont {Wang}}, \bibinfo {author} {\bibfnamefont
  {Y.}~\bibnamefont {Ma}}, \bibinfo {author} {\bibfnamefont {D.}~\bibnamefont
  {Li}}, \bibinfo {author} {\bibfnamefont {X.}~\bibnamefont {Huang}}, \bibinfo
  {author} {\bibfnamefont {B.}~\bibnamefont {Liu}},\ and\ \bibinfo {author}
  {\bibfnamefont {T.}~\bibnamefont {Cui}},\ }\bibfield  {title} {\bibinfo
  {title} {Stability and properties of the ru--h system at high pressure},\
  }\href@noop {} {\bibfield  {journal} {\bibinfo  {journal} {Physical Chemistry
  Chemical Physics}\ }\textbf {\bibinfo {volume} {18}},\ \bibinfo {pages}
  {1516} (\bibinfo {year} {2016})}\BibitemShut {NoStop}%
\bibitem [{\citenamefont {Ma}\ \emph {et~al.}(2015{\natexlab{b}})\citenamefont
  {Ma}, \citenamefont {Duan}, \citenamefont {Li}, \citenamefont {Liu},
  \citenamefont {Tian}, \citenamefont {Huang}, \citenamefont {Zhao},
  \citenamefont {Yu}, \citenamefont {Liu},\ and\ \citenamefont
  {Cui}}]{ma2015unexpected}%
  \BibitemOpen
  \bibfield  {author} {\bibinfo {author} {\bibfnamefont {Y.}~\bibnamefont
  {Ma}}, \bibinfo {author} {\bibfnamefont {D.}~\bibnamefont {Duan}}, \bibinfo
  {author} {\bibfnamefont {D.}~\bibnamefont {Li}}, \bibinfo {author}
  {\bibfnamefont {Y.}~\bibnamefont {Liu}}, \bibinfo {author} {\bibfnamefont
  {F.}~\bibnamefont {Tian}}, \bibinfo {author} {\bibfnamefont {X.}~\bibnamefont
  {Huang}}, \bibinfo {author} {\bibfnamefont {Z.}~\bibnamefont {Zhao}},
  \bibinfo {author} {\bibfnamefont {H.}~\bibnamefont {Yu}}, \bibinfo {author}
  {\bibfnamefont {B.}~\bibnamefont {Liu}},\ and\ \bibinfo {author}
  {\bibfnamefont {T.}~\bibnamefont {Cui}},\ }\bibfield  {title} {\bibinfo
  {title} {The unexpected binding and superconductivity in sbh4 at high
  pressure},\ }\href@noop {} {\bibfield  {journal} {\bibinfo  {journal} {arXiv
  preprint arXiv:1506.03889}\ } (\bibinfo {year}
  {2015}{\natexlab{b}})}\BibitemShut {NoStop}%
\bibitem [{\citenamefont {Ye}\ \emph {et~al.}(2018)\citenamefont {Ye},
  \citenamefont {Zarifi}, \citenamefont {Zurek}, \citenamefont {Hoffmann},\
  and\ \citenamefont {Ashcroft}}]{ye2018high}%
  \BibitemOpen
  \bibfield  {author} {\bibinfo {author} {\bibfnamefont {X.}~\bibnamefont
  {Ye}}, \bibinfo {author} {\bibfnamefont {N.}~\bibnamefont {Zarifi}}, \bibinfo
  {author} {\bibfnamefont {E.}~\bibnamefont {Zurek}}, \bibinfo {author}
  {\bibfnamefont {R.}~\bibnamefont {Hoffmann}},\ and\ \bibinfo {author}
  {\bibfnamefont {N.}~\bibnamefont {Ashcroft}},\ }\bibfield  {title} {\bibinfo
  {title} {High hydrides of scandium under pressure: potential
  superconductors},\ }\href@noop {} {\bibfield  {journal} {\bibinfo  {journal}
  {The Journal of Physical Chemistry C}\ }\textbf {\bibinfo {volume} {122}},\
  \bibinfo {pages} {6298} (\bibinfo {year} {2018})}\BibitemShut {NoStop}%
\bibitem [{\citenamefont {Jin}\ \emph {et~al.}(2010)\citenamefont {Jin},
  \citenamefont {Meng}, \citenamefont {He}, \citenamefont {Ma}, \citenamefont
  {Liu}, \citenamefont {Cui}, \citenamefont {Zou},\ and\ \citenamefont
  {Mao}}]{jin2010superconducting}%
  \BibitemOpen
  \bibfield  {author} {\bibinfo {author} {\bibfnamefont {X.}~\bibnamefont
  {Jin}}, \bibinfo {author} {\bibfnamefont {X.}~\bibnamefont {Meng}}, \bibinfo
  {author} {\bibfnamefont {Z.}~\bibnamefont {He}}, \bibinfo {author}
  {\bibfnamefont {Y.}~\bibnamefont {Ma}}, \bibinfo {author} {\bibfnamefont
  {B.}~\bibnamefont {Liu}}, \bibinfo {author} {\bibfnamefont {T.}~\bibnamefont
  {Cui}}, \bibinfo {author} {\bibfnamefont {G.}~\bibnamefont {Zou}},\ and\
  \bibinfo {author} {\bibfnamefont {H.-k.}\ \bibnamefont {Mao}},\ }\bibfield
  {title} {\bibinfo {title} {Superconducting high-pressure phases of
  disilane},\ }\href@noop {} {\bibfield  {journal} {\bibinfo  {journal}
  {Proceedings of the National Academy of Sciences}\ }\textbf {\bibinfo
  {volume} {107}},\ \bibinfo {pages} {9969} (\bibinfo {year}
  {2010})}\BibitemShut {NoStop}%
\bibitem [{\citenamefont {Esfahani}\ \emph {et~al.}(2016)\citenamefont
  {Esfahani}, \citenamefont {Wang}, \citenamefont {Oganov}, \citenamefont
  {Dong}, \citenamefont {Zhu}, \citenamefont {Wang}, \citenamefont {Rakitin},\
  and\ \citenamefont {Zhou}}]{esfahani2016superconductivity}%
  \BibitemOpen
  \bibfield  {author} {\bibinfo {author} {\bibfnamefont {M.~M.~D.}\
  \bibnamefont {Esfahani}}, \bibinfo {author} {\bibfnamefont {Z.}~\bibnamefont
  {Wang}}, \bibinfo {author} {\bibfnamefont {A.~R.}\ \bibnamefont {Oganov}},
  \bibinfo {author} {\bibfnamefont {H.}~\bibnamefont {Dong}}, \bibinfo {author}
  {\bibfnamefont {Q.}~\bibnamefont {Zhu}}, \bibinfo {author} {\bibfnamefont
  {S.}~\bibnamefont {Wang}}, \bibinfo {author} {\bibfnamefont {M.~S.}\
  \bibnamefont {Rakitin}},\ and\ \bibinfo {author} {\bibfnamefont {X.-F.}\
  \bibnamefont {Zhou}},\ }\bibfield  {title} {\bibinfo {title}
  {Superconductivity of novel tin hydrides (sn n h m) under pressure},\
  }\href@noop {} {\bibfield  {journal} {\bibinfo  {journal} {Scientific
  reports}\ }\textbf {\bibinfo {volume} {6}},\ \bibinfo {pages} {22873}
  (\bibinfo {year} {2016})}\BibitemShut {NoStop}%
\bibitem [{\citenamefont {Zhuang}\ \emph {et~al.}(2017)\citenamefont {Zhuang},
  \citenamefont {Jin}, \citenamefont {Cui}, \citenamefont {Ma}, \citenamefont
  {Lv}, \citenamefont {Li}, \citenamefont {Zhang}, \citenamefont {Meng},\ and\
  \citenamefont {Bao}}]{zhuang2017pressure}%
  \BibitemOpen
  \bibfield  {author} {\bibinfo {author} {\bibfnamefont {Q.}~\bibnamefont
  {Zhuang}}, \bibinfo {author} {\bibfnamefont {X.}~\bibnamefont {Jin}},
  \bibinfo {author} {\bibfnamefont {T.}~\bibnamefont {Cui}}, \bibinfo {author}
  {\bibfnamefont {Y.}~\bibnamefont {Ma}}, \bibinfo {author} {\bibfnamefont
  {Q.}~\bibnamefont {Lv}}, \bibinfo {author} {\bibfnamefont {Y.}~\bibnamefont
  {Li}}, \bibinfo {author} {\bibfnamefont {H.}~\bibnamefont {Zhang}}, \bibinfo
  {author} {\bibfnamefont {X.}~\bibnamefont {Meng}},\ and\ \bibinfo {author}
  {\bibfnamefont {K.}~\bibnamefont {Bao}},\ }\bibfield  {title} {\bibinfo
  {title} {Pressure-stabilized superconductive ionic tantalum hydrides},\
  }\href@noop {} {\bibfield  {journal} {\bibinfo  {journal} {Inorganic
  chemistry}\ }\textbf {\bibinfo {volume} {56}},\ \bibinfo {pages} {3901}
  (\bibinfo {year} {2017})}\BibitemShut {NoStop}%
\bibitem [{\citenamefont {Li}\ \emph {et~al.}(2016)\citenamefont {Li},
  \citenamefont {Liu},\ and\ \citenamefont {Peng}}]{li2016crystal}%
  \BibitemOpen
  \bibfield  {author} {\bibinfo {author} {\bibfnamefont {X.}~\bibnamefont
  {Li}}, \bibinfo {author} {\bibfnamefont {H.}~\bibnamefont {Liu}},\ and\
  \bibinfo {author} {\bibfnamefont {F.}~\bibnamefont {Peng}},\ }\bibfield
  {title} {\bibinfo {title} {Crystal structures and superconductivity of
  technetium hydrides under pressure},\ }\href@noop {} {\bibfield  {journal}
  {\bibinfo  {journal} {Physical Chemistry Chemical Physics}\ }\textbf
  {\bibinfo {volume} {18}},\ \bibinfo {pages} {28791} (\bibinfo {year}
  {2016})}\BibitemShut {NoStop}%
\bibitem [{\citenamefont {Kvashnin}\ \emph {et~al.}(2018)\citenamefont
  {Kvashnin}, \citenamefont {Semenok}, \citenamefont {Kruglov}, \citenamefont
  {Wrona},\ and\ \citenamefont {Oganov}}]{kvashnin2018high}%
  \BibitemOpen
  \bibfield  {author} {\bibinfo {author} {\bibfnamefont {A.~G.}\ \bibnamefont
  {Kvashnin}}, \bibinfo {author} {\bibfnamefont {D.~V.}\ \bibnamefont
  {Semenok}}, \bibinfo {author} {\bibfnamefont {I.~A.}\ \bibnamefont
  {Kruglov}}, \bibinfo {author} {\bibfnamefont {I.~A.}\ \bibnamefont {Wrona}},\
  and\ \bibinfo {author} {\bibfnamefont {A.~R.}\ \bibnamefont {Oganov}},\
  }\bibfield  {title} {\bibinfo {title} {High-temperature superconductivity in
  a th--h system under pressure conditions},\ }\href@noop {} {\bibfield
  {journal} {\bibinfo  {journal} {ACS applied materials \& interfaces}\
  }\textbf {\bibinfo {volume} {10}},\ \bibinfo {pages} {43809} (\bibinfo {year}
  {2018})}\BibitemShut {NoStop}%
\bibitem [{\citenamefont {Semenok}\ \emph
  {et~al.}(2020{\natexlab{a}})\citenamefont {Semenok}, \citenamefont
  {Kvashnin}, \citenamefont {Ivanova}, \citenamefont {Svitlyk}, \citenamefont
  {Fominski}, \citenamefont {Sadakov}, \citenamefont {Sobolevskiy},
  \citenamefont {Pudalov}, \citenamefont {Troyan},\ and\ \citenamefont
  {Oganov}}]{semenok2019superconductivity}%
  \BibitemOpen
  \bibfield  {author} {\bibinfo {author} {\bibfnamefont {D.~V.}\ \bibnamefont
  {Semenok}}, \bibinfo {author} {\bibfnamefont {A.~G.}\ \bibnamefont
  {Kvashnin}}, \bibinfo {author} {\bibfnamefont {A.~G.}\ \bibnamefont
  {Ivanova}}, \bibinfo {author} {\bibfnamefont {V.}~\bibnamefont {Svitlyk}},
  \bibinfo {author} {\bibfnamefont {V.~Y.}\ \bibnamefont {Fominski}}, \bibinfo
  {author} {\bibfnamefont {A.~V.}\ \bibnamefont {Sadakov}}, \bibinfo {author}
  {\bibfnamefont {O.~A.}\ \bibnamefont {Sobolevskiy}}, \bibinfo {author}
  {\bibfnamefont {V.~M.}\ \bibnamefont {Pudalov}}, \bibinfo {author}
  {\bibfnamefont {I.~A.}\ \bibnamefont {Troyan}},\ and\ \bibinfo {author}
  {\bibfnamefont {A.~R.}\ \bibnamefont {Oganov}},\ }\bibfield  {title}
  {\bibinfo {title} {Superconductivity at 161k in thorium hydride thh10:
  Synthesis and properties},\ }\href
  {https://doi.org/10.1016/j.mattod.2019.10.005} {\bibfield  {journal}
  {\bibinfo  {journal} {Materials Today}\ }\textbf {\bibinfo {volume} {33}},\
  \bibinfo {pages} {36} (\bibinfo {year} {2020}{\natexlab{a}})}\BibitemShut
  {NoStop}%
\bibitem [{\citenamefont {Kruglov}\ \emph {et~al.}(2018)\citenamefont
  {Kruglov}, \citenamefont {Kvashnin}, \citenamefont {Goncharov}, \citenamefont
  {Oganov}, \citenamefont {Lobanov}, \citenamefont {Holtgrewe}, \citenamefont
  {Jiang}, \citenamefont {Prakapenka}, \citenamefont {Greenberg},\ and\
  \citenamefont {Yanilkin}}]{kruglov2018uranium}%
  \BibitemOpen
  \bibfield  {author} {\bibinfo {author} {\bibfnamefont {I.~A.}\ \bibnamefont
  {Kruglov}}, \bibinfo {author} {\bibfnamefont {A.~G.}\ \bibnamefont
  {Kvashnin}}, \bibinfo {author} {\bibfnamefont {A.~F.}\ \bibnamefont
  {Goncharov}}, \bibinfo {author} {\bibfnamefont {A.~R.}\ \bibnamefont
  {Oganov}}, \bibinfo {author} {\bibfnamefont {S.~S.}\ \bibnamefont {Lobanov}},
  \bibinfo {author} {\bibfnamefont {N.}~\bibnamefont {Holtgrewe}}, \bibinfo
  {author} {\bibfnamefont {S.}~\bibnamefont {Jiang}}, \bibinfo {author}
  {\bibfnamefont {V.~B.}\ \bibnamefont {Prakapenka}}, \bibinfo {author}
  {\bibfnamefont {E.}~\bibnamefont {Greenberg}},\ and\ \bibinfo {author}
  {\bibfnamefont {A.~V.}\ \bibnamefont {Yanilkin}},\ }\bibfield  {title}
  {\bibinfo {title} {Uranium polyhydrides at moderate pressures: Prediction,
  synthesis, and expected superconductivity},\ }\href
  {https://advances.sciencemag.org/content/4/10/eaat9776} {\bibfield  {journal}
  {\bibinfo  {journal} {Science advances}\ }\textbf {\bibinfo {volume} {4}}
  (\bibinfo {year} {2018})}\BibitemShut {NoStop}%
\bibitem [{\citenamefont {Li}\ and\ \citenamefont
  {Peng}(2017)}]{li2017superconductivity}%
  \BibitemOpen
  \bibfield  {author} {\bibinfo {author} {\bibfnamefont {X.}~\bibnamefont
  {Li}}\ and\ \bibinfo {author} {\bibfnamefont {F.}~\bibnamefont {Peng}},\
  }\bibfield  {title} {\bibinfo {title} {Superconductivity of
  pressure-stabilized vanadium hydrides},\ }\href@noop {} {\bibfield  {journal}
  {\bibinfo  {journal} {Inorganic chemistry}\ }\textbf {\bibinfo {volume}
  {56}},\ \bibinfo {pages} {13759} (\bibinfo {year} {2017})}\BibitemShut
  {NoStop}%
\bibitem [{\citenamefont {Zheng}\ \emph {et~al.}(2018)\citenamefont {Zheng},
  \citenamefont {Zhang}, \citenamefont {Sun}, \citenamefont {Zhang},
  \citenamefont {Lin}, \citenamefont {Yang},\ and\ \citenamefont
  {Bergara}}]{zheng2018structural}%
  \BibitemOpen
  \bibfield  {author} {\bibinfo {author} {\bibfnamefont {S.}~\bibnamefont
  {Zheng}}, \bibinfo {author} {\bibfnamefont {S.}~\bibnamefont {Zhang}},
  \bibinfo {author} {\bibfnamefont {Y.}~\bibnamefont {Sun}}, \bibinfo {author}
  {\bibfnamefont {J.}~\bibnamefont {Zhang}}, \bibinfo {author} {\bibfnamefont
  {J.}~\bibnamefont {Lin}}, \bibinfo {author} {\bibfnamefont {G.}~\bibnamefont
  {Yang}},\ and\ \bibinfo {author} {\bibfnamefont {A.}~\bibnamefont
  {Bergara}},\ }\bibfield  {title} {\bibinfo {title} {Structural and
  superconducting properties of tungsten hydrides under high pressure},\
  }\href@noop {} {\bibfield  {journal} {\bibinfo  {journal} {Frontiers in
  Physics}\ }\textbf {\bibinfo {volume} {6}},\ \bibinfo {pages} {101} (\bibinfo
  {year} {2018})}\BibitemShut {NoStop}%
\bibitem [{\citenamefont {Liu}\ \emph {et~al.}(2017{\natexlab{a}})\citenamefont
  {Liu}, \citenamefont {Sun}, \citenamefont {Wang},\ and\ \citenamefont
  {Lu}}]{liu2017high}%
  \BibitemOpen
  \bibfield  {author} {\bibinfo {author} {\bibfnamefont {L.-L.}\ \bibnamefont
  {Liu}}, \bibinfo {author} {\bibfnamefont {H.-J.}\ \bibnamefont {Sun}},
  \bibinfo {author} {\bibfnamefont {C.}~\bibnamefont {Wang}},\ and\ \bibinfo
  {author} {\bibfnamefont {W.-C.}\ \bibnamefont {Lu}},\ }\bibfield  {title}
  {\bibinfo {title} {High-pressure structures of yttrium hydrides},\
  }\href@noop {} {\bibfield  {journal} {\bibinfo  {journal} {Journal of
  Physics: Condensed Matter}\ }\textbf {\bibinfo {volume} {29}},\ \bibinfo
  {pages} {325401} (\bibinfo {year} {2017}{\natexlab{a}})}\BibitemShut
  {NoStop}%
\bibitem [{\citenamefont {Li}\ \emph {et~al.}(2017)\citenamefont {Li},
  \citenamefont {Hu},\ and\ \citenamefont {Huang}}]{li2017phase}%
  \BibitemOpen
  \bibfield  {author} {\bibinfo {author} {\bibfnamefont {X.-F.}\ \bibnamefont
  {Li}}, \bibinfo {author} {\bibfnamefont {Z.-Y.}\ \bibnamefont {Hu}},\ and\
  \bibinfo {author} {\bibfnamefont {B.}~\bibnamefont {Huang}},\ }\bibfield
  {title} {\bibinfo {title} {Phase diagram and superconductivity of compressed
  zirconium hydrides},\ }\href@noop {} {\bibfield  {journal} {\bibinfo
  {journal} {Physical Chemistry Chemical Physics}\ }\textbf {\bibinfo {volume}
  {19}},\ \bibinfo {pages} {3538} (\bibinfo {year} {2017})}\BibitemShut
  {NoStop}%
\bibitem [{\citenamefont {Abe}(2018)}]{abe2018high}%
  \BibitemOpen
  \bibfield  {author} {\bibinfo {author} {\bibfnamefont {K.}~\bibnamefont
  {Abe}},\ }\bibfield  {title} {\bibinfo {title} {High-pressure properties of
  dense metallic zirconium hydrides studied by ab initio calculations},\
  }\href@noop {} {\bibfield  {journal} {\bibinfo  {journal} {Physical Review
  B}\ }\textbf {\bibinfo {volume} {98}},\ \bibinfo {pages} {134103} (\bibinfo
  {year} {2018})}\BibitemShut {NoStop}%
\bibitem [{\citenamefont {Semenok}\ \emph
  {et~al.}(2020{\natexlab{b}})\citenamefont {Semenok}, \citenamefont {Kruglov},
  \citenamefont {Savkin}, \citenamefont {Kvashnin},\ and\ \citenamefont
  {Oganov}}]{semenok2018distribution}%
  \BibitemOpen
  \bibfield  {author} {\bibinfo {author} {\bibfnamefont {D.~V.}\ \bibnamefont
  {Semenok}}, \bibinfo {author} {\bibfnamefont {I.~A.}\ \bibnamefont
  {Kruglov}}, \bibinfo {author} {\bibfnamefont {I.~A.}\ \bibnamefont {Savkin}},
  \bibinfo {author} {\bibfnamefont {A.~G.}\ \bibnamefont {Kvashnin}},\ and\
  \bibinfo {author} {\bibfnamefont {A.~R.}\ \bibnamefont {Oganov}},\ }\bibfield
   {title} {\bibinfo {title} {On distribution of superconductivity in metal
  hydrides},\ }\href {https://doi.org/10.1016/j.cossms.2020.100808} {\bibfield
  {journal} {\bibinfo  {journal} {Current Opinion in Solid State and Materials
  Science}\ ,\ \bibinfo {pages} {100808}} (\bibinfo {year}
  {2020}{\natexlab{b}})}\BibitemShut {NoStop}%
\bibitem [{\citenamefont {Gu}\ \emph {et~al.}(2017)\citenamefont {Gu},
  \citenamefont {Lu}, \citenamefont {Xia}, \citenamefont {Sun},\ and\
  \citenamefont {Xing}}]{gu2017high}%
  \BibitemOpen
  \bibfield  {author} {\bibinfo {author} {\bibfnamefont {Q.}~\bibnamefont
  {Gu}}, \bibinfo {author} {\bibfnamefont {P.}~\bibnamefont {Lu}}, \bibinfo
  {author} {\bibfnamefont {K.}~\bibnamefont {Xia}}, \bibinfo {author}
  {\bibfnamefont {J.}~\bibnamefont {Sun}},\ and\ \bibinfo {author}
  {\bibfnamefont {D.}~\bibnamefont {Xing}},\ }\bibfield  {title} {\bibinfo
  {title} {High-temperature superconducting phase of hbr under pressure
  predicted by first-principles calculations},\ }\href@noop {} {\bibfield
  {journal} {\bibinfo  {journal} {Physical Review B}\ }\textbf {\bibinfo
  {volume} {96}},\ \bibinfo {pages} {064517} (\bibinfo {year}
  {2017})}\BibitemShut {NoStop}%
\bibitem [{\citenamefont {Bi}\ \emph {et~al.}(2019)\citenamefont {Bi},
  \citenamefont {Zarifi}, \citenamefont {Terpstra},\ and\ \citenamefont
  {Zurek}}]{bi2018search}%
  \BibitemOpen
  \bibfield  {author} {\bibinfo {author} {\bibfnamefont {T.}~\bibnamefont
  {Bi}}, \bibinfo {author} {\bibfnamefont {N.}~\bibnamefont {Zarifi}}, \bibinfo
  {author} {\bibfnamefont {T.}~\bibnamefont {Terpstra}},\ and\ \bibinfo
  {author} {\bibfnamefont {E.}~\bibnamefont {Zurek}},\ }\bibfield  {title}
  {\bibinfo {title} {The search for superconductivity in high pressure
  hydrides},\ }in\ \href
  {https://www.sciencedirect.com/science/article/pii/B9780124095472114350}
  {\emph {\bibinfo {booktitle} {Reference Module in Chemistry, Molecular
  Sciences and Chemical Engineering}}}\ (\bibinfo  {publisher} {Elsevier},\
  \bibinfo {year} {2019})\BibitemShut {NoStop}%
\bibitem [{mat()}]{matlab_GPR}%
  \BibitemOpen
  \href@noop {} {}\bibinfo {howpublished}
  {https://uk.mathworks.com/help/stats/fitrgp.html},\ \bibinfo {note}
  {\textsc{matlab} GPR fitting documentation}\BibitemShut {NoStop}%
\bibitem [{\citenamefont {Drozdov}\ \emph {et~al.}(2019)\citenamefont
  {Drozdov}, \citenamefont {Kong}, \citenamefont {Minkov}, \citenamefont
  {Besedin}, \citenamefont {Kuzovnikov}, \citenamefont {Mozaffari},
  \citenamefont {Balicas}, \citenamefont {Balakirev}, \citenamefont {Graf},
  \citenamefont {Prakapenka}, \citenamefont {Greenberg}, \citenamefont
  {Knyazev}, \citenamefont {Tkacz},\ and\ \citenamefont
  {Eremets}}]{drozdov2019superconductivity}%
  \BibitemOpen
  \bibfield  {author} {\bibinfo {author} {\bibfnamefont {A.}~\bibnamefont
  {Drozdov}}, \bibinfo {author} {\bibfnamefont {P.}~\bibnamefont {Kong}},
  \bibinfo {author} {\bibfnamefont {V.}~\bibnamefont {Minkov}}, \bibinfo
  {author} {\bibfnamefont {S.}~\bibnamefont {Besedin}}, \bibinfo {author}
  {\bibfnamefont {M.}~\bibnamefont {Kuzovnikov}}, \bibinfo {author}
  {\bibfnamefont {S.}~\bibnamefont {Mozaffari}}, \bibinfo {author}
  {\bibfnamefont {L.}~\bibnamefont {Balicas}}, \bibinfo {author} {\bibfnamefont
  {F.}~\bibnamefont {Balakirev}}, \bibinfo {author} {\bibfnamefont
  {D.}~\bibnamefont {Graf}}, \bibinfo {author} {\bibfnamefont {V.}~\bibnamefont
  {Prakapenka}}, \bibinfo {author} {\bibfnamefont {E.}~\bibnamefont
  {Greenberg}}, \bibinfo {author} {\bibfnamefont {D.}~\bibnamefont {Knyazev}},
  \bibinfo {author} {\bibfnamefont {M.}~\bibnamefont {Tkacz}},\ and\ \bibinfo
  {author} {\bibfnamefont {M.}~\bibnamefont {Eremets}},\ }\bibfield  {title}
  {\bibinfo {title} {Superconductivity at 250 k in lanthanum hydride under high
  pressures},\ }\href@noop {} {\bibfield  {journal} {\bibinfo  {journal}
  {Nature}\ }\textbf {\bibinfo {volume} {569}},\ \bibinfo {pages} {528}
  (\bibinfo {year} {2019})}\BibitemShut {NoStop}%
\bibitem [{\citenamefont {Drozdov}\ \emph {et~al.}(2015)\citenamefont
  {Drozdov}, \citenamefont {Eremets}, \citenamefont {Troyan}, \citenamefont
  {Ksenofontov},\ and\ \citenamefont {Shylin}}]{drozdov2015conventional}%
  \BibitemOpen
  \bibfield  {author} {\bibinfo {author} {\bibfnamefont {A.}~\bibnamefont
  {Drozdov}}, \bibinfo {author} {\bibfnamefont {M.}~\bibnamefont {Eremets}},
  \bibinfo {author} {\bibfnamefont {I.}~\bibnamefont {Troyan}}, \bibinfo
  {author} {\bibfnamefont {V.}~\bibnamefont {Ksenofontov}},\ and\ \bibinfo
  {author} {\bibfnamefont {S.}~\bibnamefont {Shylin}},\ }\bibfield  {title}
  {\bibinfo {title} {Conventional superconductivity at 203 kelvin at high
  pressures in the sulfur hydride system},\ }\href@noop {} {\bibfield
  {journal} {\bibinfo  {journal} {Nature}\ }\textbf {\bibinfo {volume} {525}},\
  \bibinfo {pages} {73} (\bibinfo {year} {2015})}\BibitemShut {NoStop}%
\bibitem [{\citenamefont {Peng}\ \emph {et~al.}(2017)\citenamefont {Peng},
  \citenamefont {Sun}, \citenamefont {Pickard}, \citenamefont {Needs},
  \citenamefont {Wu},\ and\ \citenamefont {Ma}}]{peng2017hydrogen}%
  \BibitemOpen
  \bibfield  {author} {\bibinfo {author} {\bibfnamefont {F.}~\bibnamefont
  {Peng}}, \bibinfo {author} {\bibfnamefont {Y.}~\bibnamefont {Sun}}, \bibinfo
  {author} {\bibfnamefont {C.~J.}\ \bibnamefont {Pickard}}, \bibinfo {author}
  {\bibfnamefont {R.~J.}\ \bibnamefont {Needs}}, \bibinfo {author}
  {\bibfnamefont {Q.}~\bibnamefont {Wu}},\ and\ \bibinfo {author}
  {\bibfnamefont {Y.}~\bibnamefont {Ma}},\ }\bibfield  {title} {\bibinfo
  {title} {Hydrogen clathrate structures in rare earth hydrides at high
  pressures: possible route to room-temperature superconductivity},\
  }\href@noop {} {\bibfield  {journal} {\bibinfo  {journal} {Physical review
  letters}\ }\textbf {\bibinfo {volume} {119}},\ \bibinfo {pages} {107001}
  (\bibinfo {year} {2017})}\BibitemShut {NoStop}%
\bibitem [{\citenamefont {Xie}\ \emph {et~al.}(2014)\citenamefont {Xie},
  \citenamefont {Li}, \citenamefont {Oganov},\ and\ \citenamefont
  {Wang}}]{xie2014lithium}%
  \BibitemOpen
  \bibfield  {author} {\bibinfo {author} {\bibfnamefont {Y.}~\bibnamefont
  {Xie}}, \bibinfo {author} {\bibfnamefont {Q.}~\bibnamefont {Li}}, \bibinfo
  {author} {\bibfnamefont {A.}~\bibnamefont {Oganov}},\ and\ \bibinfo {author}
  {\bibfnamefont {H.}~\bibnamefont {Wang}},\ }\bibfield  {title} {\bibinfo
  {title} {Superconductivity of lithium-doped hydrogen under high pressure},\
  }\href@noop {} {\bibfield  {journal} {\bibinfo  {journal} {Acta
  Crystallographica Section C: Structural Chemistry}\ }\textbf {\bibinfo
  {volume} {70}},\ \bibinfo {pages} {104} (\bibinfo {year} {2014})}\BibitemShut
  {NoStop}%
\bibitem [{\citenamefont {Howie}\ \emph {et~al.}(2012)\citenamefont {Howie},
  \citenamefont {Narygina}, \citenamefont {Guillaume}, \citenamefont {Evans},\
  and\ \citenamefont {Gregoryanz}}]{howie2012high}%
  \BibitemOpen
  \bibfield  {author} {\bibinfo {author} {\bibfnamefont {R.~T.}\ \bibnamefont
  {Howie}}, \bibinfo {author} {\bibfnamefont {O.}~\bibnamefont {Narygina}},
  \bibinfo {author} {\bibfnamefont {C.~L.}\ \bibnamefont {Guillaume}}, \bibinfo
  {author} {\bibfnamefont {S.}~\bibnamefont {Evans}},\ and\ \bibinfo {author}
  {\bibfnamefont {E.}~\bibnamefont {Gregoryanz}},\ }\bibfield  {title}
  {\bibinfo {title} {High-pressure synthesis of lithium hydride},\ }\href@noop
  {} {\bibfield  {journal} {\bibinfo  {journal} {Physical Review B}\ }\textbf
  {\bibinfo {volume} {86}},\ \bibinfo {pages} {064108} (\bibinfo {year}
  {2012})}\BibitemShut {NoStop}%
\bibitem [{\citenamefont {Liu}\ \emph {et~al.}(2017{\natexlab{b}})\citenamefont
  {Liu}, \citenamefont {Naumov}, \citenamefont {Hoffmann}, \citenamefont
  {Ashcroft},\ and\ \citenamefont {Hemley}}]{liu2017potential}%
  \BibitemOpen
  \bibfield  {author} {\bibinfo {author} {\bibfnamefont {H.}~\bibnamefont
  {Liu}}, \bibinfo {author} {\bibfnamefont {I.~I.}\ \bibnamefont {Naumov}},
  \bibinfo {author} {\bibfnamefont {R.}~\bibnamefont {Hoffmann}}, \bibinfo
  {author} {\bibfnamefont {N.}~\bibnamefont {Ashcroft}},\ and\ \bibinfo
  {author} {\bibfnamefont {R.~J.}\ \bibnamefont {Hemley}},\ }\bibfield  {title}
  {\bibinfo {title} {Potential high-tc superconducting lanthanum and yttrium
  hydrides at high pressure},\ }\href@noop {} {\bibfield  {journal} {\bibinfo
  {journal} {Proceedings of the National Academy of Sciences}\ }\textbf
  {\bibinfo {volume} {114}},\ \bibinfo {pages} {6990} (\bibinfo {year}
  {2017}{\natexlab{b}})}\BibitemShut {NoStop}%
\bibitem [{\citenamefont {Somayazulu}\ \emph {et~al.}(2019)\citenamefont
  {Somayazulu}, \citenamefont {Ahart}, \citenamefont {Mishra}, \citenamefont
  {Geballe}, \citenamefont {Baldini}, \citenamefont {Meng}, \citenamefont
  {Struzhkin},\ and\ \citenamefont {Hemley}}]{somayazulu2019evidence}%
  \BibitemOpen
  \bibfield  {author} {\bibinfo {author} {\bibfnamefont {M.}~\bibnamefont
  {Somayazulu}}, \bibinfo {author} {\bibfnamefont {M.}~\bibnamefont {Ahart}},
  \bibinfo {author} {\bibfnamefont {A.~K.}\ \bibnamefont {Mishra}}, \bibinfo
  {author} {\bibfnamefont {Z.~M.}\ \bibnamefont {Geballe}}, \bibinfo {author}
  {\bibfnamefont {M.}~\bibnamefont {Baldini}}, \bibinfo {author} {\bibfnamefont
  {Y.}~\bibnamefont {Meng}}, \bibinfo {author} {\bibfnamefont {V.~V.}\
  \bibnamefont {Struzhkin}},\ and\ \bibinfo {author} {\bibfnamefont {R.~J.}\
  \bibnamefont {Hemley}},\ }\bibfield  {title} {\bibinfo {title} {Evidence for
  superconductivity above 260 k in lanthanum superhydride at megabar
  pressures},\ }\href@noop {} {\bibfield  {journal} {\bibinfo  {journal}
  {Physical review letters}\ }\textbf {\bibinfo {volume} {122}},\ \bibinfo
  {pages} {027001} (\bibinfo {year} {2019})}\BibitemShut {NoStop}%
\bibitem [{\citenamefont {Abe}(2017)}]{abe2017hydrogen}%
  \BibitemOpen
  \bibfield  {author} {\bibinfo {author} {\bibfnamefont {K.}~\bibnamefont
  {Abe}},\ }\bibfield  {title} {\bibinfo {title} {Hydrogen-rich scandium
  compounds at high pressures},\ }\href@noop {} {\bibfield  {journal} {\bibinfo
   {journal} {Physical Review B}\ }\textbf {\bibinfo {volume} {96}},\ \bibinfo
  {pages} {144108} (\bibinfo {year} {2017})}\BibitemShut {NoStop}%
\bibitem [{\citenamefont {Tanaka}\ \emph {et~al.}(2017)\citenamefont {Tanaka},
  \citenamefont {Tse},\ and\ \citenamefont {Liu}}]{tanaka2017electron}%
  \BibitemOpen
  \bibfield  {author} {\bibinfo {author} {\bibfnamefont {K.}~\bibnamefont
  {Tanaka}}, \bibinfo {author} {\bibfnamefont {J.}~\bibnamefont {Tse}},\ and\
  \bibinfo {author} {\bibfnamefont {H.}~\bibnamefont {Liu}},\ }\bibfield
  {title} {\bibinfo {title} {Electron-phonon coupling mechanisms for
  hydrogen-rich metals at high pressure},\ }\href@noop {} {\bibfield  {journal}
  {\bibinfo  {journal} {Physical Review B}\ }\textbf {\bibinfo {volume} {96}},\
  \bibinfo {pages} {100502} (\bibinfo {year} {2017})}\BibitemShut {NoStop}%
\bibitem [{\citenamefont {Feng}\ \emph {et~al.}(2015)\citenamefont {Feng},
  \citenamefont {Zhang}, \citenamefont {Gao}, \citenamefont {Liu},\ and\
  \citenamefont {Wang}}]{feng2015compressed}%
  \BibitemOpen
  \bibfield  {author} {\bibinfo {author} {\bibfnamefont {X.}~\bibnamefont
  {Feng}}, \bibinfo {author} {\bibfnamefont {J.}~\bibnamefont {Zhang}},
  \bibinfo {author} {\bibfnamefont {G.}~\bibnamefont {Gao}}, \bibinfo {author}
  {\bibfnamefont {H.}~\bibnamefont {Liu}},\ and\ \bibinfo {author}
  {\bibfnamefont {H.}~\bibnamefont {Wang}},\ }\bibfield  {title} {\bibinfo
  {title} {Compressed sodalite-like mgh6 as a potential high-temperature
  superconductor},\ }\href@noop {} {\bibfield  {journal} {\bibinfo  {journal}
  {RSC Advances}\ }\textbf {\bibinfo {volume} {5}},\ \bibinfo {pages} {59292}
  (\bibinfo {year} {2015})}\BibitemShut {NoStop}%
\bibitem [{\citenamefont {Lonie}\ \emph {et~al.}(2013)\citenamefont {Lonie},
  \citenamefont {Hooper}, \citenamefont {Altintas},\ and\ \citenamefont
  {Zurek}}]{lonie2013metallization}%
  \BibitemOpen
  \bibfield  {author} {\bibinfo {author} {\bibfnamefont {D.~C.}\ \bibnamefont
  {Lonie}}, \bibinfo {author} {\bibfnamefont {J.}~\bibnamefont {Hooper}},
  \bibinfo {author} {\bibfnamefont {B.}~\bibnamefont {Altintas}},\ and\
  \bibinfo {author} {\bibfnamefont {E.}~\bibnamefont {Zurek}},\ }\bibfield
  {title} {\bibinfo {title} {Metallization of magnesium polyhydrides under
  pressure},\ }\href@noop {} {\bibfield  {journal} {\bibinfo  {journal}
  {Physical Review B}\ }\textbf {\bibinfo {volume} {87}},\ \bibinfo {pages}
  {054107} (\bibinfo {year} {2013})}\BibitemShut {NoStop}%
\bibitem [{\citenamefont {Kong}\ \emph {et~al.}(2019)\citenamefont {Kong},
  \citenamefont {Minkov}, \citenamefont {Kuzovnikov}, \citenamefont {Besedin},
  \citenamefont {Drozdov}, \citenamefont {Mozaffari}, \citenamefont {Balicas},
  \citenamefont {Balakirev}, \citenamefont {Prakapenka}, \citenamefont
  {Greenberg}, \citenamefont {Knyazev},\ and\ \citenamefont
  {MI}}]{kong2019superconductivity}%
  \BibitemOpen
  \bibfield  {author} {\bibinfo {author} {\bibfnamefont {P.}~\bibnamefont
  {Kong}}, \bibinfo {author} {\bibfnamefont {V.}~\bibnamefont {Minkov}},
  \bibinfo {author} {\bibfnamefont {M.}~\bibnamefont {Kuzovnikov}}, \bibinfo
  {author} {\bibfnamefont {S.}~\bibnamefont {Besedin}}, \bibinfo {author}
  {\bibfnamefont {A.}~\bibnamefont {Drozdov}}, \bibinfo {author} {\bibfnamefont
  {S.}~\bibnamefont {Mozaffari}}, \bibinfo {author} {\bibfnamefont
  {L.}~\bibnamefont {Balicas}}, \bibinfo {author} {\bibfnamefont
  {F.}~\bibnamefont {Balakirev}}, \bibinfo {author} {\bibfnamefont
  {V.}~\bibnamefont {Prakapenka}}, \bibinfo {author} {\bibfnamefont
  {E.}~\bibnamefont {Greenberg}}, \bibinfo {author} {\bibfnamefont
  {D.}~\bibnamefont {Knyazev}},\ and\ \bibinfo {author} {\bibfnamefont
  {E.}~\bibnamefont {MI}},\ }\bibfield  {title} {\bibinfo {title}
  {Superconductivity up to 243 k in yttrium hydrides under high pressure},\
  }\href@noop {} {\bibfield  {journal} {\bibinfo  {journal} {arXiv preprint
  arXiv:1909.10482}\ } (\bibinfo {year} {2019})}\BibitemShut {NoStop}%
\bibitem [{\citenamefont {Nagamatsu}\ \emph {et~al.}(2001)\citenamefont
  {Nagamatsu}, \citenamefont {Nakagawa}, \citenamefont {Muranaka},
  \citenamefont {Zenitani},\ and\ \citenamefont {Akimitsu}}]{Nagamatsu2001}%
  \BibitemOpen
  \bibfield  {author} {\bibinfo {author} {\bibfnamefont {J.}~\bibnamefont
  {Nagamatsu}}, \bibinfo {author} {\bibfnamefont {N.}~\bibnamefont {Nakagawa}},
  \bibinfo {author} {\bibfnamefont {T.}~\bibnamefont {Muranaka}}, \bibinfo
  {author} {\bibfnamefont {Y.}~\bibnamefont {Zenitani}},\ and\ \bibinfo
  {author} {\bibfnamefont {J.}~\bibnamefont {Akimitsu}},\ }\bibfield  {title}
  {\bibinfo {title} {Superconductivity at 39{\thinspace}k in magnesium
  diboride},\ }\href {https://doi.org/10.1038/35065039} {\bibfield  {journal}
  {\bibinfo  {journal} {Nature}\ }\textbf {\bibinfo {volume} {410}},\ \bibinfo
  {pages} {63} (\bibinfo {year} {2001})}\BibitemShut {NoStop}%
\bibitem [{\citenamefont {Wu}\ \emph {et~al.}(2009)\citenamefont {Wu},
  \citenamefont {Xie}, \citenamefont {Chen}, \citenamefont {Zhong},
  \citenamefont {Liu}, \citenamefont {Shi}, \citenamefont {Li}, \citenamefont
  {Wang}, \citenamefont {Wu}, \citenamefont {Yan}, \citenamefont {Ying},\ and\
  \citenamefont {Chen}}]{Wu_2009}%
  \BibitemOpen
  \bibfield  {author} {\bibinfo {author} {\bibfnamefont {G.}~\bibnamefont
  {Wu}}, \bibinfo {author} {\bibfnamefont {Y.~L.}\ \bibnamefont {Xie}},
  \bibinfo {author} {\bibfnamefont {H.}~\bibnamefont {Chen}}, \bibinfo {author}
  {\bibfnamefont {M.}~\bibnamefont {Zhong}}, \bibinfo {author} {\bibfnamefont
  {R.~H.}\ \bibnamefont {Liu}}, \bibinfo {author} {\bibfnamefont {B.~C.}\
  \bibnamefont {Shi}}, \bibinfo {author} {\bibfnamefont {Q.~J.}\ \bibnamefont
  {Li}}, \bibinfo {author} {\bibfnamefont {X.~F.}\ \bibnamefont {Wang}},
  \bibinfo {author} {\bibfnamefont {T.}~\bibnamefont {Wu}}, \bibinfo {author}
  {\bibfnamefont {Y.~J.}\ \bibnamefont {Yan}}, \bibinfo {author} {\bibfnamefont
  {J.~J.}\ \bibnamefont {Ying}},\ and\ \bibinfo {author} {\bibfnamefont
  {X.~H.}\ \bibnamefont {Chen}},\ }\bibfield  {title} {\bibinfo {title}
  {Superconductivity at 56 k in samarium-doped {SrFeAsF}},\ }\href
  {https://doi.org/10.1088/0953-8984/21/14/142203} {\bibfield  {journal}
  {\bibinfo  {journal} {Journal of Physics: Condensed Matter}\ }\textbf
  {\bibinfo {volume} {21}},\ \bibinfo {pages} {142203} (\bibinfo {year}
  {2009})}\BibitemShut {NoStop}%
\bibitem [{\citenamefont {Schilling}\ \emph {et~al.}(1993)\citenamefont
  {Schilling}, \citenamefont {Cantoni}, \citenamefont {Guo},\ and\
  \citenamefont {Ott}}]{Schilling1993}%
  \BibitemOpen
  \bibfield  {author} {\bibinfo {author} {\bibfnamefont {A.}~\bibnamefont
  {Schilling}}, \bibinfo {author} {\bibfnamefont {M.}~\bibnamefont {Cantoni}},
  \bibinfo {author} {\bibfnamefont {J.~D.}\ \bibnamefont {Guo}},\ and\ \bibinfo
  {author} {\bibfnamefont {H.~R.}\ \bibnamefont {Ott}},\ }\bibfield  {title}
  {\bibinfo {title} {Superconductivity above 130 k in the hg--ba--ca--cu--o
  system},\ }\href {https://doi.org/10.1038/363056a0} {\bibfield  {journal}
  {\bibinfo  {journal} {Nature}\ }\textbf {\bibinfo {volume} {363}},\ \bibinfo
  {pages} {56} (\bibinfo {year} {1993})}\BibitemShut {NoStop}%
\bibitem [{\citenamefont {Snider}\ \emph {et~al.}(2020)\citenamefont {Snider},
  \citenamefont {Dasenbrock-Gammon}, \citenamefont {McBride}, \citenamefont
  {Debessai}, \citenamefont {Vindana}, \citenamefont {Vencatasamy},
  \citenamefont {Lawler}, \citenamefont {Salamat},\ and\ \citenamefont
  {Dias}}]{Snider2020}%
  \BibitemOpen
  \bibfield  {author} {\bibinfo {author} {\bibfnamefont {E.}~\bibnamefont
  {Snider}}, \bibinfo {author} {\bibfnamefont {N.}~\bibnamefont
  {Dasenbrock-Gammon}}, \bibinfo {author} {\bibfnamefont {R.}~\bibnamefont
  {McBride}}, \bibinfo {author} {\bibfnamefont {M.}~\bibnamefont {Debessai}},
  \bibinfo {author} {\bibfnamefont {H.}~\bibnamefont {Vindana}}, \bibinfo
  {author} {\bibfnamefont {K.}~\bibnamefont {Vencatasamy}}, \bibinfo {author}
  {\bibfnamefont {K.~V.}\ \bibnamefont {Lawler}}, \bibinfo {author}
  {\bibfnamefont {A.}~\bibnamefont {Salamat}},\ and\ \bibinfo {author}
  {\bibfnamefont {R.~P.}\ \bibnamefont {Dias}},\ }\bibfield  {title} {\bibinfo
  {title} {Room-temperature superconductivity in a carbonaceous sulfur
  hydride},\ }\href {https://doi.org/10.1038/s41586-020-2801-z} {\bibfield
  {journal} {\bibinfo  {journal} {Nature}\ }\textbf {\bibinfo {volume} {586}},\
  \bibinfo {pages} {373} (\bibinfo {year} {2020})}\BibitemShut {NoStop}%
\bibitem [{sup()}]{supplement}%
  \BibitemOpen
  \href@noop {} {\bibinfo {title} {See supplementary information, which
  includes refs.\ \cite{xie2019high, ying2019ternary,
  szczesniak2016superconductivity, chen2017prediction, Camargo_Mart_nez_2020,
  wei2016pressure, smith2009high, shao2019unique, PhysRevLett.125.217001,
  hooper2014composition}.}}\BibitemShut {Stop}%
\bibitem [{tab()}]{table_1_structures}%
  \BibitemOpen
  \href {https://doi.org/10.17863/CAM.72574} {\bibinfo {title} {Structures from
  table 1 (cambridge online data repository)}},\ \bibinfo {note}
  {\url{https://doi.org/10.17863/CAM.72574}}\BibitemShut {NoStop}%
\bibitem [{\citenamefont {Baettig}\ and\ \citenamefont
  {Zurek}(2011)}]{baettig2011}%
  \BibitemOpen
  \bibfield  {author} {\bibinfo {author} {\bibfnamefont {P.}~\bibnamefont
  {Baettig}}\ and\ \bibinfo {author} {\bibfnamefont {E.}~\bibnamefont
  {Zurek}},\ }\bibfield  {title} {\bibinfo {title} {Pressure-stabilized sodium
  polyhydrides: ${\mathrm{nah}}_{n}$ ($n>1$)},\ }\href
  {https://doi.org/10.1103/PhysRevLett.106.237002} {\bibfield  {journal}
  {\bibinfo  {journal} {Phys. Rev. Lett.}\ }\textbf {\bibinfo {volume} {106}},\
  \bibinfo {pages} {237002} (\bibinfo {year} {2011})}\BibitemShut {NoStop}%
\bibitem [{\citenamefont {Struzhkin}\ \emph {et~al.}(2016)\citenamefont
  {Struzhkin}, \citenamefont {Kim}, \citenamefont {Stavrou}, \citenamefont
  {Muramatsu}, \citenamefont {Mao}, \citenamefont {Pickard}, \citenamefont
  {Needs}, \citenamefont {Prakapenka},\ and\ \citenamefont
  {Goncharov}}]{struzhkin2016}%
  \BibitemOpen
  \bibfield  {author} {\bibinfo {author} {\bibfnamefont {V.~V.}\ \bibnamefont
  {Struzhkin}}, \bibinfo {author} {\bibfnamefont {D.~Y.}\ \bibnamefont {Kim}},
  \bibinfo {author} {\bibfnamefont {E.}~\bibnamefont {Stavrou}}, \bibinfo
  {author} {\bibfnamefont {T.}~\bibnamefont {Muramatsu}}, \bibinfo {author}
  {\bibfnamefont {H.-k.}\ \bibnamefont {Mao}}, \bibinfo {author} {\bibfnamefont
  {C.~J.}\ \bibnamefont {Pickard}}, \bibinfo {author} {\bibfnamefont {R.~J.}\
  \bibnamefont {Needs}}, \bibinfo {author} {\bibfnamefont {V.~B.}\ \bibnamefont
  {Prakapenka}},\ and\ \bibinfo {author} {\bibfnamefont {A.~F.}\ \bibnamefont
  {Goncharov}},\ }\bibfield  {title} {\bibinfo {title} {Synthesis of sodium
  polyhydrides at high pressures},\ }\href
  {https://doi.org/10.1038/ncomms12267} {\bibfield  {journal} {\bibinfo
  {journal} {Nature Communications}\ }\textbf {\bibinfo {volume} {7}},\
  \bibinfo {pages} {12267} (\bibinfo {year} {2016})}\BibitemShut {NoStop}%
\bibitem [{\citenamefont {Wang}\ \emph {et~al.}(2019)\citenamefont {Wang},
  \citenamefont {Yi},\ and\ \citenamefont {Cho}}]{1907.07820}%
  \BibitemOpen
  \bibfield  {author} {\bibinfo {author} {\bibfnamefont {C.}~\bibnamefont
  {Wang}}, \bibinfo {author} {\bibfnamefont {S.}~\bibnamefont {Yi}},\ and\
  \bibinfo {author} {\bibfnamefont {J.-H.}\ \bibnamefont {Cho}},\ }\bibfield
  {title} {\bibinfo {title} {Pressure dependence of the superconducting
  transition temperature of compressed ${\mathrm{lah}}_{10}$},\ }\href
  {https://doi.org/10.1103/PhysRevB.100.060502} {\bibfield  {journal} {\bibinfo
   {journal} {Phys. Rev. B}\ }\textbf {\bibinfo {volume} {100}},\ \bibinfo
  {pages} {060502} (\bibinfo {year} {2019})}\BibitemShut {NoStop}%
\bibitem [{\citenamefont {Bergmann}\ and\ \citenamefont
  {Rainer}(1973)}]{Bergmann1973}%
  \BibitemOpen
  \bibfield  {author} {\bibinfo {author} {\bibfnamefont {G.}~\bibnamefont
  {Bergmann}}\ and\ \bibinfo {author} {\bibfnamefont {D.}~\bibnamefont
  {Rainer}},\ }\bibfield  {title} {\bibinfo {title} {The sensitivity of the
  transition temperature to changes in $\alpha^2f(\omega)$},\ }\href
  {https://doi.org/10.1007/BF02351862} {\bibfield  {journal} {\bibinfo
  {journal} {Zeitschrift f{\"u}r Physik}\ }\textbf {\bibinfo {volume} {263}},\
  \bibinfo {pages} {59} (\bibinfo {year} {1973})}\BibitemShut {NoStop}%
\bibitem [{\citenamefont {Hooper}\ and\ \citenamefont
  {Zurek}(2012)}]{hooper2012potassium}%
  \BibitemOpen
  \bibfield  {author} {\bibinfo {author} {\bibfnamefont {J.}~\bibnamefont
  {Hooper}}\ and\ \bibinfo {author} {\bibfnamefont {E.}~\bibnamefont {Zurek}},\
  }\bibfield  {title} {\bibinfo {title} {High pressure potassium polyhydrides:
  A chemical perspective},\ }\href {https://doi.org/10.1021/jp303024h}
  {\bibfield  {journal} {\bibinfo  {journal} {The Journal of Physical Chemistry
  C}\ }\textbf {\bibinfo {volume} {116}},\ \bibinfo {pages} {13322} (\bibinfo
  {year} {2012})}\BibitemShut {NoStop}%
\bibitem [{\citenamefont {Geballe}\ \emph {et~al.}(2018)\citenamefont
  {Geballe}, \citenamefont {Liu}, \citenamefont {Mishra}, \citenamefont
  {Ahart}, \citenamefont {Somayazulu}, \citenamefont {Meng}, \citenamefont
  {Baldini},\ and\ \citenamefont {Hemley}}]{geballe2018lanthanum}%
  \BibitemOpen
  \bibfield  {author} {\bibinfo {author} {\bibfnamefont {Z.~M.}\ \bibnamefont
  {Geballe}}, \bibinfo {author} {\bibfnamefont {H.}~\bibnamefont {Liu}},
  \bibinfo {author} {\bibfnamefont {A.~K.}\ \bibnamefont {Mishra}}, \bibinfo
  {author} {\bibfnamefont {M.}~\bibnamefont {Ahart}}, \bibinfo {author}
  {\bibfnamefont {M.}~\bibnamefont {Somayazulu}}, \bibinfo {author}
  {\bibfnamefont {Y.}~\bibnamefont {Meng}}, \bibinfo {author} {\bibfnamefont
  {M.}~\bibnamefont {Baldini}},\ and\ \bibinfo {author} {\bibfnamefont {R.~J.}\
  \bibnamefont {Hemley}},\ }\bibfield  {title} {\bibinfo {title} {Synthesis and
  stability of lanthanum superhydrides},\ }\href
  {https://doi.org/10.1002/anie.201709970} {\bibfield  {journal} {\bibinfo
  {journal} {Angewandte Chemie International Edition}\ }\textbf {\bibinfo
  {volume} {57}},\ \bibinfo {pages} {688} (\bibinfo {year} {2018})}\BibitemShut
  {NoStop}%
\bibitem [{\citenamefont {Liu}\ \emph {et~al.}(2018)\citenamefont {Liu},
  \citenamefont {Naumov}, \citenamefont {Geballe}, \citenamefont {Somayazulu},
  \citenamefont {John},\ and\ \citenamefont {Hemley}}]{liu2018dynamics}%
  \BibitemOpen
  \bibfield  {author} {\bibinfo {author} {\bibfnamefont {H.}~\bibnamefont
  {Liu}}, \bibinfo {author} {\bibfnamefont {I.~I.}\ \bibnamefont {Naumov}},
  \bibinfo {author} {\bibfnamefont {Z.~M.}\ \bibnamefont {Geballe}}, \bibinfo
  {author} {\bibfnamefont {M.}~\bibnamefont {Somayazulu}}, \bibinfo {author}
  {\bibfnamefont {S.~T.}\ \bibnamefont {John}},\ and\ \bibinfo {author}
  {\bibfnamefont {R.~J.}\ \bibnamefont {Hemley}},\ }\bibfield  {title}
  {\bibinfo {title} {Dynamics and superconductivity in compressed lanthanum
  superhydride},\ }\href@noop {} {\bibfield  {journal} {\bibinfo  {journal}
  {Physical Review B}\ }\textbf {\bibinfo {volume} {98}},\ \bibinfo {pages}
  {100102} (\bibinfo {year} {2018})}\BibitemShut {NoStop}%
\bibitem [{\citenamefont {Sukmas}\ \emph {et~al.}(2020)\citenamefont {Sukmas},
  \citenamefont {Tsuppayakorn-aek}, \citenamefont {Pinsook},\ and\
  \citenamefont {Bovornratanaraks}}]{sukmas2020MgCa}%
  \BibitemOpen
  \bibfield  {author} {\bibinfo {author} {\bibfnamefont {W.}~\bibnamefont
  {Sukmas}}, \bibinfo {author} {\bibfnamefont {P.}~\bibnamefont
  {Tsuppayakorn-aek}}, \bibinfo {author} {\bibfnamefont {U.}~\bibnamefont
  {Pinsook}},\ and\ \bibinfo {author} {\bibfnamefont {T.}~\bibnamefont
  {Bovornratanaraks}},\ }\bibfield  {title} {\bibinfo {title}
  {Near-room-temperature superconductivity of mg/ca substituted metal
  hexahydride under pressure},\ }\href
  {https://doi.org/https://doi.org/10.1016/j.jallcom.2020.156434} {\bibfield
  {journal} {\bibinfo  {journal} {Journal of Alloys and Compounds}\ }\textbf
  {\bibinfo {volume} {849}},\ \bibinfo {pages} {156434} (\bibinfo {year}
  {2020})}\BibitemShut {NoStop}%
\bibitem [{\citenamefont {Heil}\ \emph {et~al.}(2019)\citenamefont {Heil},
  \citenamefont {Di~Cataldo}, \citenamefont {Bachelet},\ and\ \citenamefont
  {Boeri}}]{heil2019superconductivity}%
  \BibitemOpen
  \bibfield  {author} {\bibinfo {author} {\bibfnamefont {C.}~\bibnamefont
  {Heil}}, \bibinfo {author} {\bibfnamefont {S.}~\bibnamefont {Di~Cataldo}},
  \bibinfo {author} {\bibfnamefont {G.~B.}\ \bibnamefont {Bachelet}},\ and\
  \bibinfo {author} {\bibfnamefont {L.}~\bibnamefont {Boeri}},\ }\bibfield
  {title} {\bibinfo {title} {Superconductivity in sodalite-like yttrium hydride
  clathrates},\ }\href@noop {} {\bibfield  {journal} {\bibinfo  {journal}
  {Physical Review B}\ }\textbf {\bibinfo {volume} {99}},\ \bibinfo {pages}
  {220502} (\bibinfo {year} {2019})}\BibitemShut {NoStop}%
\bibitem [{\citenamefont {Troyan}\ \emph {et~al.}(2021)\citenamefont {Troyan},
  \citenamefont {Semenok}, \citenamefont {Kvashnin}, \citenamefont {Sadakov},
  \citenamefont {Sobolevskiy}, \citenamefont {Pudalov}, \citenamefont
  {Ivanova}, \citenamefont {Prakapenka}, \citenamefont {Greenberg},
  \citenamefont {Gavriliuk}, \citenamefont {Lyubutin}, \citenamefont
  {Struzhkin}, \citenamefont {Bergara}, \citenamefont {Errea}, \citenamefont
  {Bianco}, \citenamefont {Calandra}, \citenamefont {Mauri}, \citenamefont
  {Monacelli}, \citenamefont {Akashi},\ and\ \citenamefont
  {Oganov}}]{troyan2019synthesis}%
  \BibitemOpen
  \bibfield  {author} {\bibinfo {author} {\bibfnamefont {I.~A.}\ \bibnamefont
  {Troyan}}, \bibinfo {author} {\bibfnamefont {D.~V.}\ \bibnamefont {Semenok}},
  \bibinfo {author} {\bibfnamefont {A.~G.}\ \bibnamefont {Kvashnin}}, \bibinfo
  {author} {\bibfnamefont {A.~V.}\ \bibnamefont {Sadakov}}, \bibinfo {author}
  {\bibfnamefont {O.~A.}\ \bibnamefont {Sobolevskiy}}, \bibinfo {author}
  {\bibfnamefont {V.~M.}\ \bibnamefont {Pudalov}}, \bibinfo {author}
  {\bibfnamefont {A.~G.}\ \bibnamefont {Ivanova}}, \bibinfo {author}
  {\bibfnamefont {V.~B.}\ \bibnamefont {Prakapenka}}, \bibinfo {author}
  {\bibfnamefont {E.}~\bibnamefont {Greenberg}}, \bibinfo {author}
  {\bibfnamefont {A.~G.}\ \bibnamefont {Gavriliuk}}, \bibinfo {author}
  {\bibfnamefont {I.~S.}\ \bibnamefont {Lyubutin}}, \bibinfo {author}
  {\bibfnamefont {V.~V.}\ \bibnamefont {Struzhkin}}, \bibinfo {author}
  {\bibfnamefont {A.}~\bibnamefont {Bergara}}, \bibinfo {author} {\bibfnamefont
  {I.}~\bibnamefont {Errea}}, \bibinfo {author} {\bibfnamefont
  {R.}~\bibnamefont {Bianco}}, \bibinfo {author} {\bibfnamefont
  {M.}~\bibnamefont {Calandra}}, \bibinfo {author} {\bibfnamefont
  {F.}~\bibnamefont {Mauri}}, \bibinfo {author} {\bibfnamefont
  {L.}~\bibnamefont {Monacelli}}, \bibinfo {author} {\bibfnamefont
  {R.}~\bibnamefont {Akashi}},\ and\ \bibinfo {author} {\bibfnamefont {A.~R.}\
  \bibnamefont {Oganov}},\ }\bibfield  {title} {\bibinfo {title} {Anomalous
  high-temperature superconductivity in yh6},\ }\href
  {https://onlinelibrary.wiley.com/doi/abs/10.1002/adma.202006832} {\bibfield
  {journal} {\bibinfo  {journal} {Advanced Materials}\ }\textbf {\bibinfo
  {volume} {33}},\ \bibinfo {pages} {2006832} (\bibinfo {year}
  {2021})}\BibitemShut {NoStop}%
\bibitem [{\citenamefont {Li}\ \emph {et~al.}(2015)\citenamefont {Li},
  \citenamefont {Hao}, \citenamefont {Liu}, \citenamefont {John}, \citenamefont
  {Wang},\ and\ \citenamefont {Ma}}]{li2015pressure}%
  \BibitemOpen
  \bibfield  {author} {\bibinfo {author} {\bibfnamefont {Y.}~\bibnamefont
  {Li}}, \bibinfo {author} {\bibfnamefont {J.}~\bibnamefont {Hao}}, \bibinfo
  {author} {\bibfnamefont {H.}~\bibnamefont {Liu}}, \bibinfo {author}
  {\bibfnamefont {S.~T.}\ \bibnamefont {John}}, \bibinfo {author}
  {\bibfnamefont {Y.}~\bibnamefont {Wang}},\ and\ \bibinfo {author}
  {\bibfnamefont {Y.}~\bibnamefont {Ma}},\ }\bibfield  {title} {\bibinfo
  {title} {Pressure-stabilized superconductive yttrium hydrides},\ }\href@noop
  {} {\bibfield  {journal} {\bibinfo  {journal} {Scientific reports}\ }\textbf
  {\bibinfo {volume} {5}},\ \bibinfo {pages} {9948} (\bibinfo {year}
  {2015})}\BibitemShut {NoStop}%
\bibitem [{\citenamefont {Qian}\ \emph {et~al.}(2017)\citenamefont {Qian},
  \citenamefont {Sheng}, \citenamefont {Yan}, \citenamefont {Chen},\ and\
  \citenamefont {Song}}]{qian2017theoretical}%
  \BibitemOpen
  \bibfield  {author} {\bibinfo {author} {\bibfnamefont {S.}~\bibnamefont
  {Qian}}, \bibinfo {author} {\bibfnamefont {X.}~\bibnamefont {Sheng}},
  \bibinfo {author} {\bibfnamefont {X.}~\bibnamefont {Yan}}, \bibinfo {author}
  {\bibfnamefont {Y.}~\bibnamefont {Chen}},\ and\ \bibinfo {author}
  {\bibfnamefont {B.}~\bibnamefont {Song}},\ }\bibfield  {title} {\bibinfo
  {title} {Theoretical study of stability and superconductivity of sch n (n=
  4--8) at high pressure},\ }\href@noop {} {\bibfield  {journal} {\bibinfo
  {journal} {Physical Review B}\ }\textbf {\bibinfo {volume} {96}},\ \bibinfo
  {pages} {094513} (\bibinfo {year} {2017})}\BibitemShut {NoStop}%
\bibitem [{\citenamefont {Zurek}\ \emph {et~al.}(2009)\citenamefont {Zurek},
  \citenamefont {Hoffmann}, \citenamefont {Ashcroft}, \citenamefont {Oganov},\
  and\ \citenamefont {Lyakhov}}]{zurek2009lithium}%
  \BibitemOpen
  \bibfield  {author} {\bibinfo {author} {\bibfnamefont {E.}~\bibnamefont
  {Zurek}}, \bibinfo {author} {\bibfnamefont {R.}~\bibnamefont {Hoffmann}},
  \bibinfo {author} {\bibfnamefont {N.~W.}\ \bibnamefont {Ashcroft}}, \bibinfo
  {author} {\bibfnamefont {A.~R.}\ \bibnamefont {Oganov}},\ and\ \bibinfo
  {author} {\bibfnamefont {A.~O.}\ \bibnamefont {Lyakhov}},\ }\bibfield
  {title} {\bibinfo {title} {A little bit of lithium does a lot for hydrogen},\
  }\href {https://doi.org/10.1073/pnas.0908262106} {\bibfield  {journal}
  {\bibinfo  {journal} {Proceedings of the National Academy of Sciences}\
  }\textbf {\bibinfo {volume} {106}},\ \bibinfo {pages} {17640} (\bibinfo
  {year} {2009})}\BibitemShut {NoStop}%
\bibitem [{\citenamefont {Wang}\ \emph {et~al.}(2015)\citenamefont {Wang},
  \citenamefont {Wang}, \citenamefont {John}, \citenamefont {Iitaka},\ and\
  \citenamefont {Ma}}]{wang2015structural}%
  \BibitemOpen
  \bibfield  {author} {\bibinfo {author} {\bibfnamefont {Y.}~\bibnamefont
  {Wang}}, \bibinfo {author} {\bibfnamefont {H.}~\bibnamefont {Wang}}, \bibinfo
  {author} {\bibfnamefont {S.~T.}\ \bibnamefont {John}}, \bibinfo {author}
  {\bibfnamefont {T.}~\bibnamefont {Iitaka}},\ and\ \bibinfo {author}
  {\bibfnamefont {Y.}~\bibnamefont {Ma}},\ }\bibfield  {title} {\bibinfo
  {title} {Structural morphologies of high-pressure polymorphs of strontium
  hydrides},\ }\href@noop {} {\bibfield  {journal} {\bibinfo  {journal}
  {Physical Chemistry Chemical Physics}\ }\textbf {\bibinfo {volume} {17}},\
  \bibinfo {pages} {19379} (\bibinfo {year} {2015})}\BibitemShut {NoStop}%
\bibitem [{\citenamefont {Errea}\ \emph {et~al.}(2020)\citenamefont {Errea},
  \citenamefont {Belli}, \citenamefont {Monacelli}, \citenamefont {Sanna},
  \citenamefont {Koretsune}, \citenamefont {Tadano}, \citenamefont {Bianco},
  \citenamefont {Calandra}, \citenamefont {Arita}, \citenamefont {Mauri},\ and\
  \citenamefont {Flores-Livas}}]{errea2019}%
  \BibitemOpen
  \bibfield  {author} {\bibinfo {author} {\bibfnamefont {I.}~\bibnamefont
  {Errea}}, \bibinfo {author} {\bibfnamefont {F.}~\bibnamefont {Belli}},
  \bibinfo {author} {\bibfnamefont {L.}~\bibnamefont {Monacelli}}, \bibinfo
  {author} {\bibfnamefont {A.}~\bibnamefont {Sanna}}, \bibinfo {author}
  {\bibfnamefont {T.}~\bibnamefont {Koretsune}}, \bibinfo {author}
  {\bibfnamefont {T.}~\bibnamefont {Tadano}}, \bibinfo {author} {\bibfnamefont
  {R.}~\bibnamefont {Bianco}}, \bibinfo {author} {\bibfnamefont
  {M.}~\bibnamefont {Calandra}}, \bibinfo {author} {\bibfnamefont
  {R.}~\bibnamefont {Arita}}, \bibinfo {author} {\bibfnamefont
  {F.}~\bibnamefont {Mauri}},\ and\ \bibinfo {author} {\bibfnamefont {J.~A.}\
  \bibnamefont {Flores-Livas}},\ }\bibfield  {title} {\bibinfo {title} {Quantum
  crystal structure in the 250-kelvin superconducting lanthanum hydride},\
  }\href {https://doi.org/10.1038/s41586-020-1955-z} {\bibfield  {journal}
  {\bibinfo  {journal} {Nature}\ }\textbf {\bibinfo {volume} {578}},\ \bibinfo
  {pages} {66} (\bibinfo {year} {2020})}\BibitemShut {NoStop}%
\bibitem [{\citenamefont {Errea}\ \emph {et~al.}(2016)\citenamefont {Errea},
  \citenamefont {Calandra}, \citenamefont {Pickard}, \citenamefont {Nelson},
  \citenamefont {Needs}, \citenamefont {Li}, \citenamefont {Liu}, \citenamefont
  {Zhang}, \citenamefont {Ma},\ and\ \citenamefont {Mauri}}]{Errea2016}%
  \BibitemOpen
  \bibfield  {author} {\bibinfo {author} {\bibfnamefont {I.}~\bibnamefont
  {Errea}}, \bibinfo {author} {\bibfnamefont {M.}~\bibnamefont {Calandra}},
  \bibinfo {author} {\bibfnamefont {C.~J.}\ \bibnamefont {Pickard}}, \bibinfo
  {author} {\bibfnamefont {J.~R.}\ \bibnamefont {Nelson}}, \bibinfo {author}
  {\bibfnamefont {R.~J.}\ \bibnamefont {Needs}}, \bibinfo {author}
  {\bibfnamefont {Y.}~\bibnamefont {Li}}, \bibinfo {author} {\bibfnamefont
  {H.}~\bibnamefont {Liu}}, \bibinfo {author} {\bibfnamefont {Y.}~\bibnamefont
  {Zhang}}, \bibinfo {author} {\bibfnamefont {Y.}~\bibnamefont {Ma}},\ and\
  \bibinfo {author} {\bibfnamefont {F.}~\bibnamefont {Mauri}},\ }\bibfield
  {title} {\bibinfo {title} {Quantum hydrogen-bond symmetrization in the
  superconducting hydrogen sulfide system},\ }\href
  {https://doi.org/10.1038/nature17175} {\bibfield  {journal} {\bibinfo
  {journal} {Nature}\ }\textbf {\bibinfo {volume} {532}},\ \bibinfo {pages}
  {81} (\bibinfo {year} {2016})}\BibitemShut {NoStop}%
\bibitem [{\citenamefont {Errea}\ \emph {et~al.}(2014)\citenamefont {Errea},
  \citenamefont {Calandra},\ and\ \citenamefont {Mauri}}]{PhysRevB.89.064302}%
  \BibitemOpen
  \bibfield  {author} {\bibinfo {author} {\bibfnamefont {I.}~\bibnamefont
  {Errea}}, \bibinfo {author} {\bibfnamefont {M.}~\bibnamefont {Calandra}},\
  and\ \bibinfo {author} {\bibfnamefont {F.}~\bibnamefont {Mauri}},\ }\bibfield
   {title} {\bibinfo {title} {Anharmonic free energies and phonon dispersions
  from the stochastic self-consistent harmonic approximation: Application to
  platinum and palladium hydrides},\ }\href
  {https://doi.org/10.1103/PhysRevB.89.064302} {\bibfield  {journal} {\bibinfo
  {journal} {Phys. Rev. B}\ }\textbf {\bibinfo {volume} {89}},\ \bibinfo
  {pages} {064302} (\bibinfo {year} {2014})}\BibitemShut {NoStop}%
\bibitem [{\citenamefont {Hui}\ and\ \citenamefont {Allen}(1974)}]{Hui_1974}%
  \BibitemOpen
  \bibfield  {author} {\bibinfo {author} {\bibfnamefont {J.~C.~K.}\
  \bibnamefont {Hui}}\ and\ \bibinfo {author} {\bibfnamefont {P.~B.}\
  \bibnamefont {Allen}},\ }\bibfield  {title} {\bibinfo {title} {Effect of
  lattice anharmonicity on superconductivity},\ }\href
  {https://doi.org/10.1088/0305-4608/4/3/003} {\bibfield  {journal} {\bibinfo
  {journal} {Journal of Physics F: Metal Physics}\ }\textbf {\bibinfo {volume}
  {4}},\ \bibinfo {pages} {L42} (\bibinfo {year} {1974})}\BibitemShut {NoStop}%
\bibitem [{\citenamefont {Meregalli}\ and\ \citenamefont
  {Savrasov}(1998)}]{PhysRevB.57.14453}%
  \BibitemOpen
  \bibfield  {author} {\bibinfo {author} {\bibfnamefont {V.}~\bibnamefont
  {Meregalli}}\ and\ \bibinfo {author} {\bibfnamefont {S.~Y.}\ \bibnamefont
  {Savrasov}},\ }\bibfield  {title} {\bibinfo {title} {Electron-phonon coupling
  and properties of doped ${\mathrm{babio}}_{3}$},\ }\href
  {https://doi.org/10.1103/PhysRevB.57.14453} {\bibfield  {journal} {\bibinfo
  {journal} {Phys. Rev. B}\ }\textbf {\bibinfo {volume} {57}},\ \bibinfo
  {pages} {14453} (\bibinfo {year} {1998})}\BibitemShut {NoStop}%
\bibitem [{\citenamefont {Wan}\ \emph {et~al.}(2013)\citenamefont {Wan},
  \citenamefont {Ding}, \citenamefont {Savrasov},\ and\ \citenamefont
  {Duan}}]{PhysRevB.87.115124}%
  \BibitemOpen
  \bibfield  {author} {\bibinfo {author} {\bibfnamefont {X.}~\bibnamefont
  {Wan}}, \bibinfo {author} {\bibfnamefont {H.-C.}\ \bibnamefont {Ding}},
  \bibinfo {author} {\bibfnamefont {S.~Y.}\ \bibnamefont {Savrasov}},\ and\
  \bibinfo {author} {\bibfnamefont {C.-G.}\ \bibnamefont {Duan}},\ }\bibfield
  {title} {\bibinfo {title} {Electron-phonon superconductivity near
  charge-density-wave instability in lao${}_{0.5}$f${}_{0.5}$bis${}_{2}$:
  Density-functional calculations},\ }\href
  {https://doi.org/10.1103/PhysRevB.87.115124} {\bibfield  {journal} {\bibinfo
  {journal} {Phys. Rev. B}\ }\textbf {\bibinfo {volume} {87}},\ \bibinfo
  {pages} {115124} (\bibinfo {year} {2013})}\BibitemShut {NoStop}%
\bibitem [{\citenamefont {Xie}\ \emph {et~al.}(2019)\citenamefont {Xie},
  \citenamefont {Duan}, \citenamefont {Shao}, \citenamefont {Song},
  \citenamefont {Wang}, \citenamefont {Xiao}, \citenamefont {Li}, \citenamefont
  {Tian}, \citenamefont {Liu},\ and\ \citenamefont {Cui}}]{xie2019high}%
  \BibitemOpen
  \bibfield  {author} {\bibinfo {author} {\bibfnamefont {H.}~\bibnamefont
  {Xie}}, \bibinfo {author} {\bibfnamefont {D.}~\bibnamefont {Duan}}, \bibinfo
  {author} {\bibfnamefont {Z.}~\bibnamefont {Shao}}, \bibinfo {author}
  {\bibfnamefont {H.}~\bibnamefont {Song}}, \bibinfo {author} {\bibfnamefont
  {Y.}~\bibnamefont {Wang}}, \bibinfo {author} {\bibfnamefont {X.}~\bibnamefont
  {Xiao}}, \bibinfo {author} {\bibfnamefont {D.}~\bibnamefont {Li}}, \bibinfo
  {author} {\bibfnamefont {F.}~\bibnamefont {Tian}}, \bibinfo {author}
  {\bibfnamefont {B.}~\bibnamefont {Liu}},\ and\ \bibinfo {author}
  {\bibfnamefont {T.}~\bibnamefont {Cui}},\ }\bibfield  {title} {\bibinfo
  {title} {High-temperature superconductivity in ternary clathrate ycah12 under
  high pressures},\ }\href@noop {} {\bibfield  {journal} {\bibinfo  {journal}
  {Journal of Physics: Condensed Matter}\ }\textbf {\bibinfo {volume} {31}},\
  \bibinfo {pages} {245404} (\bibinfo {year} {2019})}\BibitemShut {NoStop}%
\bibitem [{\citenamefont {Sun}\ \emph {et~al.}(2019)\citenamefont {Sun},
  \citenamefont {Lv}, \citenamefont {Xie}, \citenamefont {Liu},\ and\
  \citenamefont {Ma}}]{ying2019ternary}%
  \BibitemOpen
  \bibfield  {author} {\bibinfo {author} {\bibfnamefont {Y.}~\bibnamefont
  {Sun}}, \bibinfo {author} {\bibfnamefont {J.}~\bibnamefont {Lv}}, \bibinfo
  {author} {\bibfnamefont {Y.}~\bibnamefont {Xie}}, \bibinfo {author}
  {\bibfnamefont {H.}~\bibnamefont {Liu}},\ and\ \bibinfo {author}
  {\bibfnamefont {Y.}~\bibnamefont {Ma}},\ }\bibfield  {title} {\bibinfo
  {title} {Route to a superconducting phase above room temperature in
  electron-doped hydride compounds under high pressure},\ }\href
  {https://doi.org/10.1103/PhysRevLett.123.097001} {\bibfield  {journal}
  {\bibinfo  {journal} {Phys. Rev. Lett.}\ }\textbf {\bibinfo {volume} {123}},\
  \bibinfo {pages} {097001} (\bibinfo {year} {2019})}\BibitemShut {NoStop}%
\bibitem [{\citenamefont {Szczesniak}\ and\ \citenamefont
  {Durajski}(2016)}]{szczesniak2016superconductivity}%
  \BibitemOpen
  \bibfield  {author} {\bibinfo {author} {\bibfnamefont {R.}~\bibnamefont
  {Szczesniak}}\ and\ \bibinfo {author} {\bibfnamefont {A.}~\bibnamefont
  {Durajski}},\ }\bibfield  {title} {\bibinfo {title} {Superconductivity well
  above room temperature in compressed mgh 6},\ }\href@noop {} {\bibfield
  {journal} {\bibinfo  {journal} {Frontiers of Physics}\ }\textbf {\bibinfo
  {volume} {11}},\ \bibinfo {pages} {117406} (\bibinfo {year}
  {2016})}\BibitemShut {NoStop}%
\bibitem [{\citenamefont {Chen}\ \emph {et~al.}(2017)\citenamefont {Chen},
  \citenamefont {Geng}, \citenamefont {Yan}, \citenamefont {Sun}, \citenamefont
  {Wu},\ and\ \citenamefont {Chen}}]{chen2017prediction}%
  \BibitemOpen
  \bibfield  {author} {\bibinfo {author} {\bibfnamefont {Y.}~\bibnamefont
  {Chen}}, \bibinfo {author} {\bibfnamefont {H.~Y.}\ \bibnamefont {Geng}},
  \bibinfo {author} {\bibfnamefont {X.}~\bibnamefont {Yan}}, \bibinfo {author}
  {\bibfnamefont {Y.}~\bibnamefont {Sun}}, \bibinfo {author} {\bibfnamefont
  {Q.}~\bibnamefont {Wu}},\ and\ \bibinfo {author} {\bibfnamefont
  {X.}~\bibnamefont {Chen}},\ }\bibfield  {title} {\bibinfo {title} {Prediction
  of stable ground-state lithium polyhydrides under high pressures},\
  }\href@noop {} {\bibfield  {journal} {\bibinfo  {journal} {Inorganic
  Chemistry}\ }\textbf {\bibinfo {volume} {56}},\ \bibinfo {pages} {3867}
  (\bibinfo {year} {2017})}\BibitemShut {NoStop}%
\bibitem [{\citenamefont {Camargo-Mart{\'{\i}}nez}\ \emph
  {et~al.}(2020)\citenamefont {Camargo-Mart{\'{\i}}nez}, \citenamefont
  {Gonz{\'{a}}lez-Pedreros},\ and\ \citenamefont
  {Mesa}}]{Camargo_Mart_nez_2020}%
  \BibitemOpen
  \bibfield  {author} {\bibinfo {author} {\bibfnamefont {J.~A.}\ \bibnamefont
  {Camargo-Mart{\'{\i}}nez}}, \bibinfo {author} {\bibfnamefont {G.~I.}\
  \bibnamefont {Gonz{\'{a}}lez-Pedreros}},\ and\ \bibinfo {author}
  {\bibfnamefont {F.}~\bibnamefont {Mesa}},\ }\bibfield  {title} {\bibinfo
  {title} {The higher superconducting transition temperature \uppercase{T}$_c$
  and the functional derivative of \uppercase{T}$_c$ with
  $\alpha^2$\uppercase{F}($\omega$) for electron{\textendash}phonon
  superconductors},\ }\href {https://doi.org/10.1088/1361-648x/abb741}
  {\bibfield  {journal} {\bibinfo  {journal} {Journal of Physics: Condensed
  Matter}\ }\textbf {\bibinfo {volume} {32}},\ \bibinfo {pages} {505901}
  (\bibinfo {year} {2020})}\BibitemShut {NoStop}%
\bibitem [{\citenamefont {Wei}\ \emph {et~al.}(2016)\citenamefont {Wei},
  \citenamefont {Yuan}, \citenamefont {Khan}, \citenamefont {Ji}, \citenamefont
  {Gu},\ and\ \citenamefont {Wei}}]{wei2016pressure}%
  \BibitemOpen
  \bibfield  {author} {\bibinfo {author} {\bibfnamefont {Y.-K.}\ \bibnamefont
  {Wei}}, \bibinfo {author} {\bibfnamefont {J.-N.}\ \bibnamefont {Yuan}},
  \bibinfo {author} {\bibfnamefont {F.~I.}\ \bibnamefont {Khan}}, \bibinfo
  {author} {\bibfnamefont {G.-F.}\ \bibnamefont {Ji}}, \bibinfo {author}
  {\bibfnamefont {Z.-W.}\ \bibnamefont {Gu}},\ and\ \bibinfo {author}
  {\bibfnamefont {D.-Q.}\ \bibnamefont {Wei}},\ }\bibfield  {title} {\bibinfo
  {title} {Pressure induced superconductivity and electronic structure
  properties of scandium hydrides using first principles calculations},\
  }\href@noop {} {\bibfield  {journal} {\bibinfo  {journal} {RSC Advances}\
  }\textbf {\bibinfo {volume} {6}},\ \bibinfo {pages} {81534} (\bibinfo {year}
  {2016})}\BibitemShut {NoStop}%
\bibitem [{\citenamefont {Smith}\ \emph {et~al.}(2009)\citenamefont {Smith},
  \citenamefont {Desgreniers}, \citenamefont {Klug},\ and\ \citenamefont
  {John}}]{smith2009high}%
  \BibitemOpen
  \bibfield  {author} {\bibinfo {author} {\bibfnamefont {J.~S.}\ \bibnamefont
  {Smith}}, \bibinfo {author} {\bibfnamefont {S.}~\bibnamefont {Desgreniers}},
  \bibinfo {author} {\bibfnamefont {D.~D.}\ \bibnamefont {Klug}},\ and\
  \bibinfo {author} {\bibfnamefont {S.~T.}\ \bibnamefont {John}},\ }\bibfield
  {title} {\bibinfo {title} {High-density strontium hydride: An experimental
  and theoretical study},\ }\href@noop {} {\bibfield  {journal} {\bibinfo
  {journal} {Solid state communications}\ }\textbf {\bibinfo {volume} {149}},\
  \bibinfo {pages} {830} (\bibinfo {year} {2009})}\BibitemShut {NoStop}%
\bibitem [{\citenamefont {Shao}\ \emph {et~al.}(2019)\citenamefont {Shao},
  \citenamefont {Duan}, \citenamefont {Ma}, \citenamefont {Yu}, \citenamefont
  {Song}, \citenamefont {Xie}, \citenamefont {Li}, \citenamefont {Tian},
  \citenamefont {Liu},\ and\ \citenamefont {Cui}}]{shao2019unique}%
  \BibitemOpen
  \bibfield  {author} {\bibinfo {author} {\bibfnamefont {Z.}~\bibnamefont
  {Shao}}, \bibinfo {author} {\bibfnamefont {D.}~\bibnamefont {Duan}}, \bibinfo
  {author} {\bibfnamefont {Y.}~\bibnamefont {Ma}}, \bibinfo {author}
  {\bibfnamefont {H.}~\bibnamefont {Yu}}, \bibinfo {author} {\bibfnamefont
  {H.}~\bibnamefont {Song}}, \bibinfo {author} {\bibfnamefont {H.}~\bibnamefont
  {Xie}}, \bibinfo {author} {\bibfnamefont {D.}~\bibnamefont {Li}}, \bibinfo
  {author} {\bibfnamefont {F.}~\bibnamefont {Tian}}, \bibinfo {author}
  {\bibfnamefont {B.}~\bibnamefont {Liu}},\ and\ \bibinfo {author}
  {\bibfnamefont {T.}~\bibnamefont {Cui}},\ }\bibfield  {title} {\bibinfo
  {title} {Unique phase diagram and superconductivity of calcium hydrides at
  high pressures},\ }\href@noop {} {\bibfield  {journal} {\bibinfo  {journal}
  {Inorganic chemistry}\ }\textbf {\bibinfo {volume} {58}},\ \bibinfo {pages}
  {2558} (\bibinfo {year} {2019})}\BibitemShut {NoStop}%
\bibitem [{\citenamefont {Xie}\ \emph {et~al.}(2020)\citenamefont {Xie},
  \citenamefont {Yao}, \citenamefont {Feng}, \citenamefont {Duan},
  \citenamefont {Song}, \citenamefont {Zhang}, \citenamefont {Jiang},
  \citenamefont {Redfern}, \citenamefont {Kresin}, \citenamefont {Pickard},\
  and\ \citenamefont {Cui}}]{PhysRevLett.125.217001}%
  \BibitemOpen
  \bibfield  {author} {\bibinfo {author} {\bibfnamefont {H.}~\bibnamefont
  {Xie}}, \bibinfo {author} {\bibfnamefont {Y.}~\bibnamefont {Yao}}, \bibinfo
  {author} {\bibfnamefont {X.}~\bibnamefont {Feng}}, \bibinfo {author}
  {\bibfnamefont {D.}~\bibnamefont {Duan}}, \bibinfo {author} {\bibfnamefont
  {H.}~\bibnamefont {Song}}, \bibinfo {author} {\bibfnamefont {Z.}~\bibnamefont
  {Zhang}}, \bibinfo {author} {\bibfnamefont {S.}~\bibnamefont {Jiang}},
  \bibinfo {author} {\bibfnamefont {S.~A.~T.}\ \bibnamefont {Redfern}},
  \bibinfo {author} {\bibfnamefont {V.~Z.}\ \bibnamefont {Kresin}}, \bibinfo
  {author} {\bibfnamefont {C.~J.}\ \bibnamefont {Pickard}},\ and\ \bibinfo
  {author} {\bibfnamefont {T.}~\bibnamefont {Cui}},\ }\bibfield  {title}
  {\bibinfo {title} {Hydrogen pentagraphenelike structure stabilized by
  hafnium: A high-temperature conventional superconductor},\ }\href
  {https://doi.org/10.1103/PhysRevLett.125.217001} {\bibfield  {journal}
  {\bibinfo  {journal} {Phys. Rev. Lett.}\ }\textbf {\bibinfo {volume} {125}},\
  \bibinfo {pages} {217001} (\bibinfo {year} {2020})}\BibitemShut {NoStop}%
\bibitem [{\citenamefont {Hooper}\ \emph {et~al.}(2014)\citenamefont {Hooper},
  \citenamefont {Terpstra}, \citenamefont {Shamp},\ and\ \citenamefont
  {Zurek}}]{hooper2014composition}%
  \BibitemOpen
  \bibfield  {author} {\bibinfo {author} {\bibfnamefont {J.}~\bibnamefont
  {Hooper}}, \bibinfo {author} {\bibfnamefont {T.}~\bibnamefont {Terpstra}},
  \bibinfo {author} {\bibfnamefont {A.}~\bibnamefont {Shamp}},\ and\ \bibinfo
  {author} {\bibfnamefont {E.}~\bibnamefont {Zurek}},\ }\bibfield  {title}
  {\bibinfo {title} {Composition and constitution of compressed strontium
  polyhydrides},\ }\href@noop {} {\bibfield  {journal} {\bibinfo  {journal}
  {The Journal of Physical Chemistry C}\ }\textbf {\bibinfo {volume} {118}},\
  \bibinfo {pages} {6433} (\bibinfo {year} {2014})}\BibitemShut {NoStop}%
\end{thebibliography}%
\end{document}